\documentclass[preprint2]{aastex6}

\usepackage{ulem}
\usepackage{xcolor}
\usepackage{amsmath}
\usepackage{cases}
\usepackage[T1]{fontenc}
\usepackage[figuresright]{rotating}

\begin{document}
\title{Constraining the circumburst medium of gamma-ray bursts with X-ray afterglows}
\author{Xiao Tian\altaffilmark{1}, Ying Qin\altaffilmark{2}, Mei Du\altaffilmark{1}, Shuang-Xi Yi\altaffilmark{1} and Yan-Ke Tang\altaffilmark{3}}
\altaffiltext{1}{School of Physics and Physical Engineering, Qufu Normal University, Qufu 273165, China; yisx2015@qfnu.edu.cn}
\altaffiltext{2}{Department of Physics, Anhui Normal University, Wuhu 241000, China; yingqin2013@hotmail.com}
\altaffiltext{3}{College of Physics and Electronic, Dezhou University, Dezhou 253023, China.}

\begin{abstract}
Long gamma-ray bursts (GRBs) are considered to be originated from core collapse of massive stars. It is believed that the afterglow property is determined by the density of the material in the surrounding interstellar medium. Therefore, the circumburst density can be used to distinguish between an interstellar wind $n(R) \propto R^{\rm -k}$, and a constant density medium (ISM), $n(R) = const$. Previous studies with different afterglow samples, show that the circumburst medium of GRBs is neither simply supported by an interstellar wind, nor completely favored by an ISM. In this work, our new sample is consisted of 39 GRBs with smoothly onset bump-like features in early X-ray afterglows, in which 20 GRBs have the redshift measurements. By using a smooth broken power law function to fit the bumps of X-ray light curves, we derive the full width at half-maximum (FWHM) as the feature width ($\omega$), as well as the rise and decay time scales of the bumps ($T_{\rm r}$ and $T_{\rm d}$). The correlations between the timescales of X-ray bumps are similar to those found previously in the optical afterglows. Based on the fireball forward shock (FS) model of the thin shell case, we obtain the distribution of the electron spectral index $p$, and further constrain the medium density distribution index $k$. The new inferred $k$ is found to be concentrated at 1.0, with a range from 0.2 to 1.8. This finding is consistent with previous studies. The conclusion of our detailed investigation for X-ray afterglows suggests that the ambient medium of the selected GRBs is not homogeneous, i.e., neither ISM nor the typical interstellar wind.
\end{abstract}

\section{Introduction}
\label{Sec:A: Introduction}

Gamma-ray bursts (GRBs) are the brightest phenomena in the universe. With more observational detectors lunched successfully, the understanding of the origins for GRBs is becoming more and more profound (\citealt{2004ApJ...611.1005G, 2005SSRv..120..165B, 2005SSRv..120...95R}; \citealt{2009ApJ...697.1071A}; \citealt{2009ApJ...702..791M}; \citealt{2020SSPMA..50l9508L}; \citealt{2020SCPMA..6349502Z}). One of the most popular model of the dominant mechanism for GRBs is the fireball model \citep{2002ARA&A..40..137M,2004IJMPA..19.2385Z,2004RvMP...76.1143P}, by which the prompt $\gamma$-ray emission can be explained by the internal dissipative processes occurring in relativistic ejecta or due to internal shocks \citep{1993ApJ...405..278M,1994ApJ...430L..93R}, and by which the multi-band afterglow emission is produced by the synchrotron emission of the external shock when the fireball is decelerated by the ambient medium \citep{1997ApJ...476..232M,1998ApJ...497L..17S,2018ApJ...859..160W,2021ApJ...908..242D}.

GRBs, based on the prompt burst duration $T_{90}$, can be divided into two groups: long GRBs ($T_{\rm 90} > 2s$) and short GRBs ($T_{\rm 90} < 2s$ ) \citep{2013ApJ...763...15Q,2020ApJ...893...77W}. Studies suggest that short GRBs are originated from the merger of binary compact stars, i.e., two neutron stars or neutron star-black hole \citep{2002ApJ...571..779P}, while long GRBs are related to core collapse of massive stars \citep{1993ApJ...405..273W, 1999ApJ...524..262M}. In this standard collapsar model, massive, fast rotating Wolf-Rayet (WR) stars are believed to be the progenitors of long GRBs. In order to maintain enough angular momentum, one has to consider WR stars with a compact companion in a close binary system, or at a low metallicity for a lower mass loss rate. In case of the first scenario, the angular momentum content of a WR is determined by tides and its stellar winds. In this context, whether a WR star with a compact companion in a close binary system can be spun up to form a collapsar or not, has been systematically investigated \citep{2007A&A...465L..29C,2008A&A...484..831D,2017NewAR..79....1L,2018A&A...616A..28Q,2018ApJ...852...20L}. Alternatively, WR stars as the progenitors of long GRBs are single, massive stars, which must be born in a metal-poor environment to avoid losing a larger number of angular momentum through strong winds. The ejected matter from massive WRs through winds thus can enhance chemical enrichment of the host galaxies, which further makes the circumferential material density more complex. In the context of the second scenario, when the fireball shell slows down in collision with the circumburst medium, two types of shock waves will be generated: reverse shock (RS) and forward shock (FS). The RS pierces into the fireball ejecta, while the FS propagates into the circumburst medium. The FS model predicts that when the fireball shell is decelerated by the circumferential environment, the afterglow light curves will show a smooth onset bump. A large number of early optical afterglows with smooth onset bumps have been collected in the $Swift$ era, which are in a good agreement with the predictions by the external FS model \citep{1999ApJ...520..641S,2007A&A...469L..13M,2009ApJ...702..489R,2010ApJ...725.2209L,2013ApJ...774...13L,2013ApJ...776..120Y,2020ApJ...900..112Z}. Based on the external FS model, it is believed that the temporal and spectral index of the multi-band afterglows can be used to constrain the radiation mechanism and the properties of the circumburst medium (\citealt{1998ApJ...497L..17S}; \citealt{2020ApJ...900..176L}).

\cite{1998MNRAS.298...87D}, assuming that the expansion of the fireball is adiabatic and that the density in the medium varies as a power-law function of shock radius, $n \propto R^{\rm -k}$, found that the X-ray afterglow of GRB 970616 is well fitted by assuming $k = 2$. By fitting the simultaneous spectra of X-ray, optical, and NIR afterglow data of 10 $BeppoSAX$ GRBs, \cite{2008ApJ...672..433S} first investigated the density structure index of the circumburst dedium. It was found that $k$ is well constrained for five GRBs, in four of which the circumburst medium is interstellar wind medium, i.e., $k$ = 2. The fifth source has $k$ in the range of $0 - 1$, consistent with ISM. With the available optical and $Swift$/X-ray late-time afterglow, \cite{2011A&A...526A..23S} found that 24\% is in favor of the wind medium, while the rest (except for the short GRBs 051221A and 060904B) supports ISM. Then a following investigation for a subsample of GRB optical afterglow with onset bump features shows the index $k$ is centered at 1 with a range of $0.4 - 1.4$ \citep{2013ApJ...776..120Y}. This result further indicates that the GRBs ambient medium is neither ISM nor stellar wind environment, which is also consistent with the results of \cite{2013ApJ...774...13L}.

Recently, \cite{2020ApJ...895...94Y} also collected some optical afterglows with significant RS features, and inferred the $k$ value in the range of $0 - 1.5$ under the assumption of the external RS model. Therefore, GRBs may have different type ambient medium. Interestingly, we find that some GRB X-ray afterglows show the onset deceleration feature. It is believed that the X-ray afterglows with smooth onset bumps, can be explained by the external FS model. Hence the investigation of the circumferential environment for GRBs with X-ray onset bump is needed. Furthermore, the X-ray afterglow is simpler when compared with the optical afterglow, and it can be clearly described in five different segments \citep{2006ApJ...642..354Z,2006ApJ...642..389N}.

Following the previous works, we here continue to constrain GRBs circumburst medium with X-ray afterglows. The paper is organized as follows. We introduce the selection criteria of X-ray onset bump in Section 2. In the Section 3, we present the characteristics of X-ray bump and their correlations. The results for the application of the external FS model to our selected X-ray sample are shown in Section 4. We finally give our Dicussions and Conclusions in Sections 5 and 6, respectively.

\section{SAMPLE SELECTION AND LIGHT CURVE FITTING}
\label{Sec: BPdistribution}

The optical onset afterglows are considered as a good probe of testing the fireball external shock model since it is generally less polluted by the prompt emission tail or by the later emission of engine activity \citep{2009MNRAS.395..490O,2010ApJ...725.2209L,2013ApJ...774...13L,2012MNRAS.420..483G,2017JHEAp..13....1Y,2021MNRAS.507.1047Y,2021arXiv211101041Y}. In general, an onset bump is made of two components, i.e., a slow rise and decay. \cite{2010ApJ...725.2209L} found that the rise slope is less than 3, typically around 1 and 2. However, for the X-ray afterglows, some emissions from GRB central engine can appear in early time, including the erratic flares, plateaus or the GRB tail emission. The deceleration features in X-ray afterglows can be affected by those emission components, which are related to the activity from the central engine. Hence the sample of X-ray light curves with a clear deceleration onset bump signature is limited.

We first show three criteria for collecting the X-ray sample with the onset bump features: (1) a complete component of rise and decay feature in X-ray afterglows; (2) the rise slope less than 3; and (3) without the superposition of erratic flares or plateaus in the smooth bumps. Our target sample is consisted of 39 GRBs with a smooth bump feature in X-ray afterglows, collected from the official website of $Swift$ \footnote{http://www.swift.ac.uk/xrt\_curves/.} in the time period of July 2005 - October 2020. In order to derive the temporal parameters of X-ray afterglow onset bumps, we follow the previous work (\citealt{2016ApJS..224...20Y,2018ApJ...863...50S}) and use an empirical smooth broken power-law function (SBPL, \citealt{1999A&A...352L..26B}) to fit X-ray bumps,

\begin{equation}
\\F_{\rm model}(t) = F_0 [(\frac{{{t}}}{{{T_{\rm p}}}})^{\alpha_{1}{\rm b}}+(\frac{{{t}}}{{{T_{\rm p}}}})^{\alpha_{2}{\rm b}}]^{-\frac{1}{\rm b}},
\label{Eq: BP_SNe}
\end{equation}
where $\alpha_1$ and $\alpha_2$ are the rise and decay indexes, respectively; $F_{\rm p}=F_{\rm 0} \,2^{\rm -1/b}$ is the flux of peak time and $b$ represents the peak sharpness of the
light-curve components. We take 3 as a typical value of the parameter $b$. Some X-ray afterglows are usually combined of several power-law segments along with some X-ray flares, or steep decays. Therefore, we only fit the data around the X-ray onset bumps, without including mixed components. The fitting results of selected GRBs are showed in Figure 1, and the corresponding fitting parameters for each GRB are presented in Tables 1 and 2, respectively. The data of redshift ($z$), duration ($T_{\rm 90}$) and the photon index ($\Gamma$) are collected from the official website of $Swift$ and are listed in Tables 1 and 2.

\begin{table*}[thp]\footnotesize%
\tiny
  \caption{Properties of the GRB Sample with Well-sampled X-ray Light Curves   ($\nu>max\{\nu_c^f,\nu_m^f\}$) \label{Tab: sample}}
  \setlength{\tabcolsep}{0.5mm}{
 \begin{tabular}{ccccccccccccccc}
  \hline
  \hline
    GRB   & $z$&${T_{\rm 90}}^b$& $\Gamma$ &${F_{\rm p}}^a$ &  $\alpha_1$  & $\alpha_2$ & ${T_{\rm p}}^b$&$\chi^2/dof$ &$\omega^b$  &${T_{\rm r}}^b$ & ${T_{\rm d}}^b$    &$k$  & $p$    \\
  \hline
    050713B &  ...  & 54.2  & 1.94$\pm$0.12 & 1.55$\pm$0.09         & 0.30$\pm$0.09 & -0.98$\pm$0.02 & 9090$\pm$780 &   162/109        & 22488$\pm$3766 & 6951$\pm$1171 & 15537$\pm$3745 & 1.71$\pm$0.09 & 1.97$\pm$0.03 \\
    070103 & 2.6208 & 18.6 &2.10$\pm$0.27   &  1.81$\pm$0.17        & 0.55$\pm$0.16 & -1.32$\pm$0.05 & 905$\pm$124 &     38/27             & 1673$\pm$447 & 610$\pm$165 & 1063$\pm$450 & 1.31$\pm$0.14 & 2.43$\pm$0.06 \\
    070208 & 1.165  & 48    &2.05$\pm$0.19 &   2.81$\pm$0.47       &1.09$\pm$0.25   & -1.29$\pm$0.06 & 968$\pm$130 &    53/29        & 2013$\pm$639 & 608$\pm$174 &1405$\pm$642 & 0.83$\pm$0.23 & 2.39$\pm$0.08 \\
    080409 & ...    &  20.2 &1.99$\pm$0.39 &   0.90$\pm$0.14        &0.50$\pm$0.34 & -0.96$\pm$0.07 & 548$\pm$171 &       12/9        &1303$\pm$772  &349$\pm$230 &953$\pm$776 & 1.52$\pm$0.34 & 1.95$\pm$0.09 \\
    100901A & 1.408 &  439  &2.11$\pm$0.06 &   1.13$\pm$0.03        & 0.80$\pm$0.07 & -1.45$\pm$0.03 & 26437$\pm$893 & 9457/480       & 43367$\pm$3301 & 13101$\pm$1363 & 30266$\pm$3261 & 1.04$\pm$0.06 & 2.60$\pm$0.04 \\
    120118B & 2.943 & 23.26 &1.98$\pm$0.15 &   1.13$\pm$0.08        & 0.79$\pm$0.20& -0.91$\pm$0.08 & 1508$\pm$256 &    310/41     &3715$\pm$1215 & 804$\pm$332 & 2911$\pm$1224    &1.25$\pm$0.21 & 1.88$\pm$0.11 \\
    120215A & ...   & 26.5  & 2.22$\pm$0.20 &    0.15$\pm$0.02      &1.30$\pm$0.42 & -0.90$\pm$0.06 & 11704$\pm$1601 &   351/45    & 32806$\pm$9547 & 5655$\pm$2221 & 27151$\pm$9557 & 0.72$\pm$0.44 & 1.87$\pm$0.07 \\
    120224A & 1.1   &  8.13 &2.10$\pm$0.13 &  2.76$\pm$0.13         & 0.80$\pm$0.11& -1.00$\pm$0.03 & 1379$\pm$121 &    167/80         &3865$\pm$623 & 953$\pm$144 & 2911$\pm$630& 1.20$\pm$0.11 & 2.00$\pm$0.03 \\
    121209A & 2.1   & 42.7  &2.11$\pm$0.13 &   15.79$\pm$1.96       & 0.92$\pm$0.21& -1.23$\pm$0.03 & 752$\pm$81 &   134/89         &1523$\pm$390 & 460$\pm$115& 1063$\pm$390& 1.00$\pm$0.20 & 2.31$\pm$0.05 \\
     ${130807A}^c$ & ...   &       &2.40$\pm$0.50 &   0.27$\pm$0.03        &1.60$\pm$0.68 & -0.93$\pm$0.06 & 9067$\pm$1721 & 7651/190 &24816$\pm$10167 & 4278$\pm$2361 & 20539$\pm$10185& 0.41$\pm$0.69 & 1.91$\pm$0.08 \\
    140719A & ...   & 48    &2.16$\pm$0.20 &    0.75$\pm$0.19     &0.83$\pm$0.37 & -0.99$\pm$0.05 & 2456$\pm$673&     106/35        &6754$\pm$4076 & 1666$\pm$853 & 5088$\pm$4097 & 1.17$\pm$0.37 & 1.99$\pm$0.06 \\
    151112A & 4.1   &19.32  &2.16$\pm$0.16 &    0.59$\pm$0.04     &1.57$\pm$0.56 & -1.04$\pm$0.04 & 5118$\pm$382 &     60/45        &9945$\pm$1790 & 2448$\pm$626 & 7497$\pm$1762& 0.42$\pm$0.55 & 2.05$\pm$0.05 \\
    151114A & ...   &4.86   & 2.01$\pm$0.20 &     2.05$\pm$0.33     &0.47$\pm$0.35 & -1.12$\pm$0.06 & 439$\pm$90 &    14/12      & 1007$\pm$432 & 285$\pm$148 & 721$\pm$426 & 1.47$\pm$0.34 & 2.16$\pm$0.08 \\
    171123A &  ...  & 58.5  & 2.06$\pm$0.23 &  0.40$\pm$0.03       &0.77$\pm$0.29 & -0.75$\pm$0.09 & 1800$\pm$463 &   90/23          &6951$\pm$3082& 1260$\pm$531& 5691$\pm$3106 & 1.34$\pm$0.32 & 1.67$\pm$0.12 \\
    190422A &  ... &   212  & 2.09$\pm$0.18 &    3.94$\pm$0.29     &1.11$\pm$0.24 & -0.79$\pm$0.04 & 2993$\pm$291 &    154/87       &8583 $\pm$1909 & 1857$\pm$377 & 6726$\pm$1916 & 0.96$\pm$0.26 & 1.72$\pm$0.05 \\
    190828B &  ... & 66.6   & 2.03$\pm$0.11 &   11.09$\pm$0.56      &0.91$\pm$0.17 & -1.09$\pm$0.02 & 694$\pm$47 &     1898/116         &1608$\pm$236 & 348$\pm$71 & 1260$\pm$ 235& 1.06$\pm$0.17 & 2.12$\pm$0.02 \\
    200409A &  ... & 17.91  & 2.06$\pm$0.17 &      2.30$\pm$0.28    &0.70$\pm$0.41 & -0.89$\pm$0.03 & 300$\pm$60 &     46/29         & 985$\pm$383 & 179$\pm$85 & 807$\pm$383& 1.35$\pm$0.42 & 1.85$\pm$0.04 \\
    200925B &  ... &  18.25 & 2.13$\pm$0.19 &      2.47$\pm$0.21    &0.47$\pm$0.17 &-1.04$\pm$0.04 & 1000$\pm$167 &      123/27           &2325$\pm$722& 659$\pm$222 & 1666$\pm$727& 1.51$\pm$0.16 & 2.05$\pm$0.05 \\
  \hline
\end{tabular}
}
\tablecomments{\\$^a$ $F_{\rm p}$ is the peak flux for the bump, and in units of $10^{-11}$ $erg$ $cm^{-2}$ $s^{-1}$. \\
$^b$  In units of seconds.\\
$^c$  The $T_{\rm 90}$ of GRB 130807A is not available, but GRB 130807A can be identified as a long burst from Swift/BAT light curve. }
\end{table*}

\begin{table*}[thp]\footnotesize
\tiny
  \caption{Properties of the GRB Sample with Well-sampled X-ray Light Curves ($\nu_m^f<\nu<\nu_c^f$) \label{Tab: sample}}
  \setlength{\tabcolsep}{0.5mm}{
  \begin{center}
  \begin{tabular}{ccccccccccccccc}
  \hline
  \hline
    GRB   & $z$& ${T_{\rm 90}}^b$ &$\Gamma$ &${F_{\rm p}}^a$ &  $\alpha_1$  & $\alpha_2$ & ${T_{\rm p}}^b$&$\chi^2/dof$ &$\omega^b$  &${T_{\rm r}}^b$ & ${T_{\rm d}}^b$    &$k$  & $p$    \\
  \hline
    060804 &  ...  & 17.8  & 1.89$\pm$0.19 &  8.83$\pm$0.71      &0.85$\pm$0.32 & -1.10$\pm$0.03 & 419$\pm$56 &   54/29              & 920$\pm$255  & 199$\pm$83 & 721$\pm$253  & 1.20$\pm$0.18 & 2.18$\pm$0.08 \\
    070714A & 1.58 &  2     & 2.05$\pm$0.34 &  1.56$\pm$0.16         &2.05$\pm$0.77 & -0.86$\pm$0.05 & 234$\pm$32 &    19/10         &659$\pm$208 & 113$\pm$45 & 545.5$\pm$208   & 0.54$\pm$0.44 & 2.04$\pm$0.12 \\
    071010A & 0.98  &  6     &2.18$\pm$0.34 &   0.14$\pm$0.02        &2.10$\pm$0.91 & -2.00$\pm$0.15 & 61364$\pm$5918 &  17/16       & 82915$\pm$22557 & 30178$\pm$9884& 52737$\pm$21935  & 0.42$\pm$0.42 & 3.59$\pm$0.22 \\
    080307 &...    & 125.9   &1.71$\pm$0.20 &  66.39$\pm$2.09      & 0.80$\pm$0.10 & -2.00$\pm$0.03 & 270$\pm$7         &209/157     &377$\pm$22  & 179$\pm$12   & 198$\pm$21   & 1.04$\pm$0.05 & 3.43$\pm$0.04 \\
    080319C & 1.95  &  34   &1.55$\pm$0.10 &  70.32$\pm$4.20        & 1.85$\pm$0.46 & -1.38$\pm$0.02 & 434$\pm$23 &   107/55         &807$\pm$107  & 198$\pm$38& 608$\pm$105  & 0.60$\pm$0.24 & 2.72$\pm$0.06 \\
    081028 & 3.038 & 260    &1.94$\pm$0.07 &  0.89$\pm$0.28  & 1.53$\pm$0.12 & -1.96$\pm$0.04 & 22951$\pm$559 &   1850/649           &27151$\pm$1815 & 9882$\pm$882  & 17269$\pm$1772 & 0.69$\pm$0.05 & 3.47$\pm$0.05 \\
    090205 & 4.6497 &  8.8   &1.95$\pm$0.15 &  0.94$\pm$0.08       & 0.98$\pm$0.32 & -1.10$\pm$0.04 & 1000$\pm$123 &     135/34      & 2013$\pm$502 & 608$\pm$166 & 1405$\pm$505  & 1.12$\pm$0.18 & 2.21$\pm$0.08 \\
    090429B & 9.3   &  5.5   &1.87$\pm$0.22 &  2.13$\pm$0.17       &1.57$\pm$0.34 & -1.40$\pm$0.06 & 574$\pm$49 &    35/16           &1066$\pm$211 & 262$\pm$71    & 804$\pm$211  & 0.74$\pm$0.17 & 2.71$\pm$0.09 \\
    101024A & ...   &  18.7   &1.77$\pm$0.12 &    10.27$\pm$0.83       &0.60$\pm$0.17 & -1.20$\pm$0.03 & 500$\pm$57 &    101/55      &957$\pm$213   & 349$\pm$75  & 608$\pm$215  & 1.32$\pm$0.10 & 2.27$\pm$0.06 \\
    110213A & 1.46  &  48   &1.86$\pm$0.06 & 34.65$\pm$1.67        & 0.60$\pm$0.09 & -1.50$\pm$0.01 & 1433$\pm$65 &   1054/245       &2923$\pm$305  & 721$\pm$125 & 2202$\pm$294   & 1.25$\pm$0.05 & 2.70$\pm$0.03 \\
    110715A & 0.82  & 13     &1.82$\pm$0.19 &  0.84$\pm$0.06   & 2.07$\pm$0.53 & -1.37$\pm$0.05 & 38689$\pm$2080 &     6388/253      &62722$\pm$9968 & 9832$\pm$3610  & 52890$\pm$9746& 0.48$\pm$0.27 & 2.74$\pm$0.09 \\
    120326A & 1.798 & 69.6    &1.85$\pm$0.07 &   1.21$\pm$0.03         &1.00$\pm$0.07 & -1.90$\pm$0.05 & 37336$\pm$983 &  2754/242   &53073$\pm$3660 & 13063$\pm$1699 & 40009$\pm$3527 & 0.96$\pm$0.03 & 3.32$\pm$0.07 \\
    130511A & 1.3033 &  5.43  &1.85$\pm$0.17 &   5.71$\pm$0.68        &1.40$\pm$0.83 & -1.11$\pm$0.03 & 150$\pm$19 &    46/25        &264$\pm$87   & 65$\pm$32 & 199$\pm$86  & 0.88$\pm$0.46 & 2.29$\pm$0.14 \\
    130831B & ...   &  37.8   &1.96$\pm$0.17 &     8.64$\pm$0.55      &2.16$\pm$0.59 &-1.03$\pm$0.03 & 455$\pm$33&    104/42         &1066$\pm$195   & 113$\pm$53  & 953$\pm$193   & 0.46$\pm$0.32 & 2.29$\pm$0.08 \\
    140515A & 6.32  & 23.4    &1.73$\pm$0.12 &   1.86$\pm$0.14        &2.55$\pm$0.28 & -1.38$\pm$0.04 & 2932$\pm$147 &  253/53       &5088$\pm$795   & 797$\pm$236 & 4290$\pm$787    & 0.23$\pm$0.14 & 2.80$\pm$0.06 \\
    150103A & ...   &  49.1  &1.82$\pm$0.40 &  75.85$\pm$2.96         & 2.00$\pm$0.27 & -2.53$\pm$0.07 & 220$\pm$6 &      98/65      &199$\pm$18   & 49$\pm$10  & 150$\pm$17   & 0.43$\pm$0.11 & 4.29$\pm$0.10 \\
    150626B & ...   & 48      &2.02$\pm$0.15 &    1.09$\pm$0.05        &0.96$\pm$0.10 & -2.23$\pm$0.12 & 21828$\pm$962&    313/79    &30370$\pm$3486  & 7475$\pm$1670  & 22895$\pm$3349 & 0.93$\pm$0.05 & 3.77$\pm$0.16 \\
    150911A & ...   &  7.2    &1.96$\pm$0.14 &   6.65$\pm$0.40        &1.64$\pm$0.12 & -1.84$\pm$0.06 & 1905$\pm$76 &   1950/107     &1857$\pm$273  & 456$\pm$142  & 1401$\pm$256  & 0.65$\pm$0.06 & 3.32$\pm$0.08 \\
    161001A & ... &  2.6      &1.97$\pm$0.16 &    40.88$\pm$1.99       &1.19$\pm$0.35 & -1.07$\pm$0.01 & 109$\pm$7 &    196/79        &199$\pm$29  & 49$\pm$11  & 151$\pm$29   & 1.01$\pm$0.19 & 2.20$\pm$0.06 \\
    171205A & 0.0368 &  189.4   &1.63$\pm$0.17 &   0.10$\pm$0.01       & 0.45$\pm$0.12 &-1.06$\pm$0.05 & 58429$\pm$7174 &  7439/211   &152975$\pm$33264 & 43367$\pm$8808 & 109608$\pm$33643  & 1.45$\pm$0.07 & 2.03$\pm$0.08 \\
    181110A & 1.505 &  138.4    &1.83$\pm$0.09 &    13.81$\pm$0.79     &0.93$\pm$0.14 & -1.79$\pm$0.05 & 1711$\pm$110 &12832/264      &2463$\pm$394 & 606$\pm$187  & 1857$\pm$380  & 1.02$\pm$0.07 & 3.16$\pm$0.08 \\
    \hline
\end{tabular}
\end{center}}
\tablecomments{\\$^a$ In units of $10^{-11}$ $erg$ $cm^{-2}$ $s^{-1}$. \\
 $^b$  In units of seconds.}
\end{table*}

\begin{figure*}
\center
\includegraphics[width=6cm,height=4.5cm]{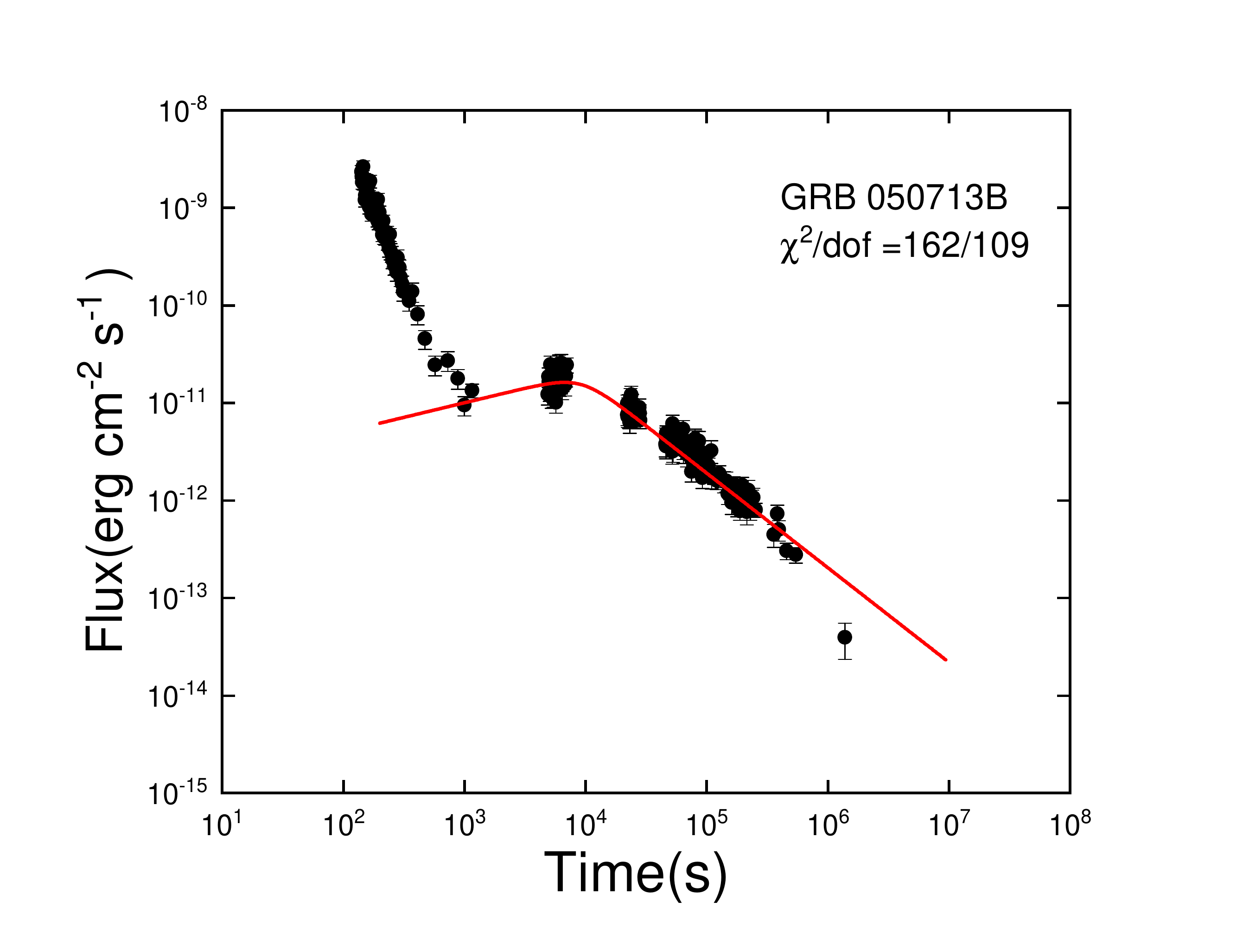}\includegraphics[width=6cm,height=4.5cm]{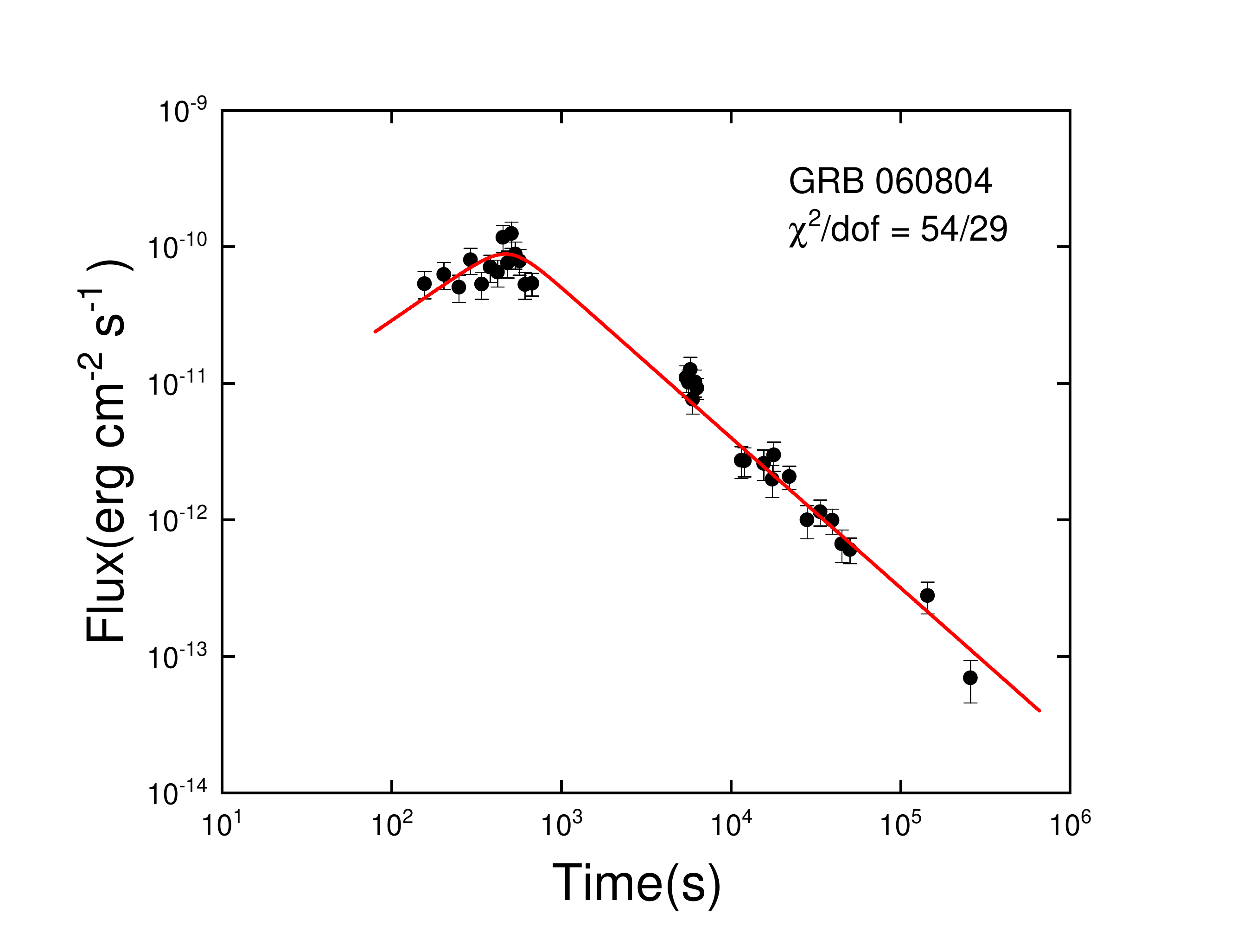}\includegraphics[width=6cm,height=4.5cm]{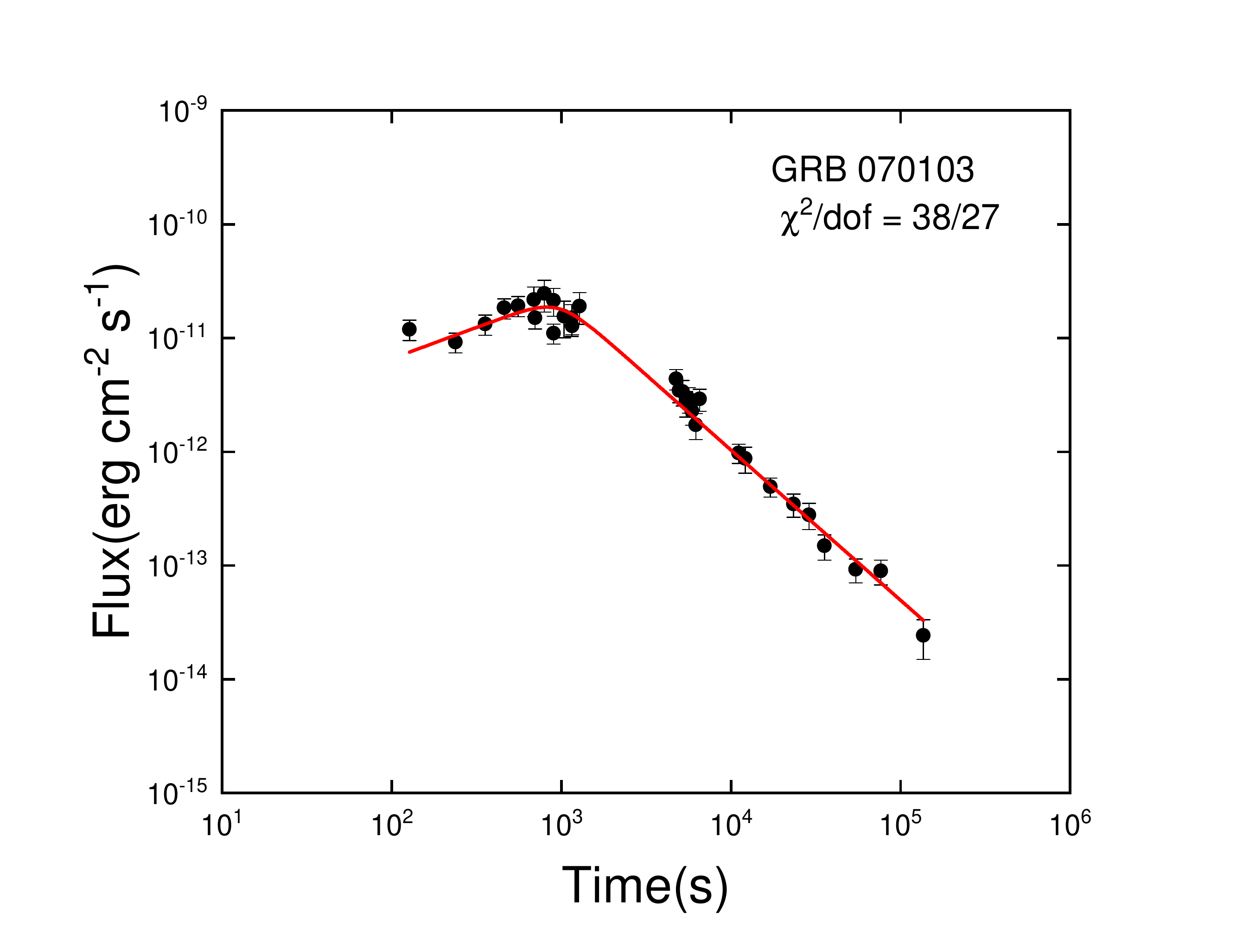}
\includegraphics[width=6cm,height=4.5cm]{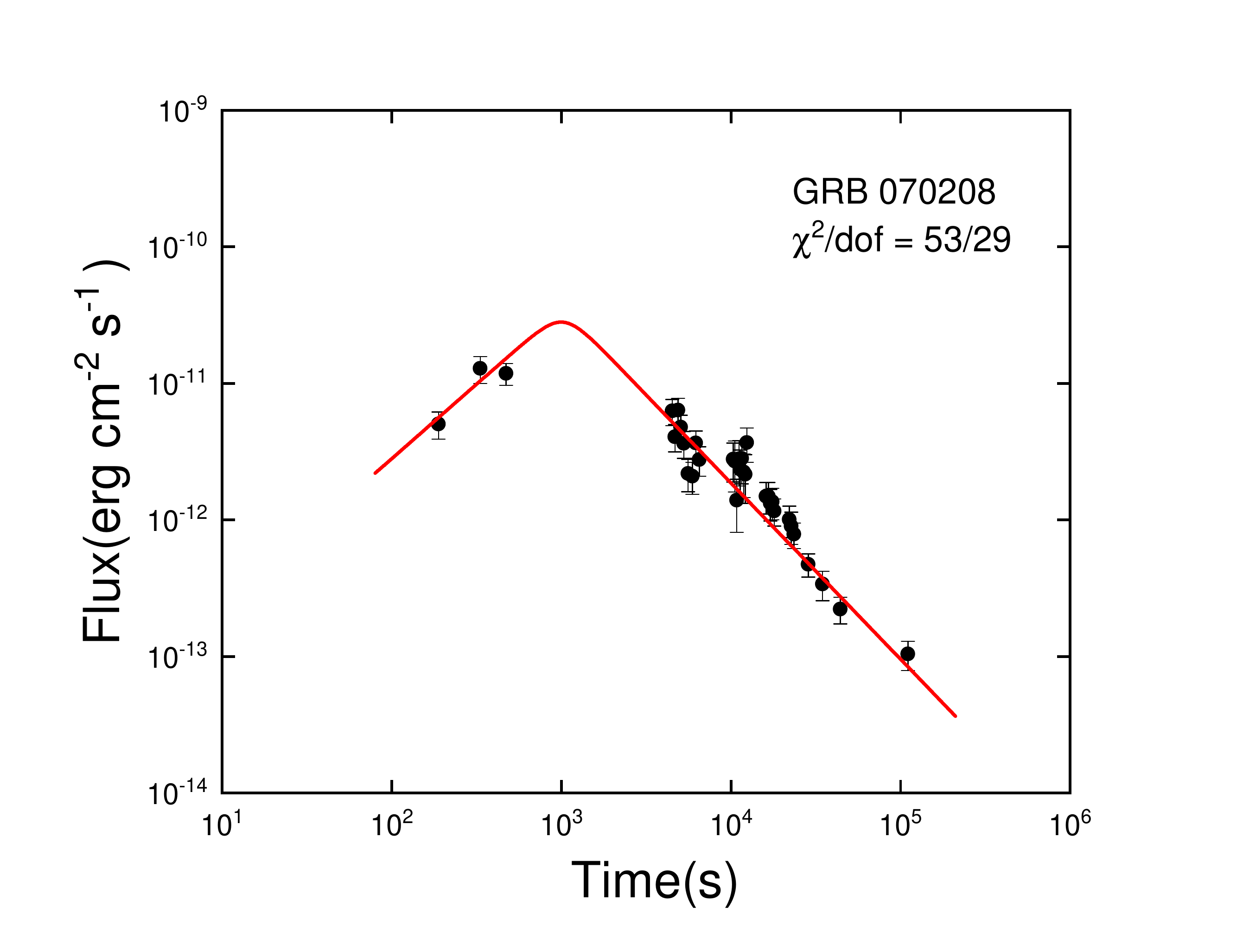}\includegraphics[width=6cm,height=4.5cm]{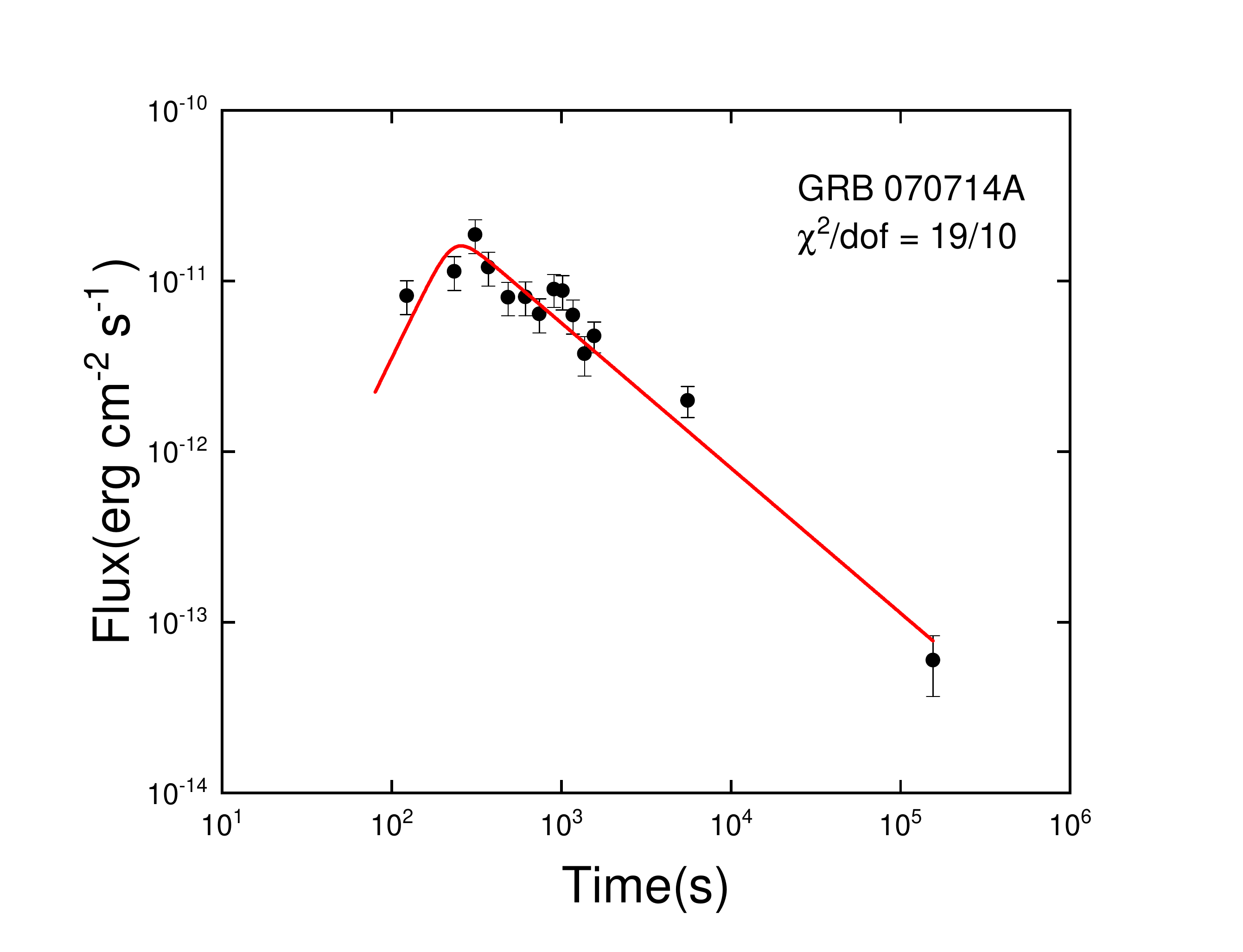}\includegraphics[width=6cm,height=4.5cm]{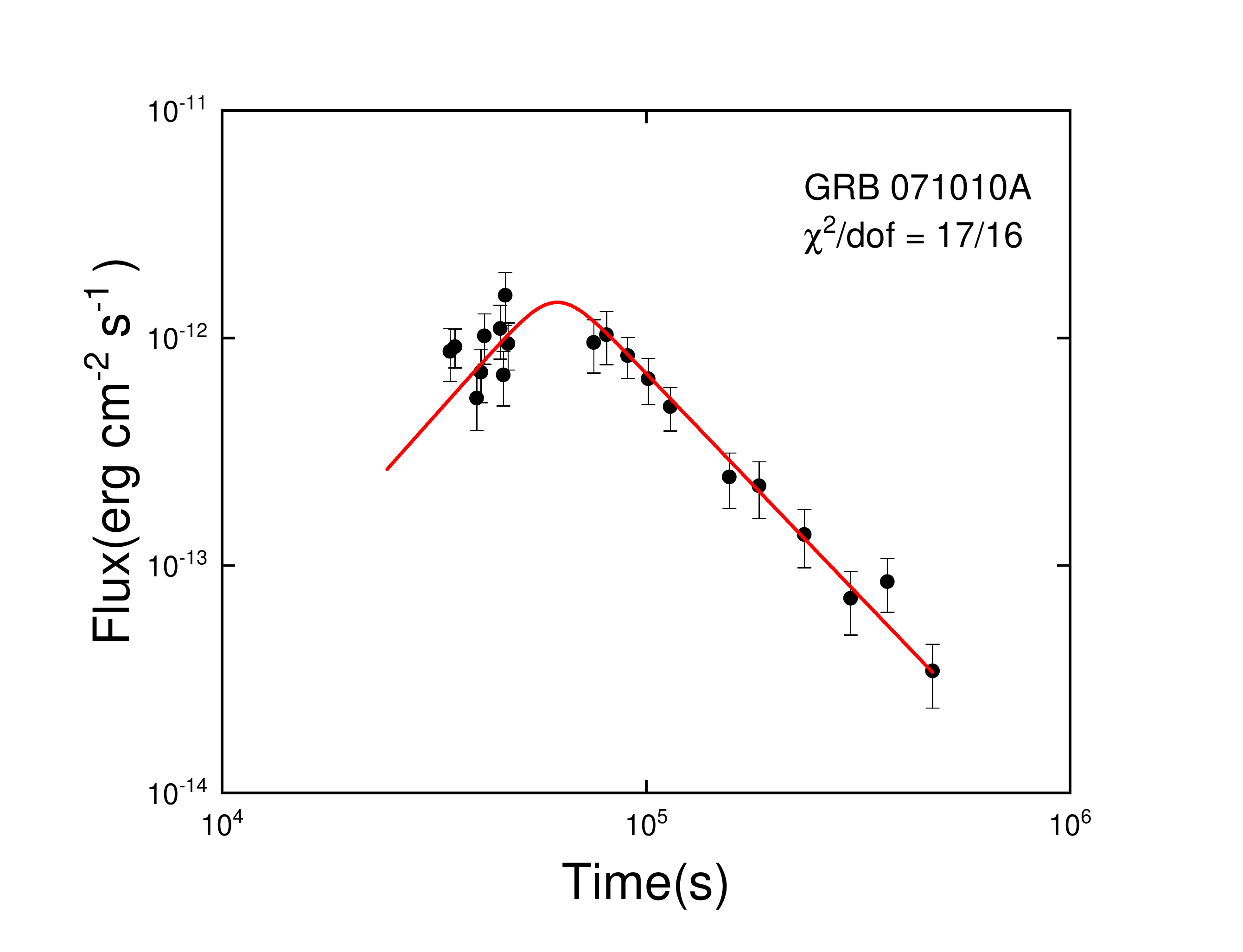}
\includegraphics[width=6cm,height=4.5cm]{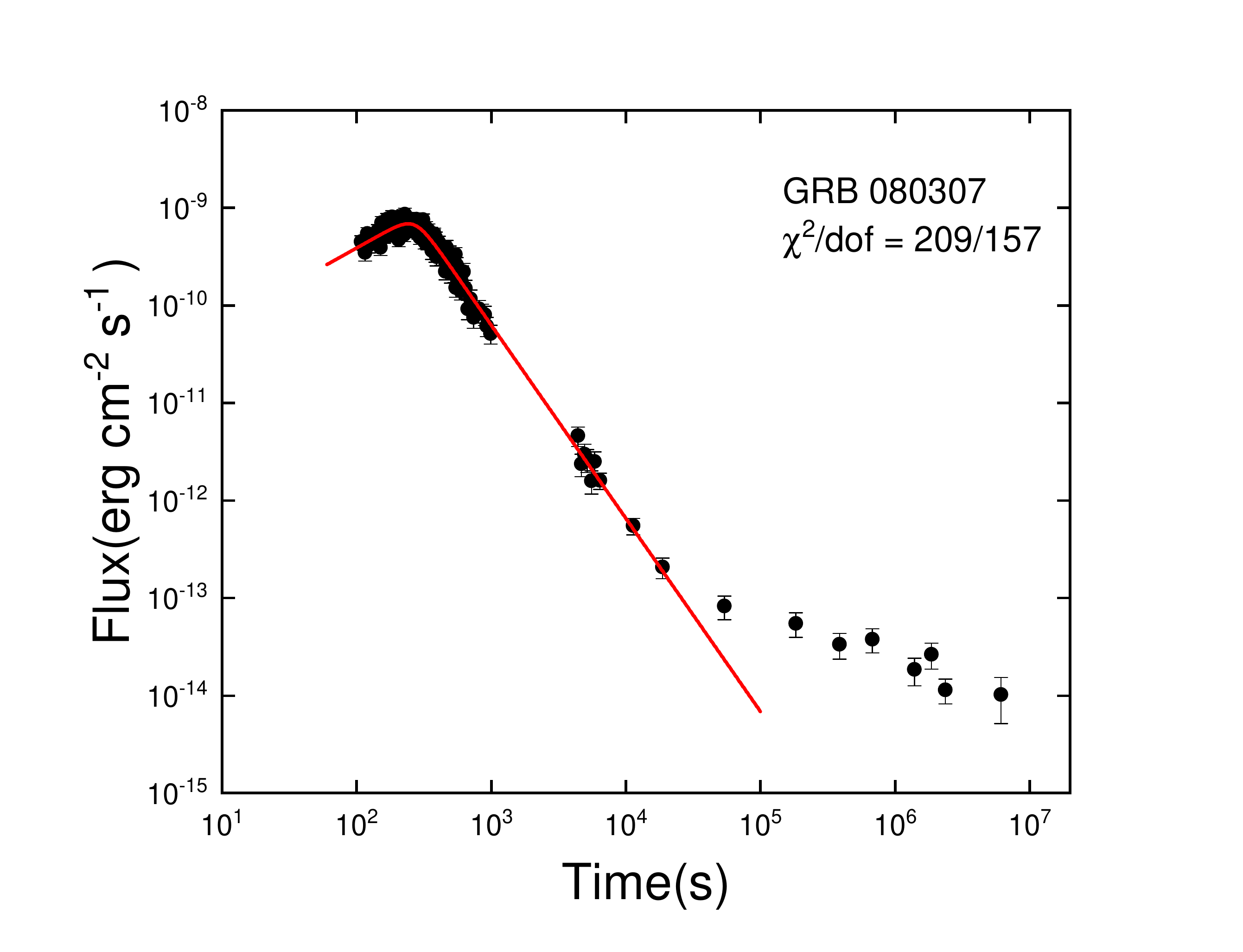}\includegraphics[width=6cm,height=4.5cm]{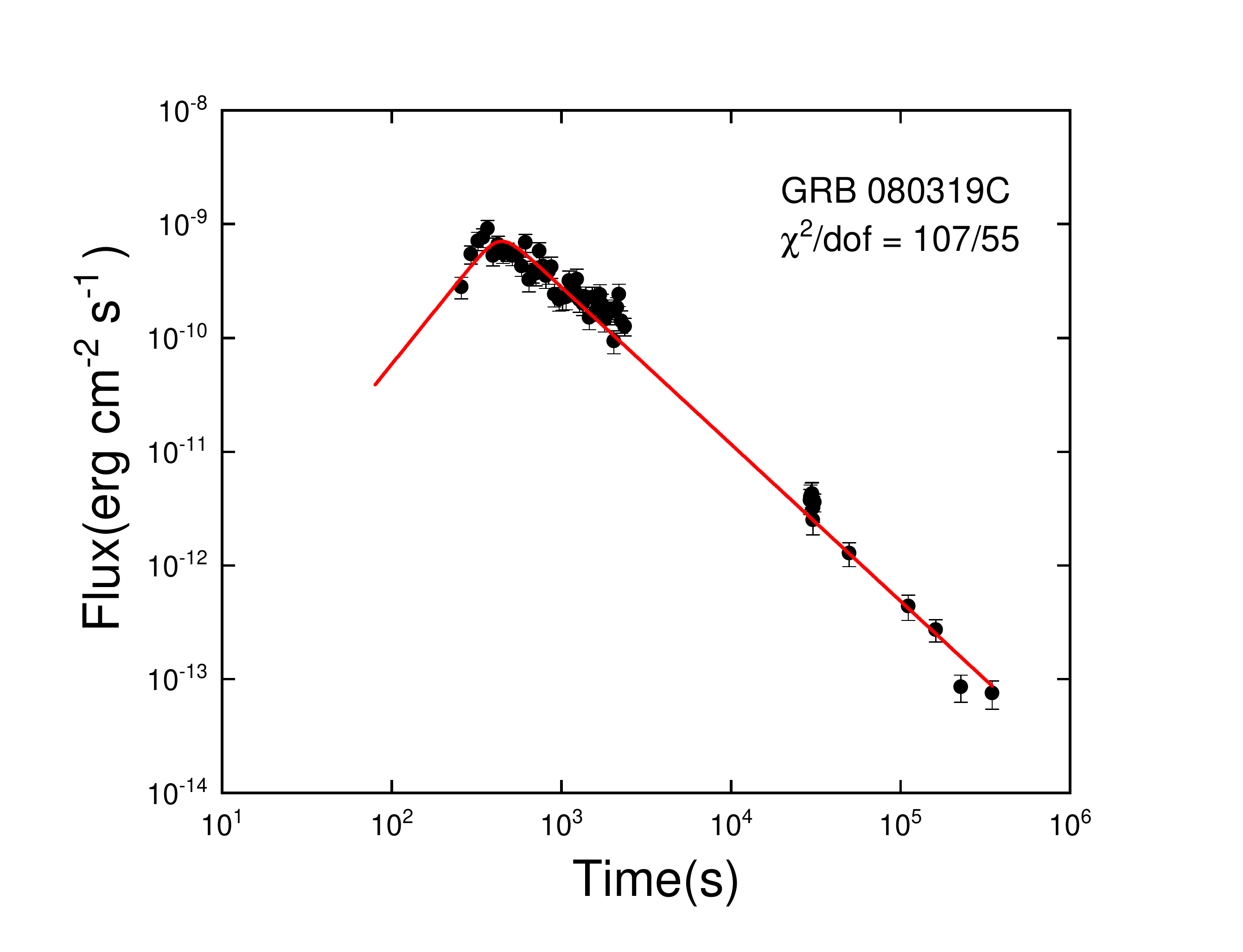}\includegraphics[width=6cm,height=4.5cm]{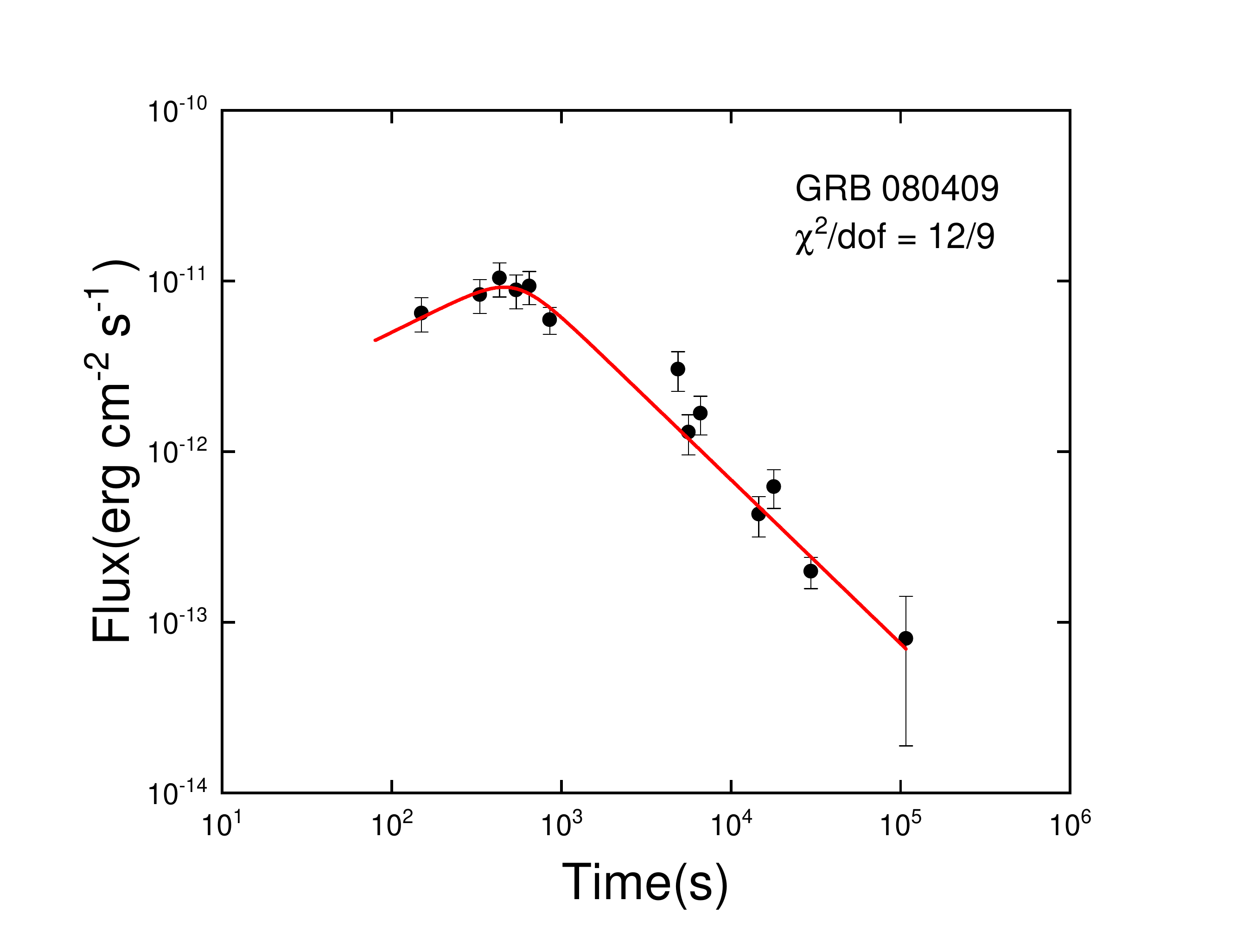}
\includegraphics[width=6cm,height=4.5cm]{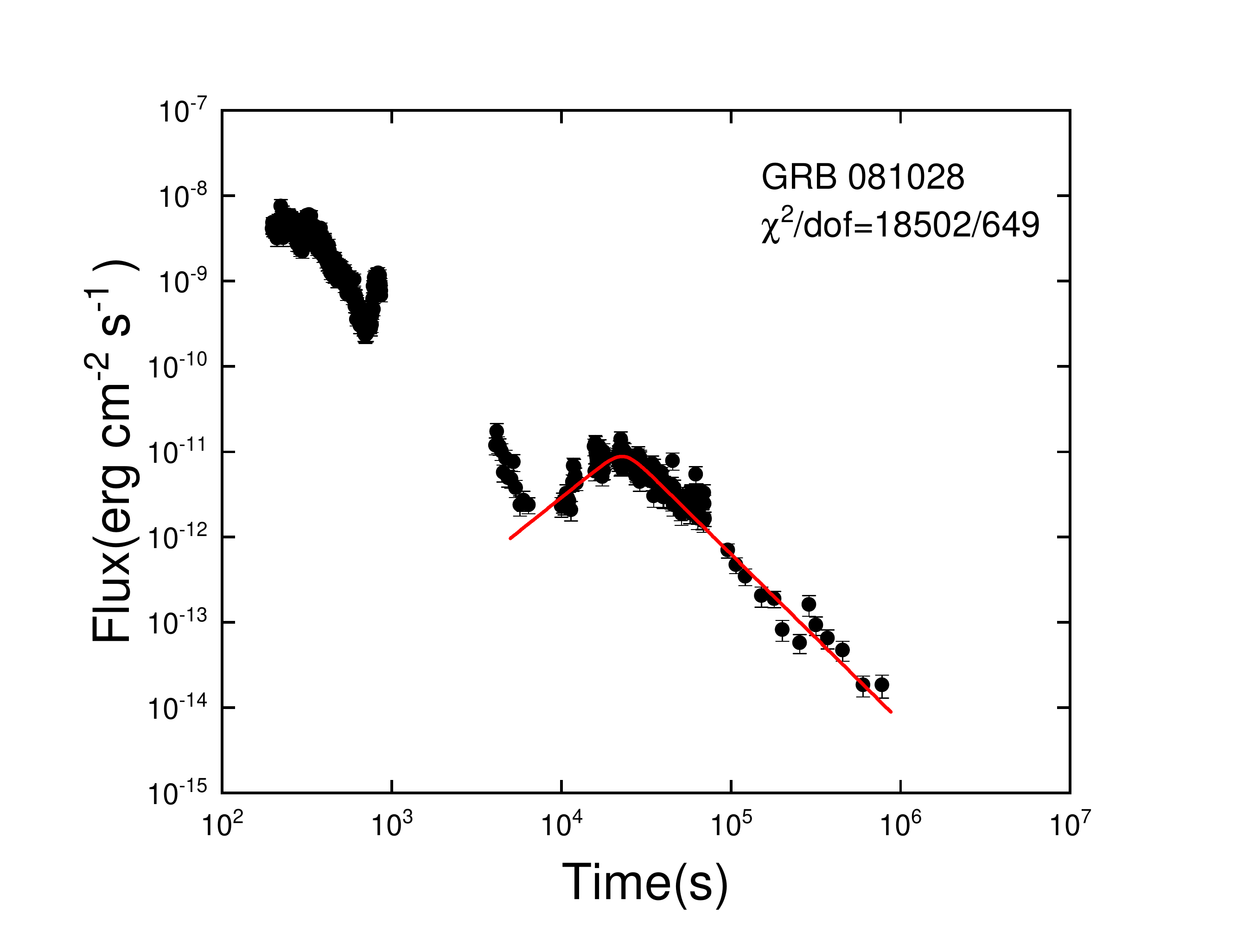}\includegraphics[width=6cm,height=4.5cm]{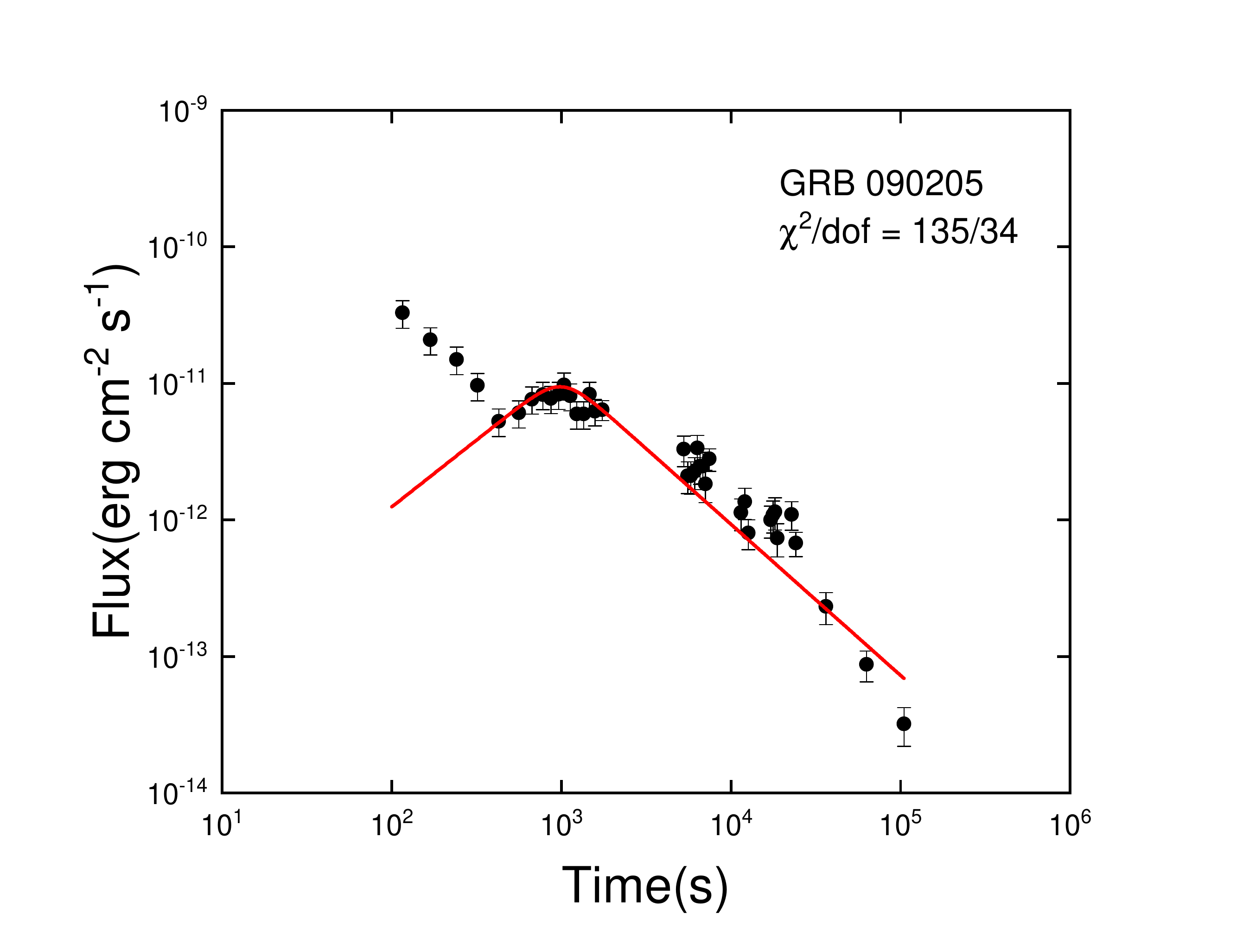}\includegraphics[width=6cm,height=4.5cm]{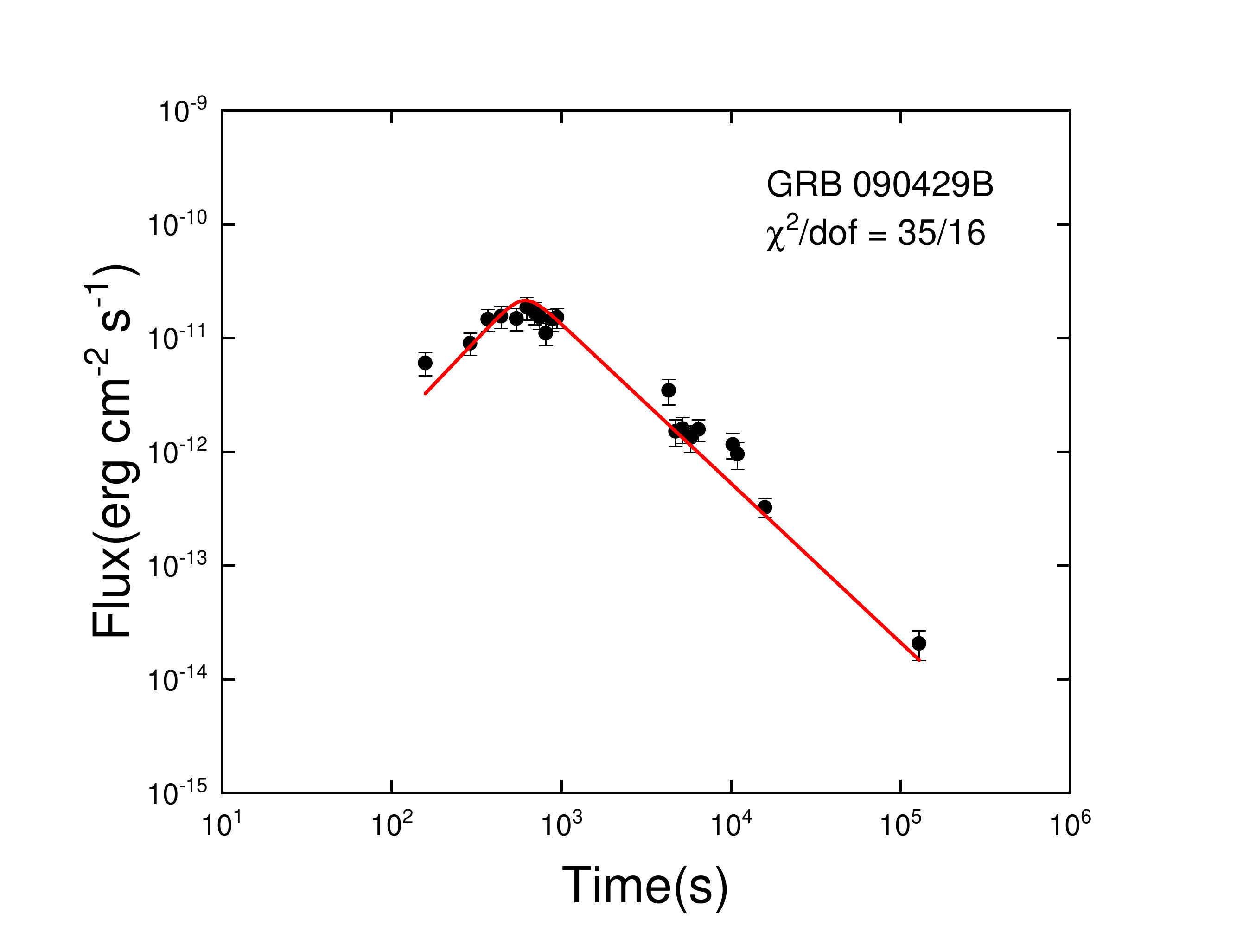}
\includegraphics[width=6cm,height=4.5cm]{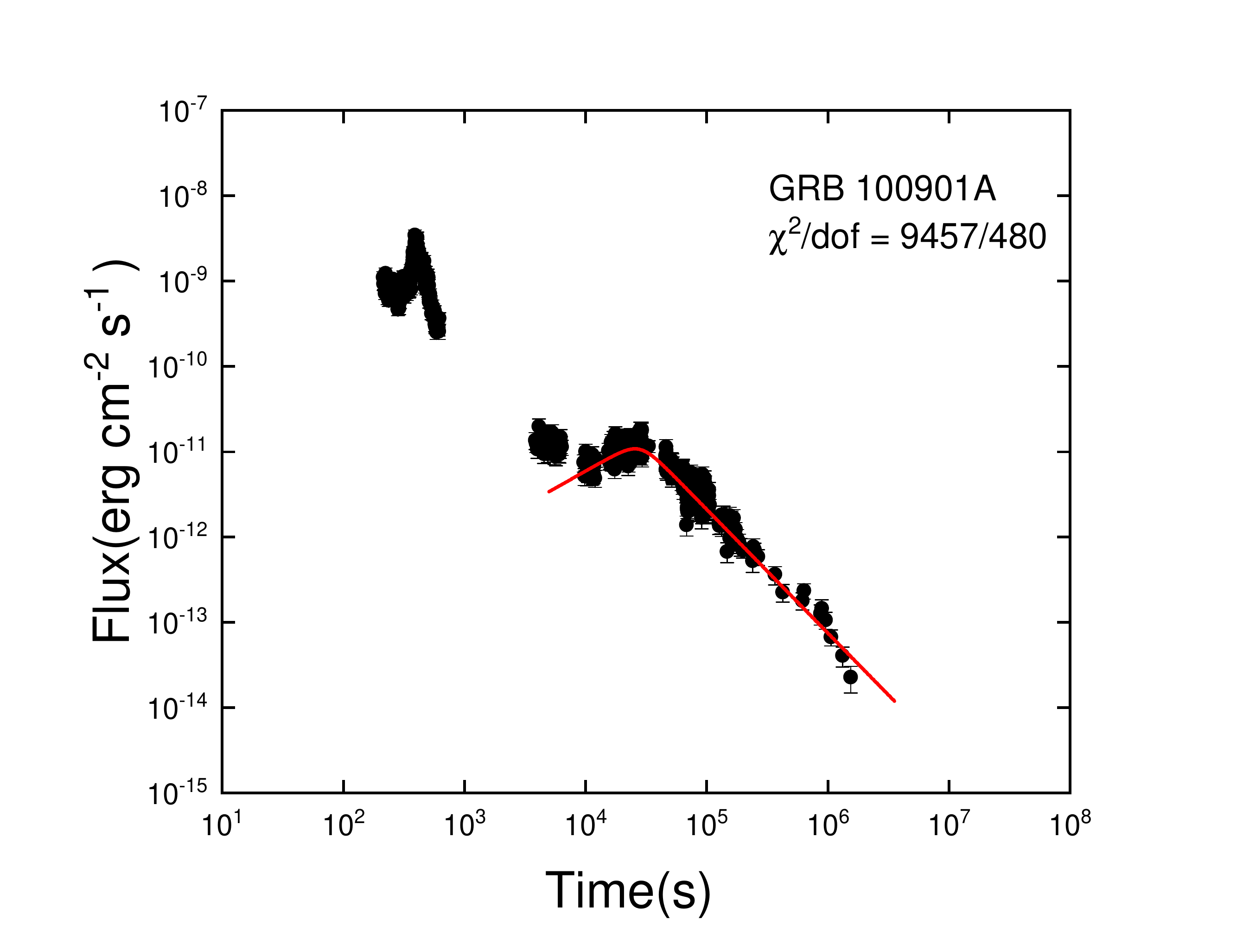}\includegraphics[width=6cm,height=4.5cm]{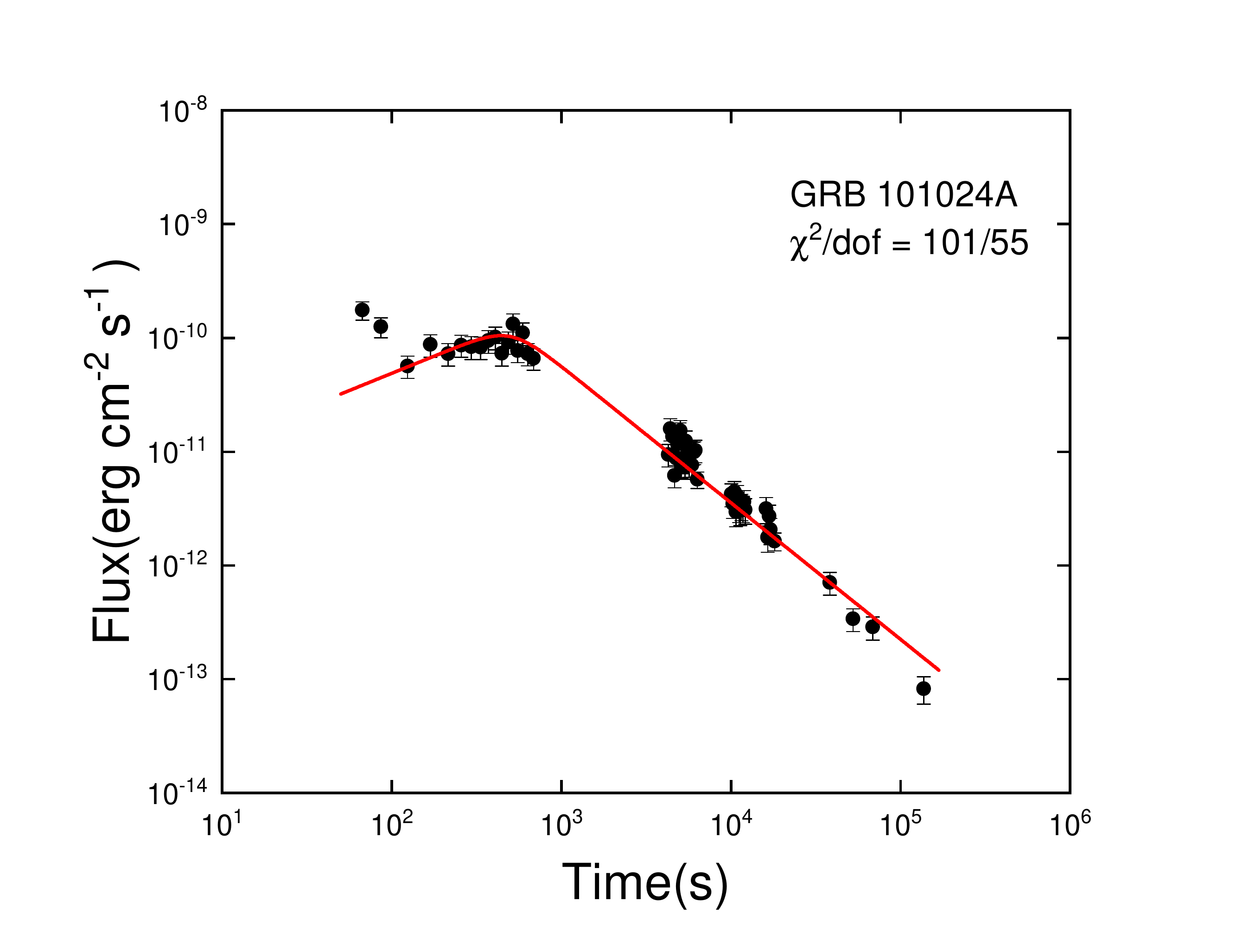}\includegraphics[width=6cm,height=4.5cm]{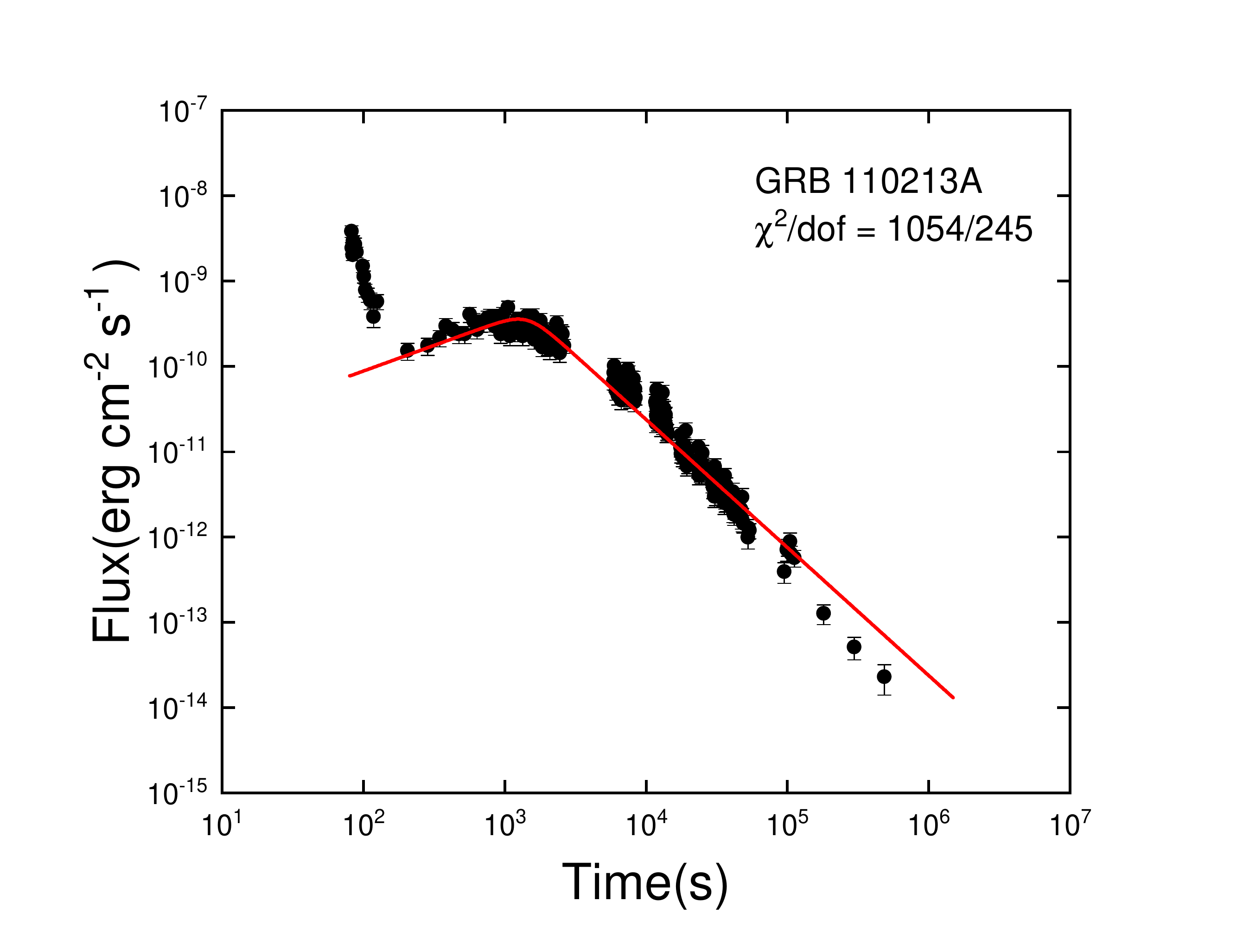}
\caption{ Light curves of the X-ray selected sample. The black dots represent the X-ray data, while the solid red lines are the fitting lines.}
\end{figure*}

\begin{figure*}
\includegraphics[width=6cm,height=4.5cm]{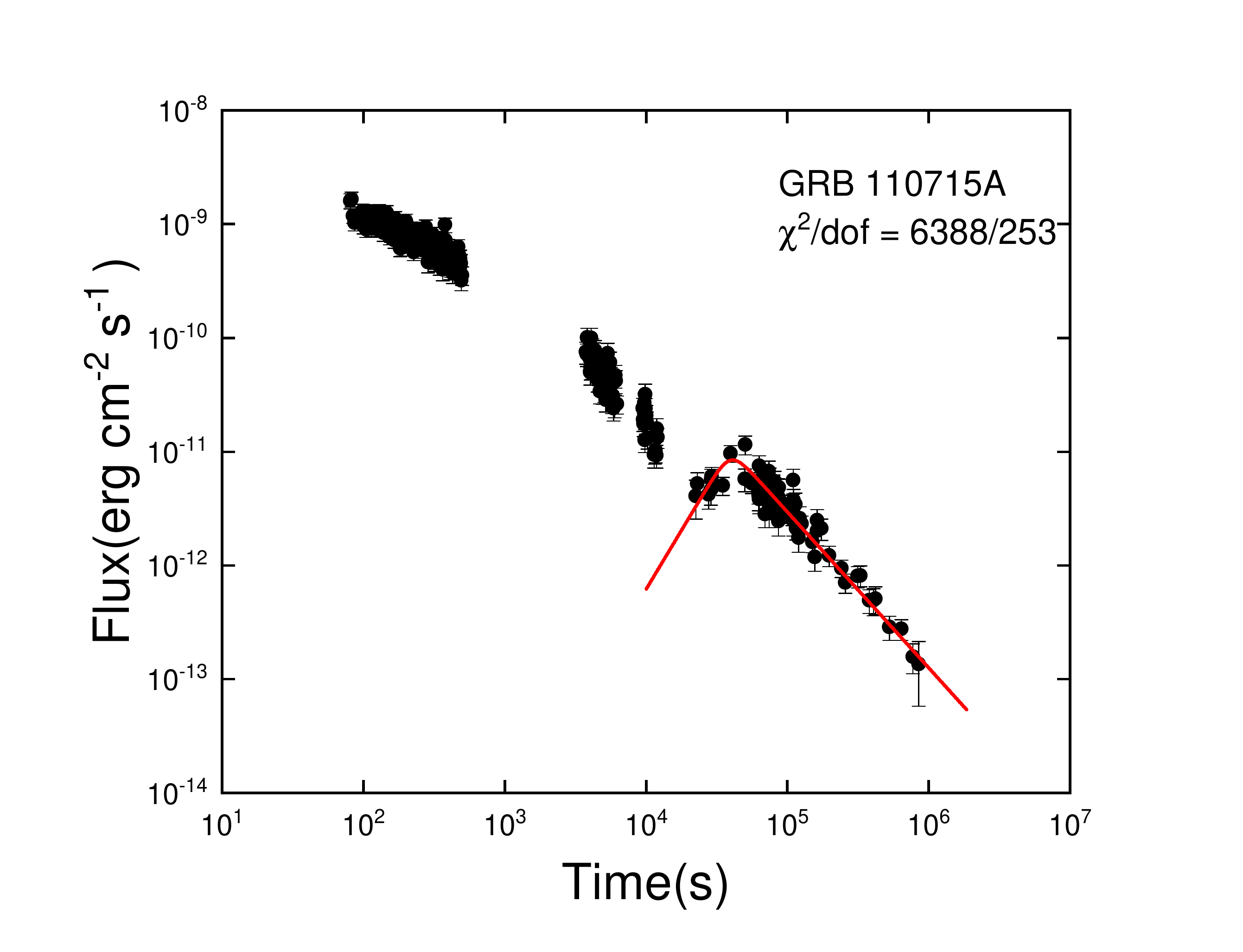}\includegraphics[width=6cm,height=4.5cm]{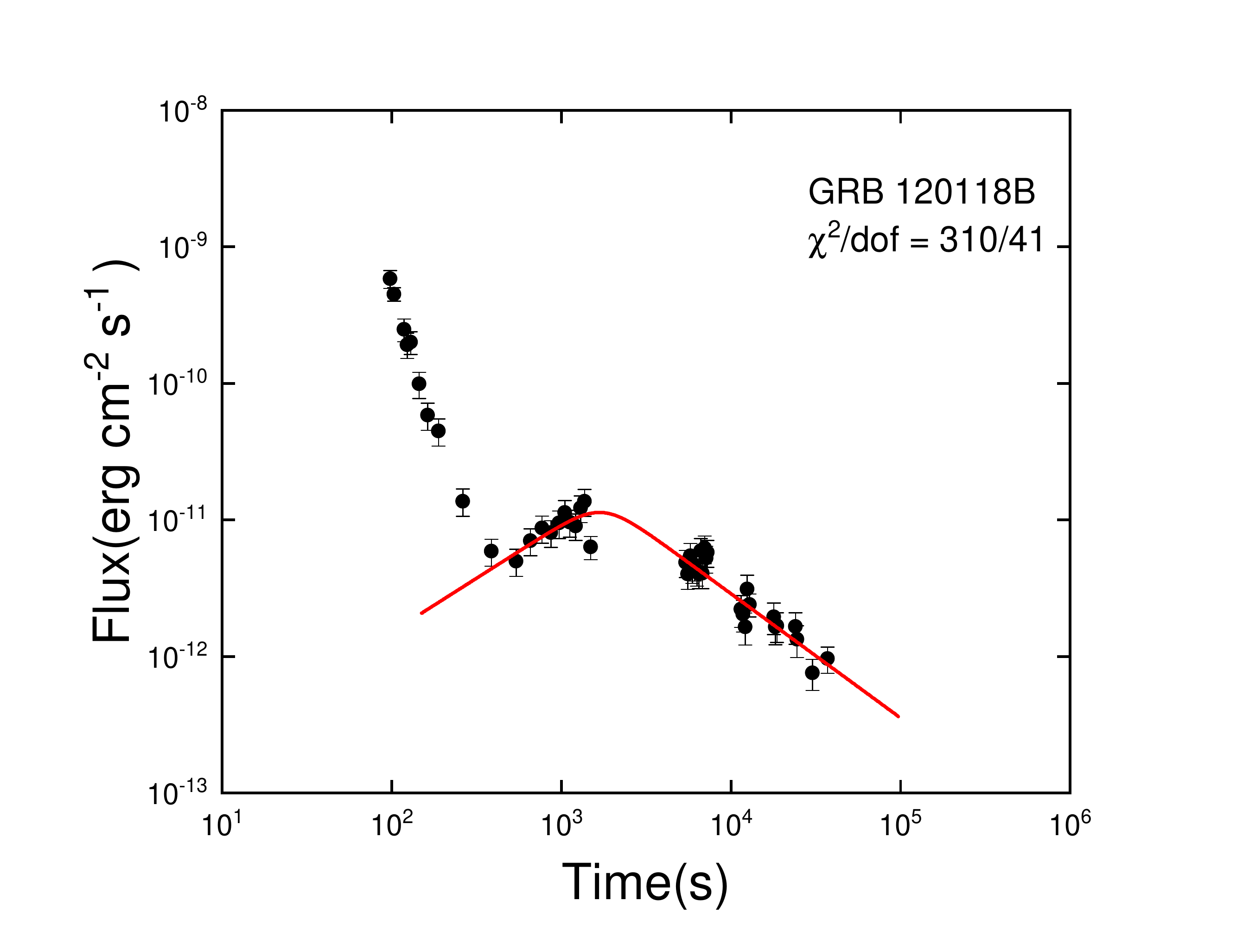}\includegraphics[width=6cm,height=4.5cm]{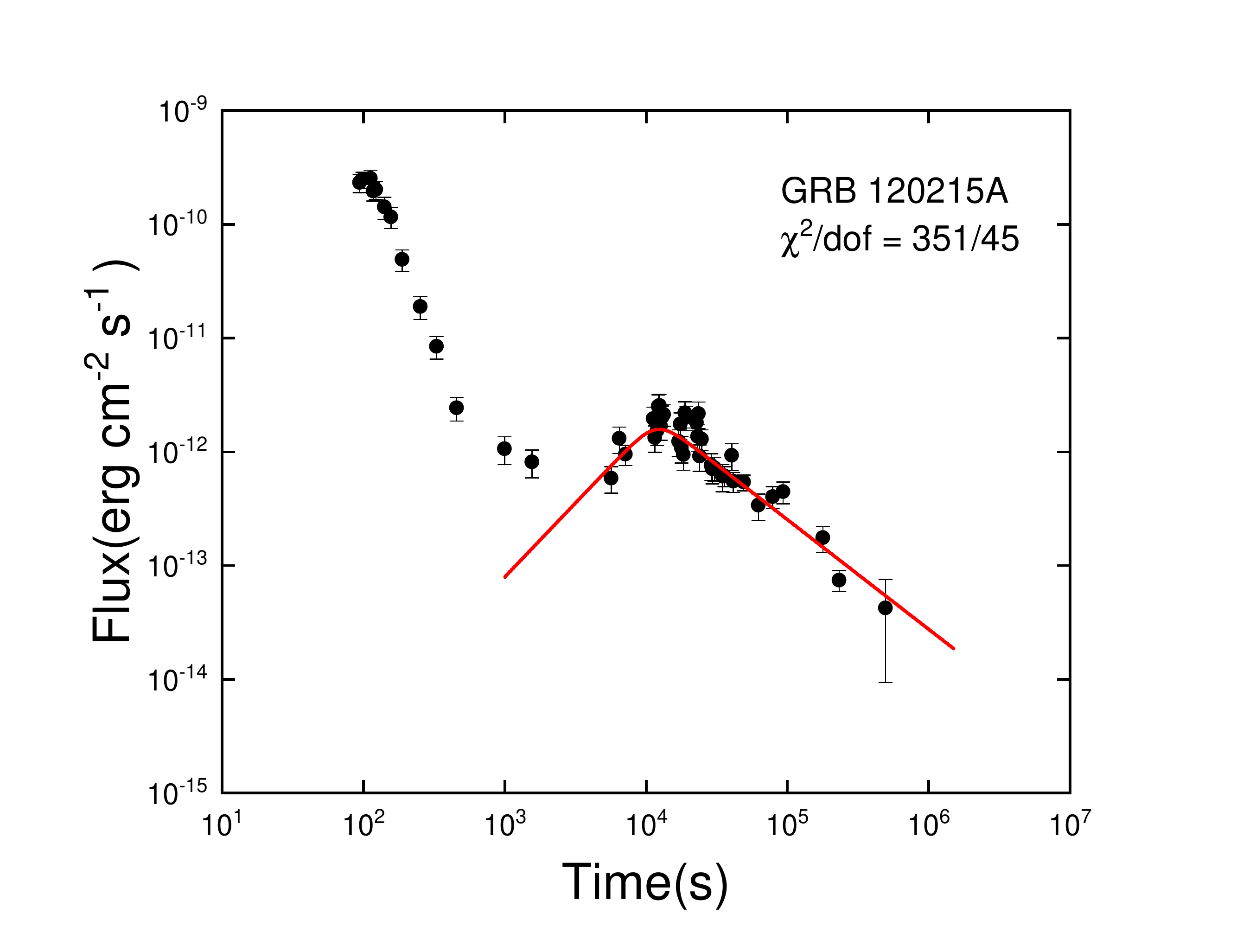}
\includegraphics[width=6cm,height=4.5cm]{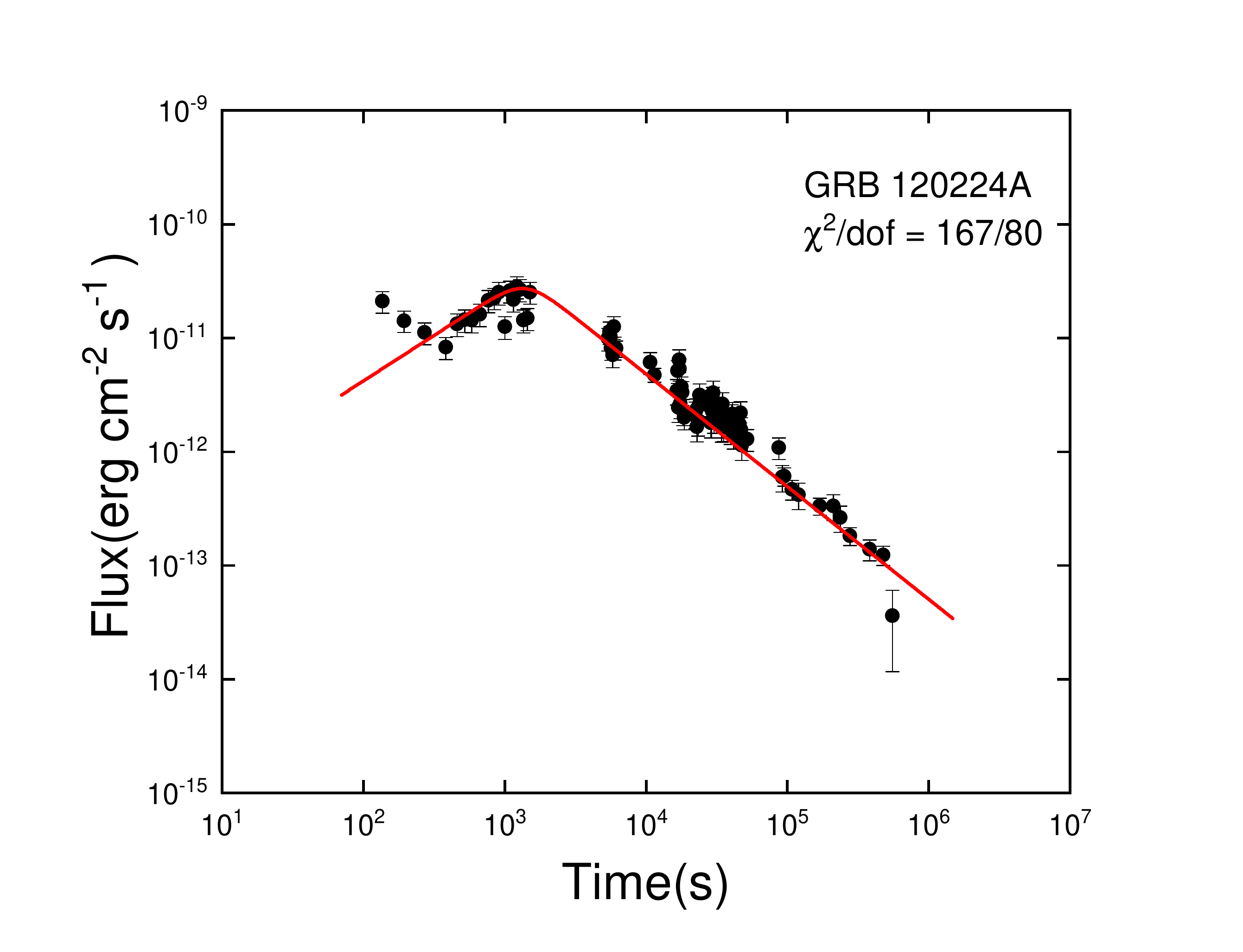}\includegraphics[width=6cm,height=4.5cm]{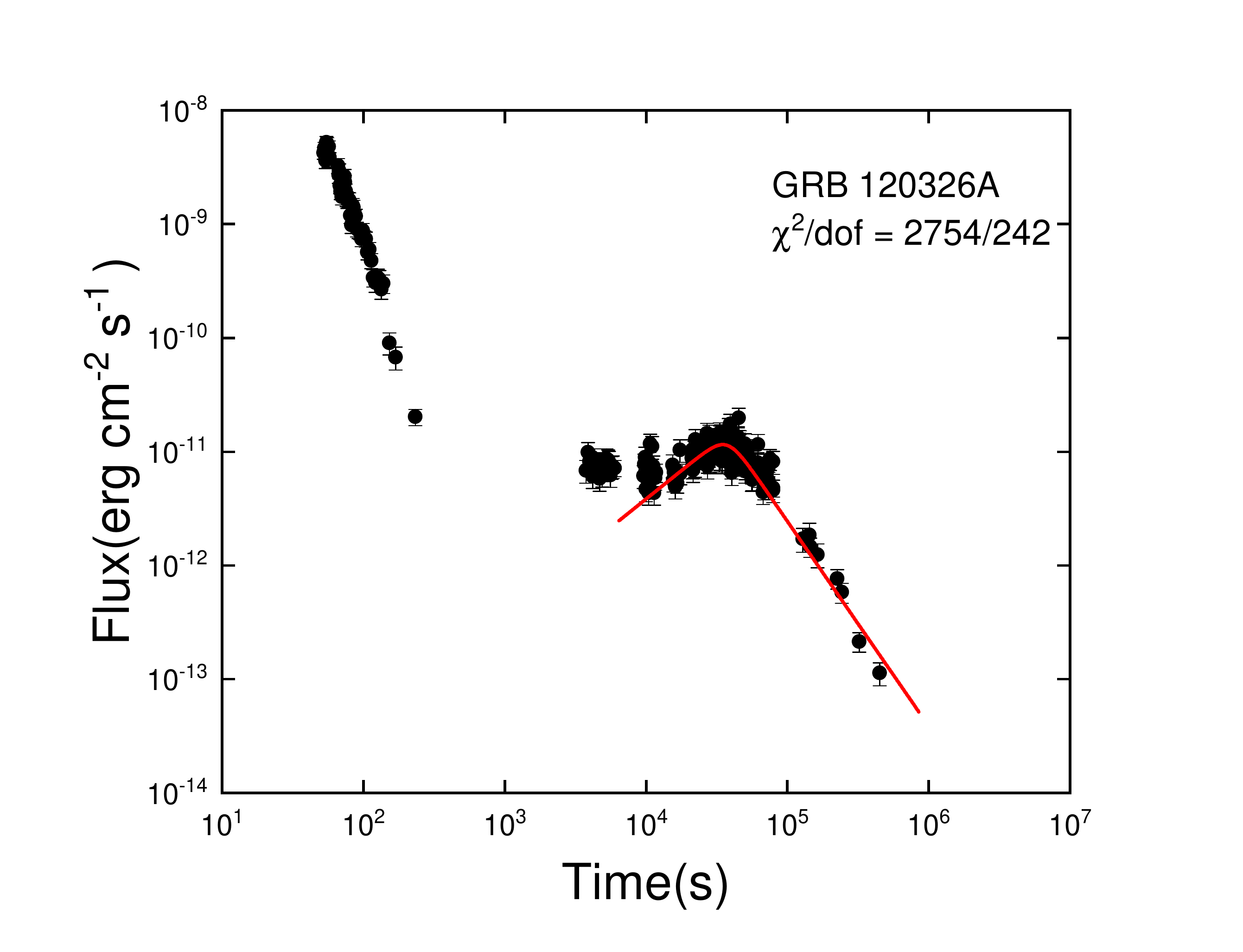}\includegraphics[width=6cm,height=4.5cm]{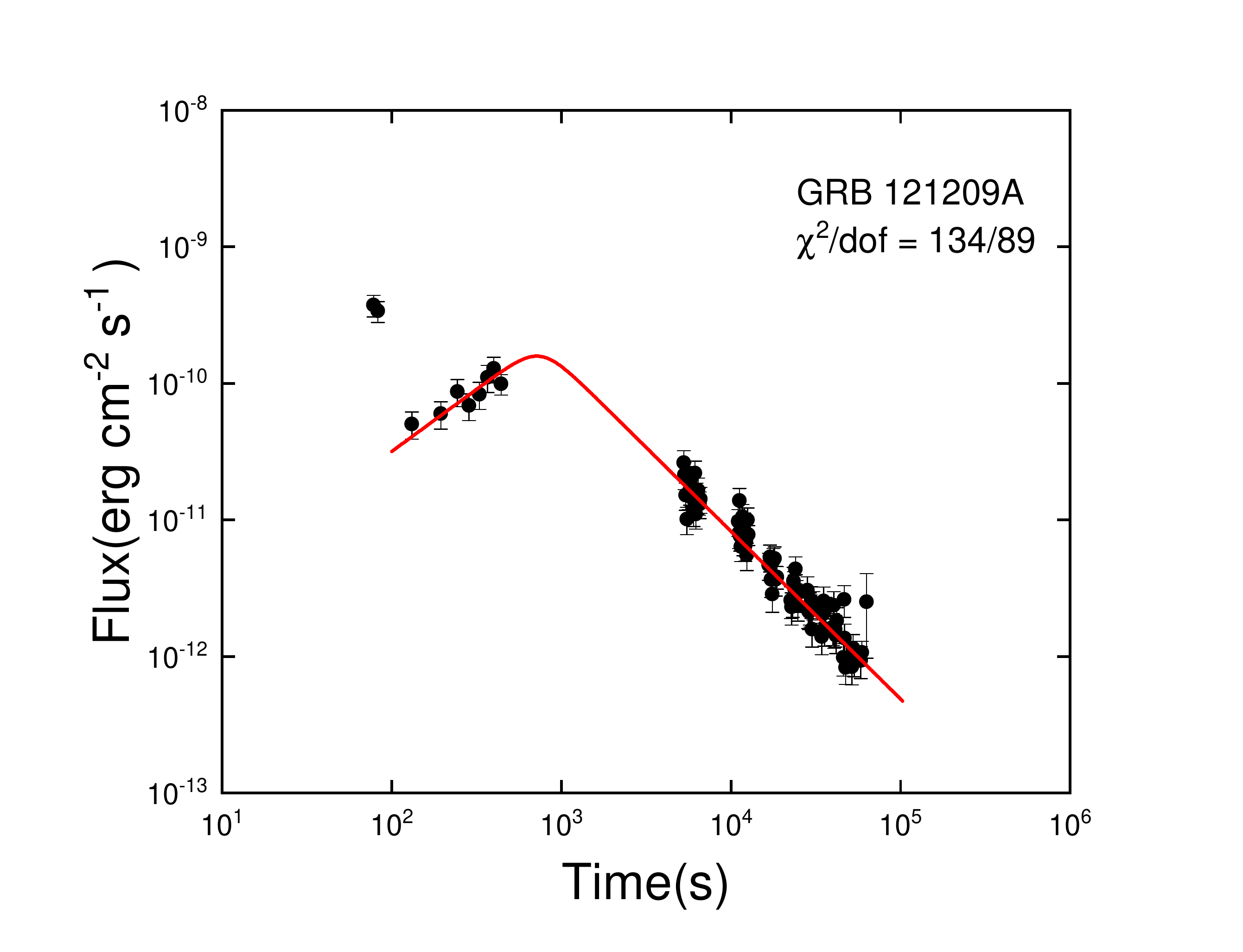}
\includegraphics[width=6cm,height=4.5cm]{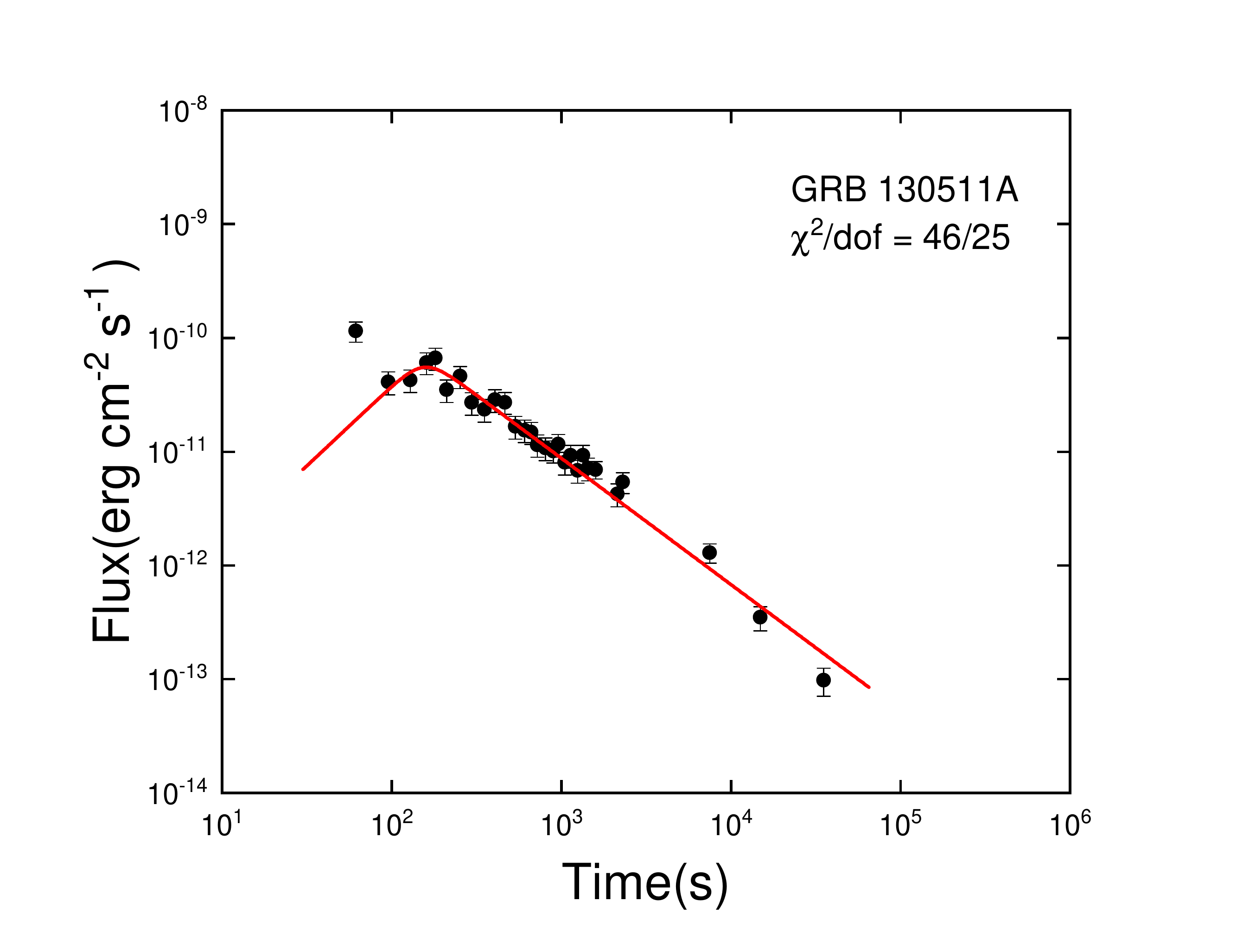}\includegraphics[width=6cm,height=4.5cm]{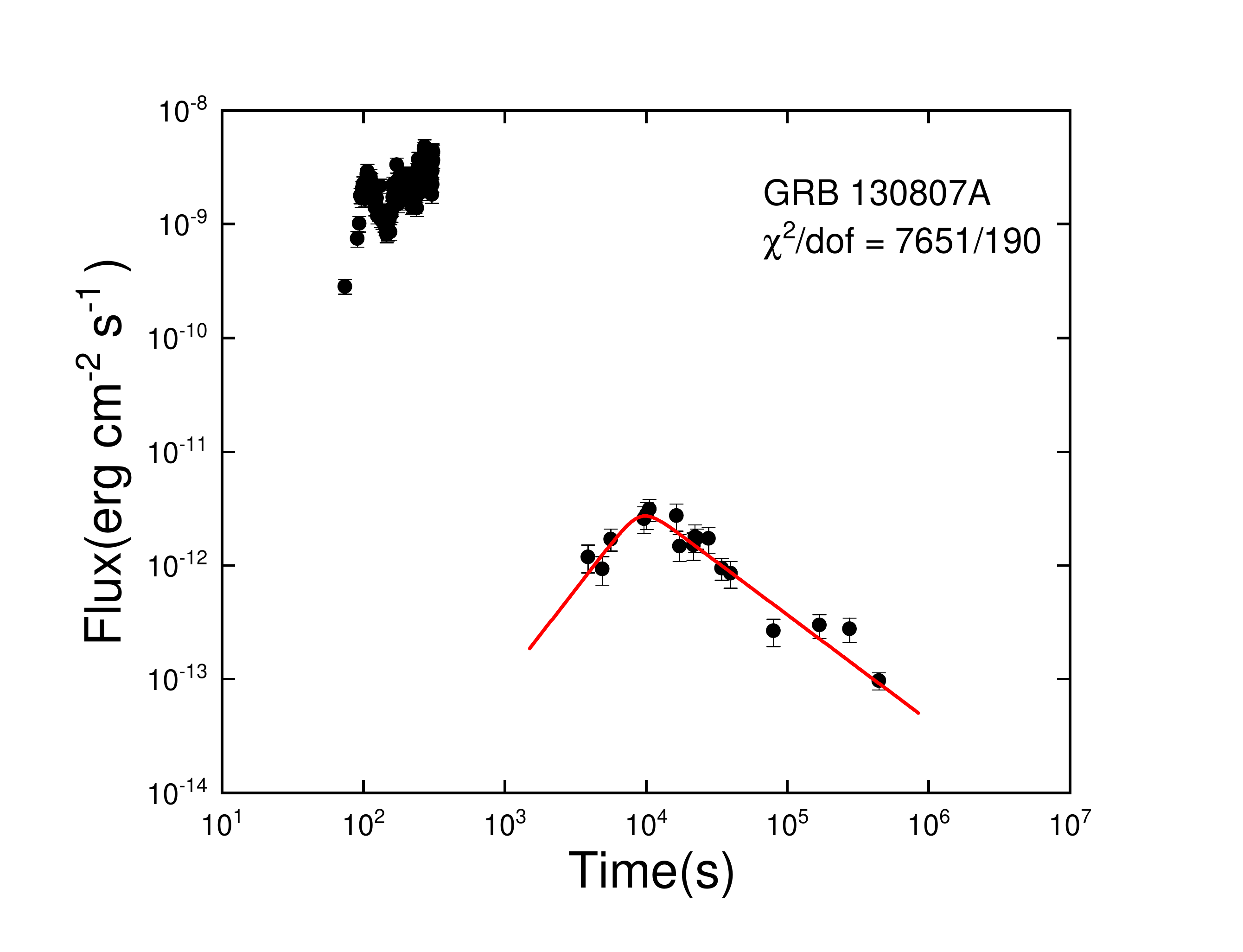}\includegraphics[width=6cm,height=4.5cm]{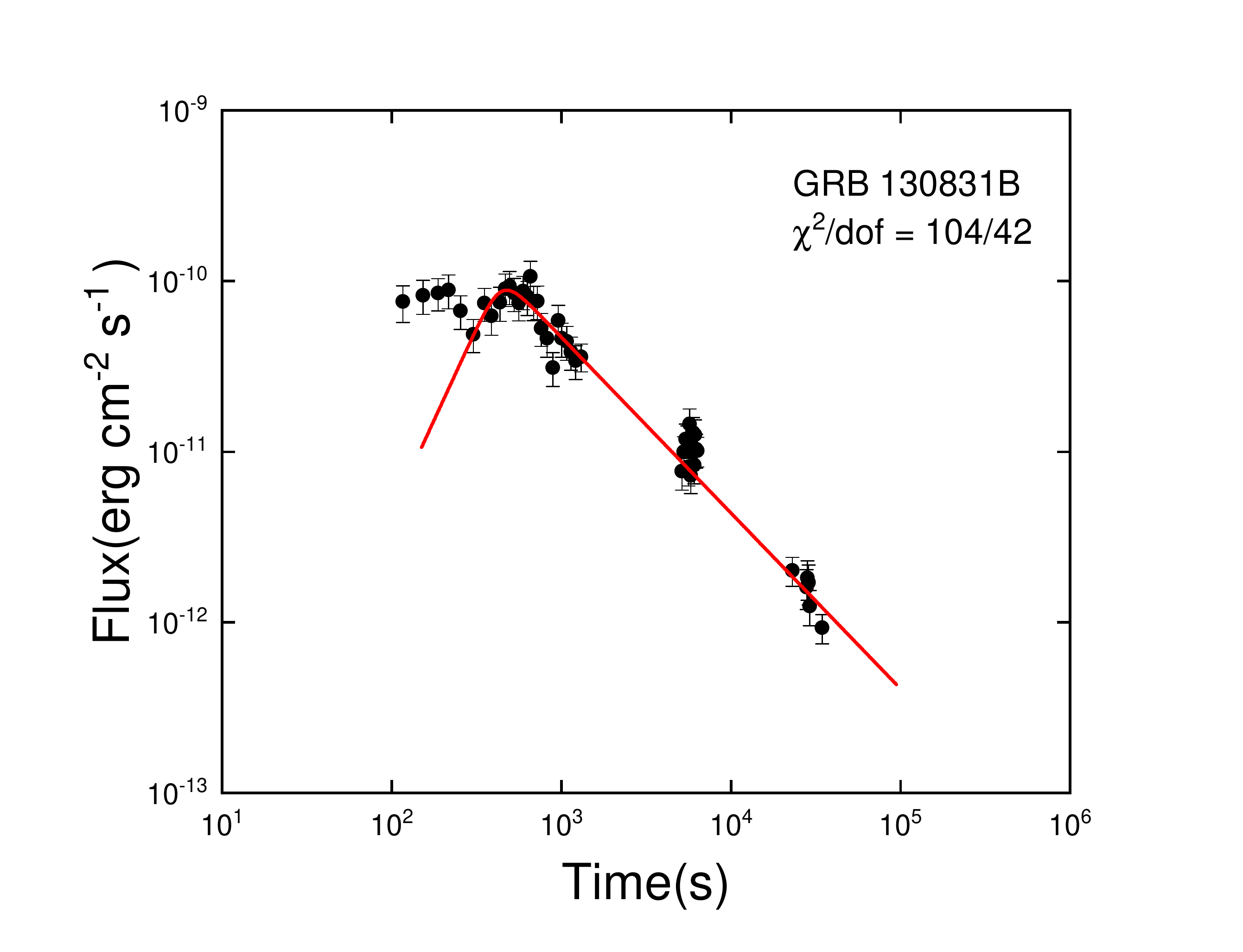}
\includegraphics[width=6cm,height=4.5cm]{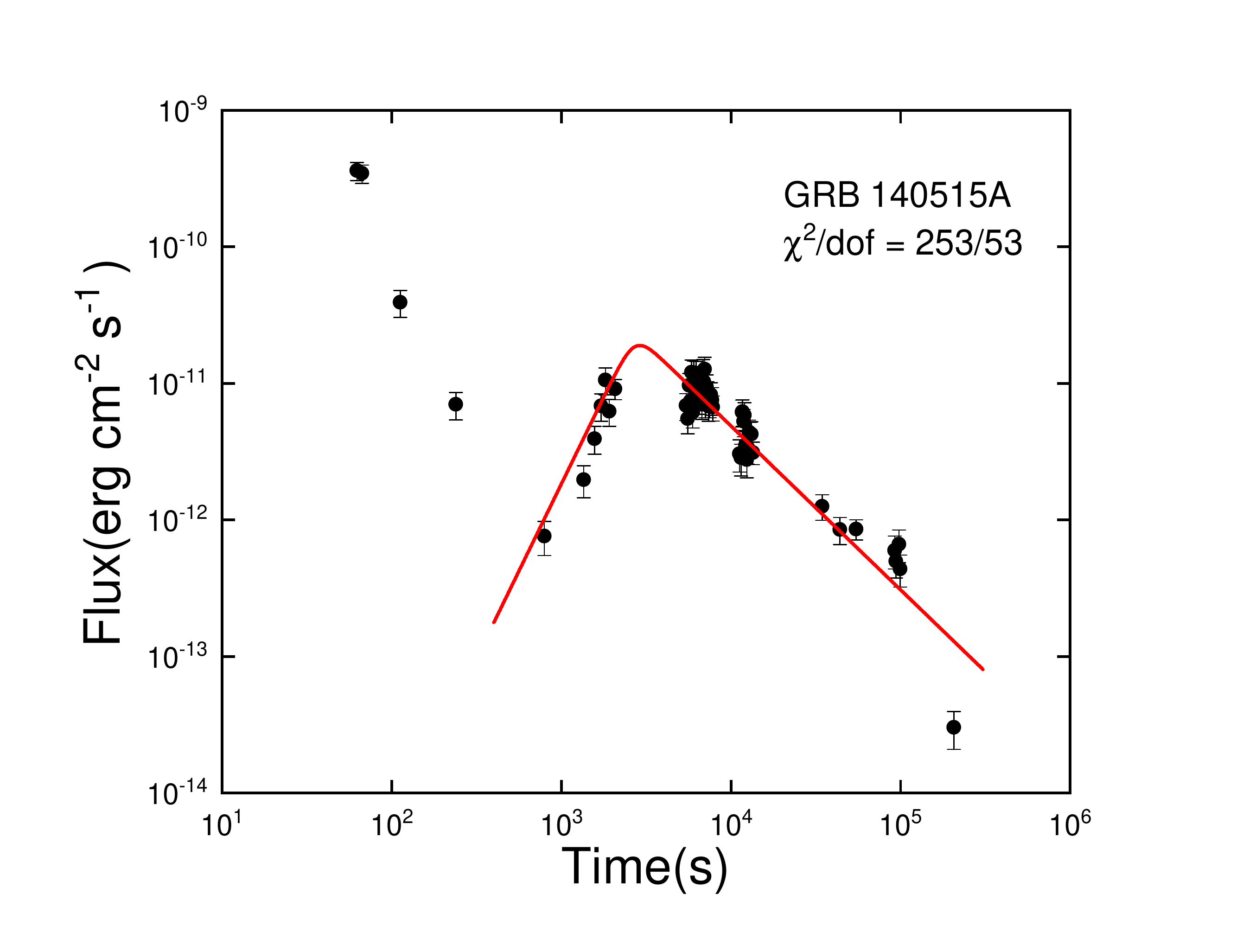}\includegraphics[width=6cm,height=4.5cm]{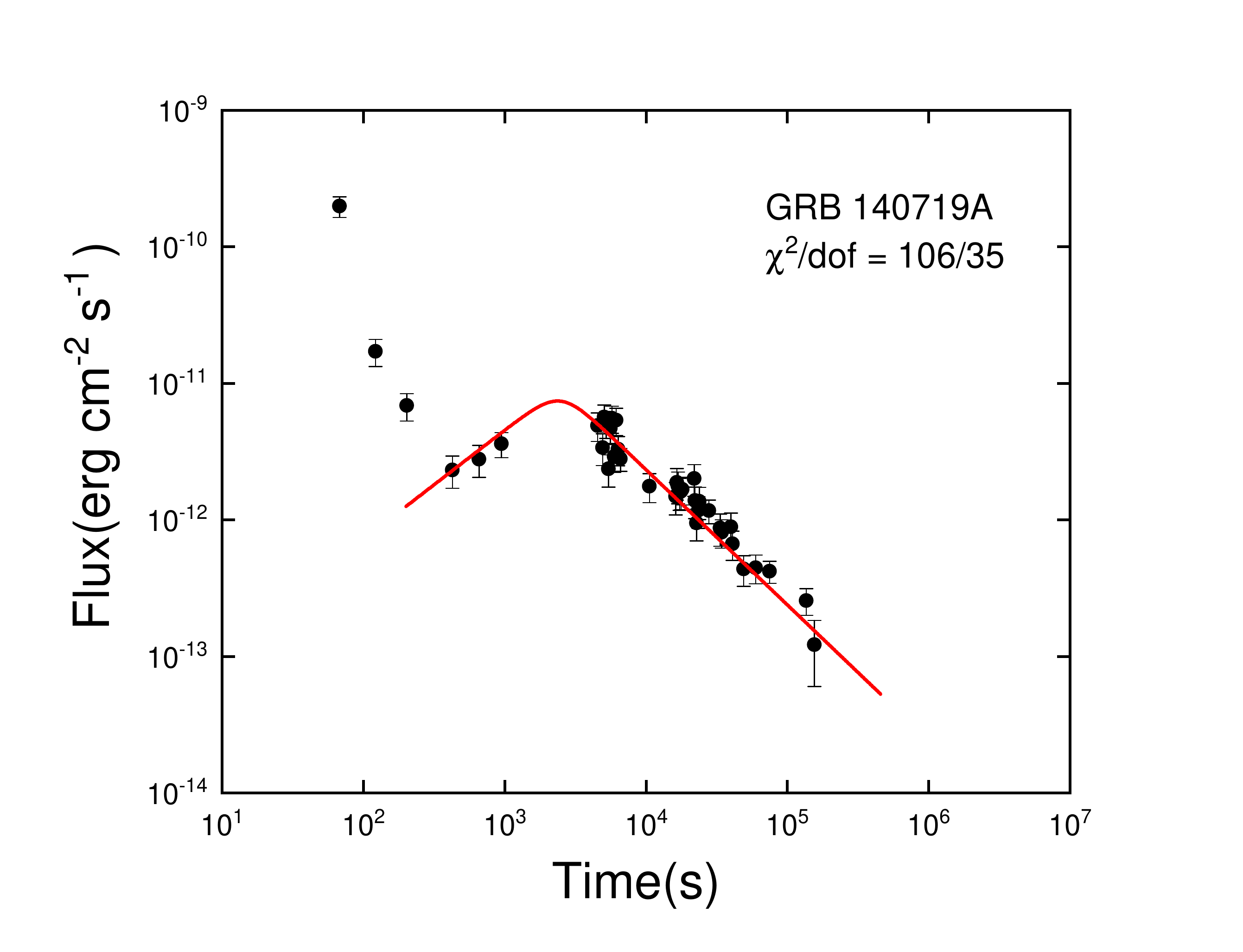}\includegraphics[width=6cm,height=4.5cm]{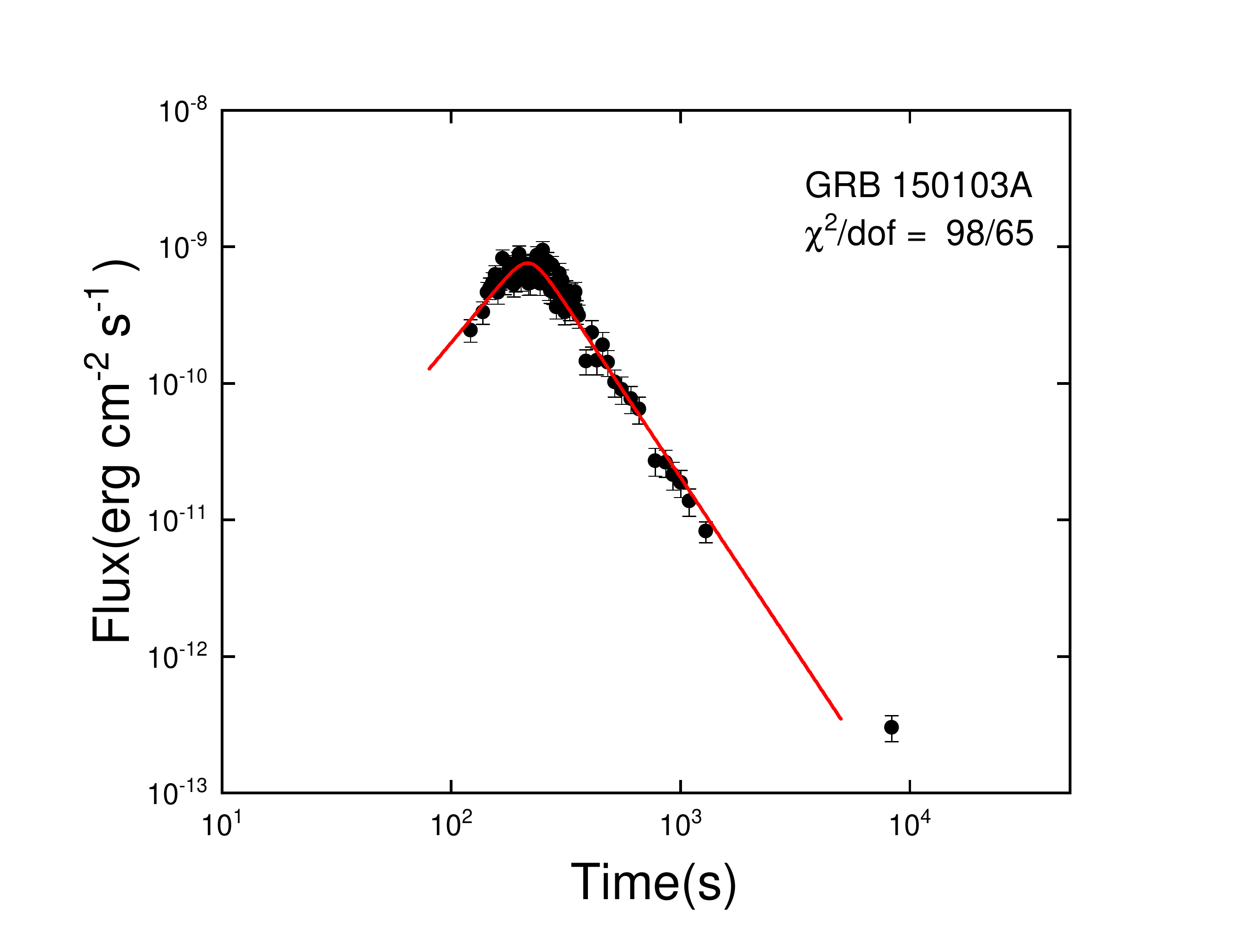}
\includegraphics[width=6cm,height=4.5cm]{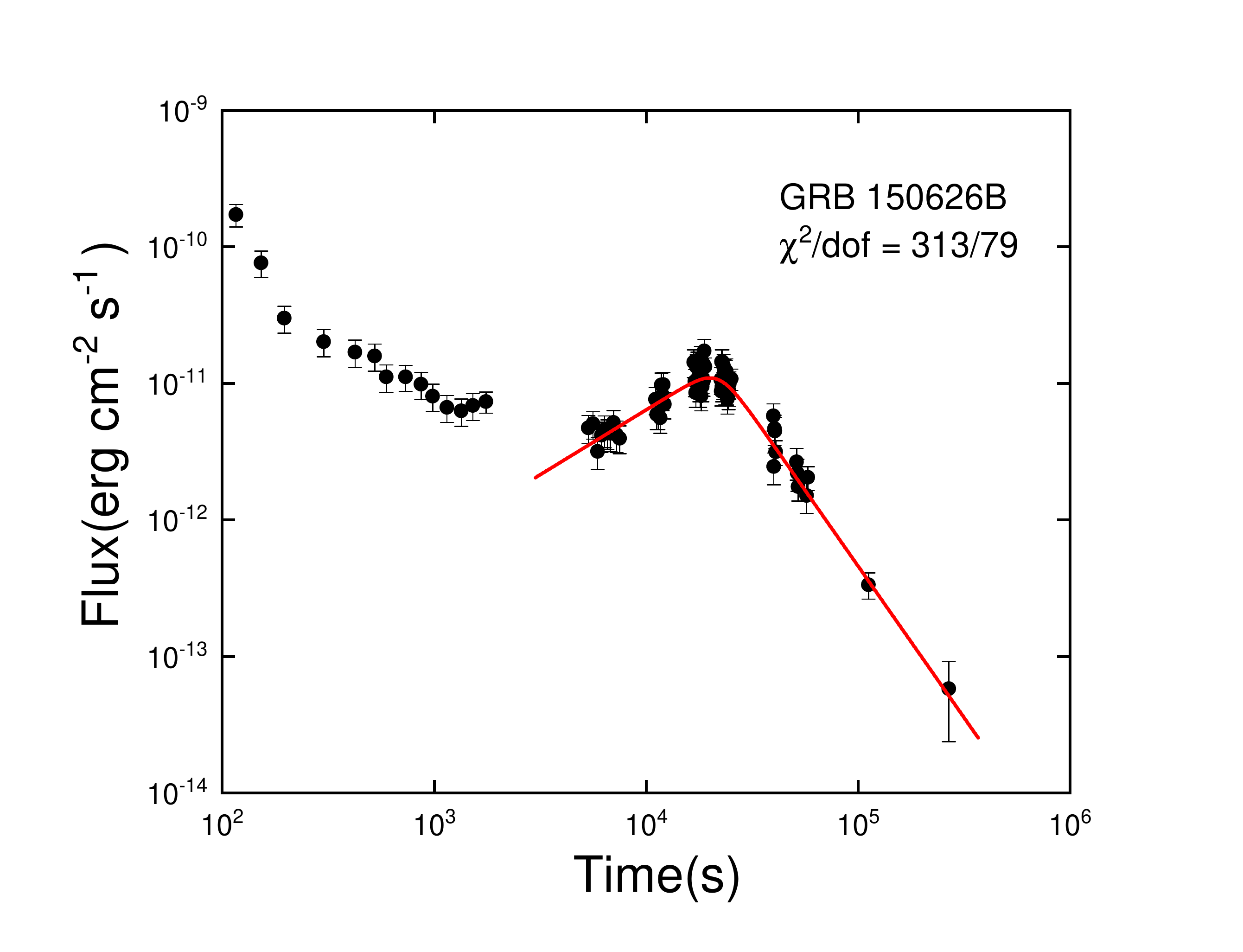}\includegraphics[width=6cm,height=4.5cm]{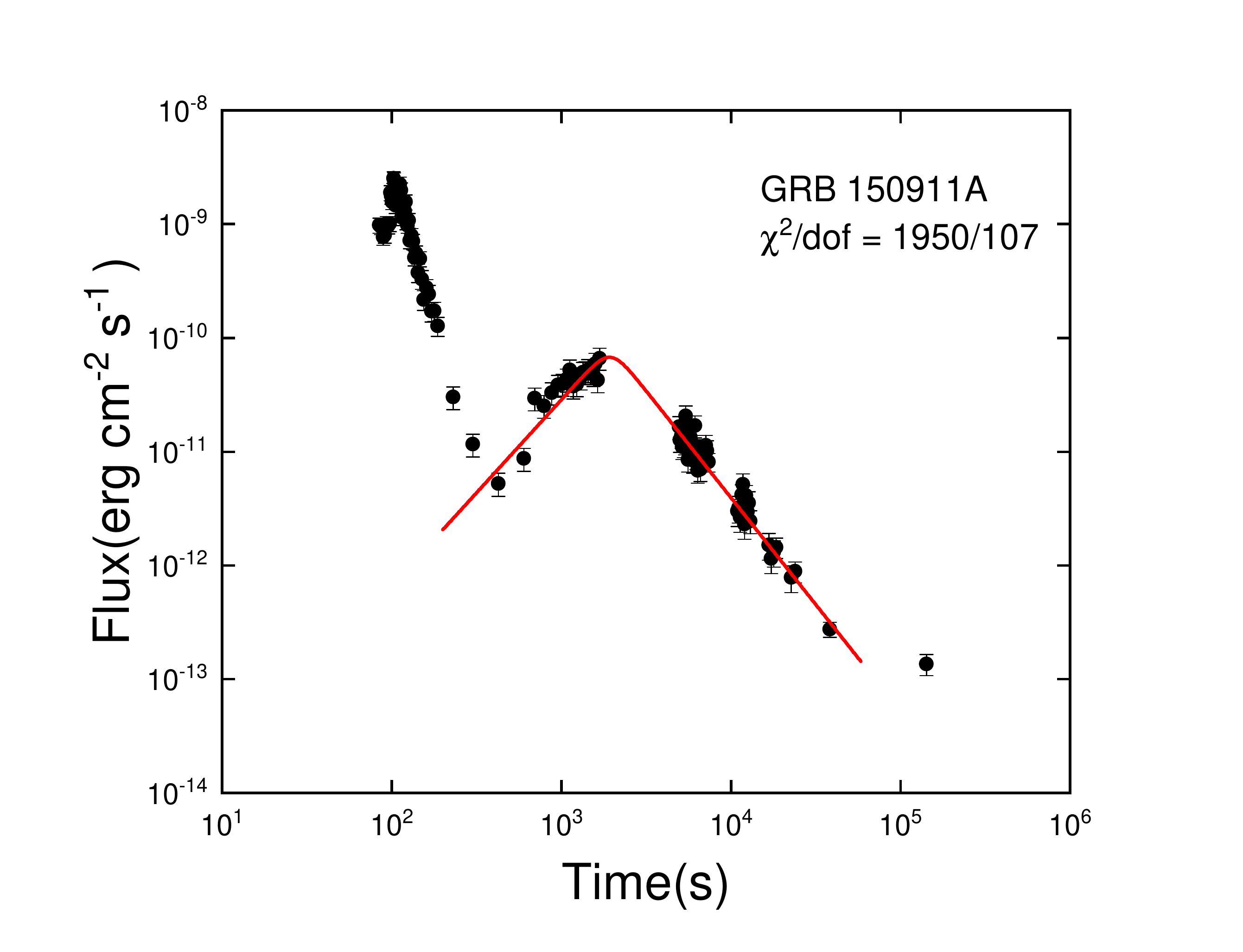}\includegraphics[width=6cm,height=4.5cm]{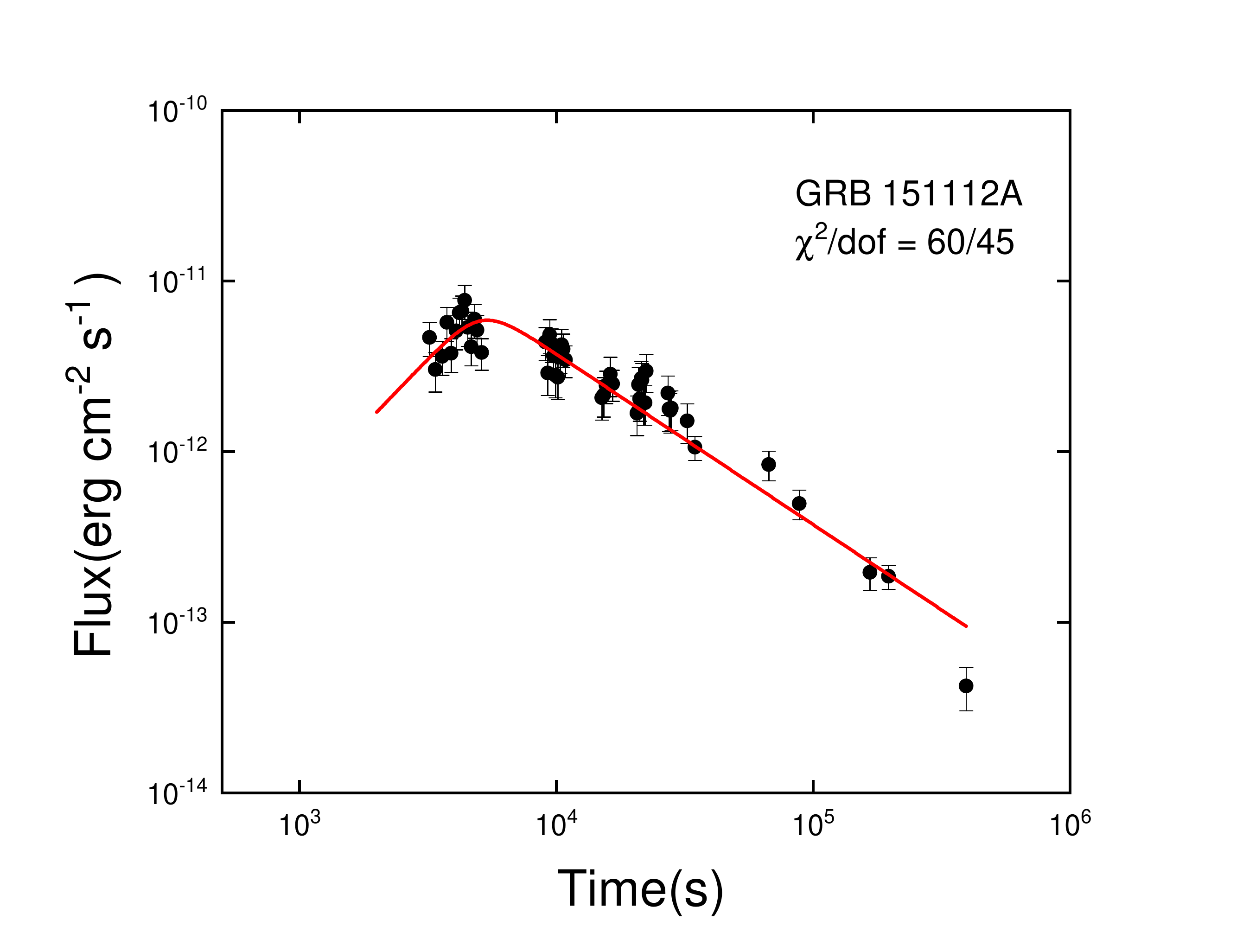}
\center{ Figure 1 (Continued)}
\end{figure*}

\begin{figure*}
\includegraphics[width=6cm,height=4.5cm]{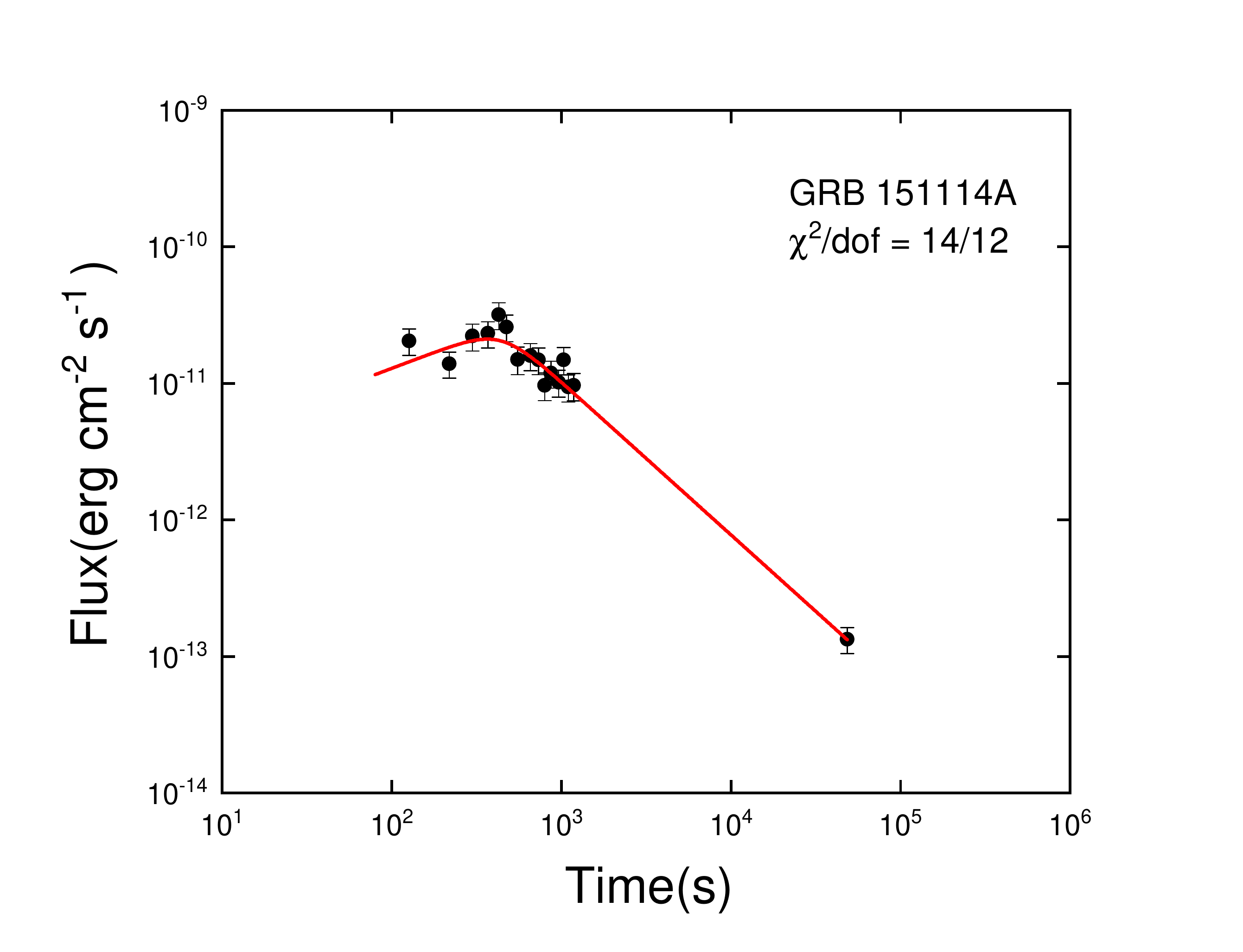}\includegraphics[width=6cm,height=4.5cm]{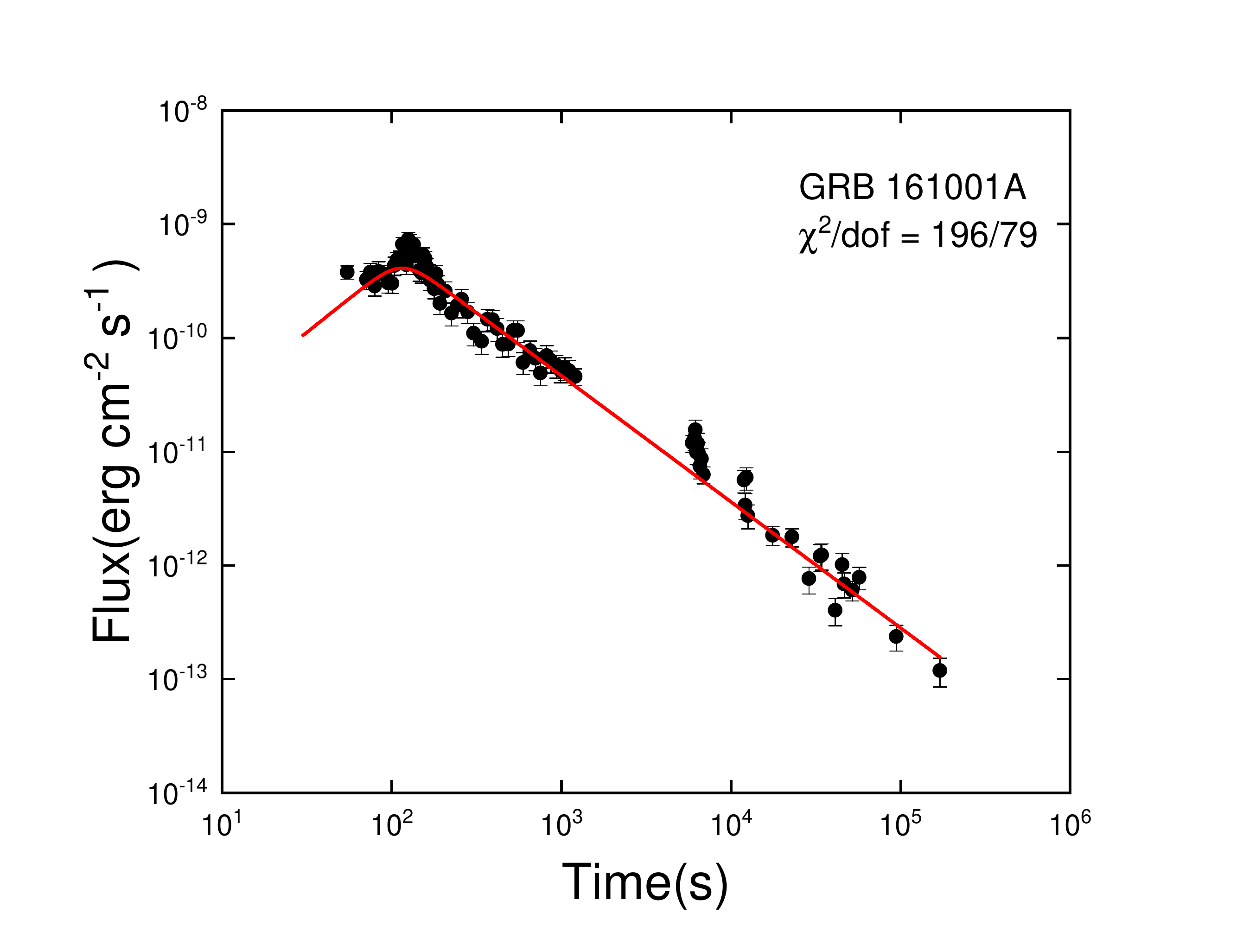}\includegraphics[width=6cm,height=4.5cm]{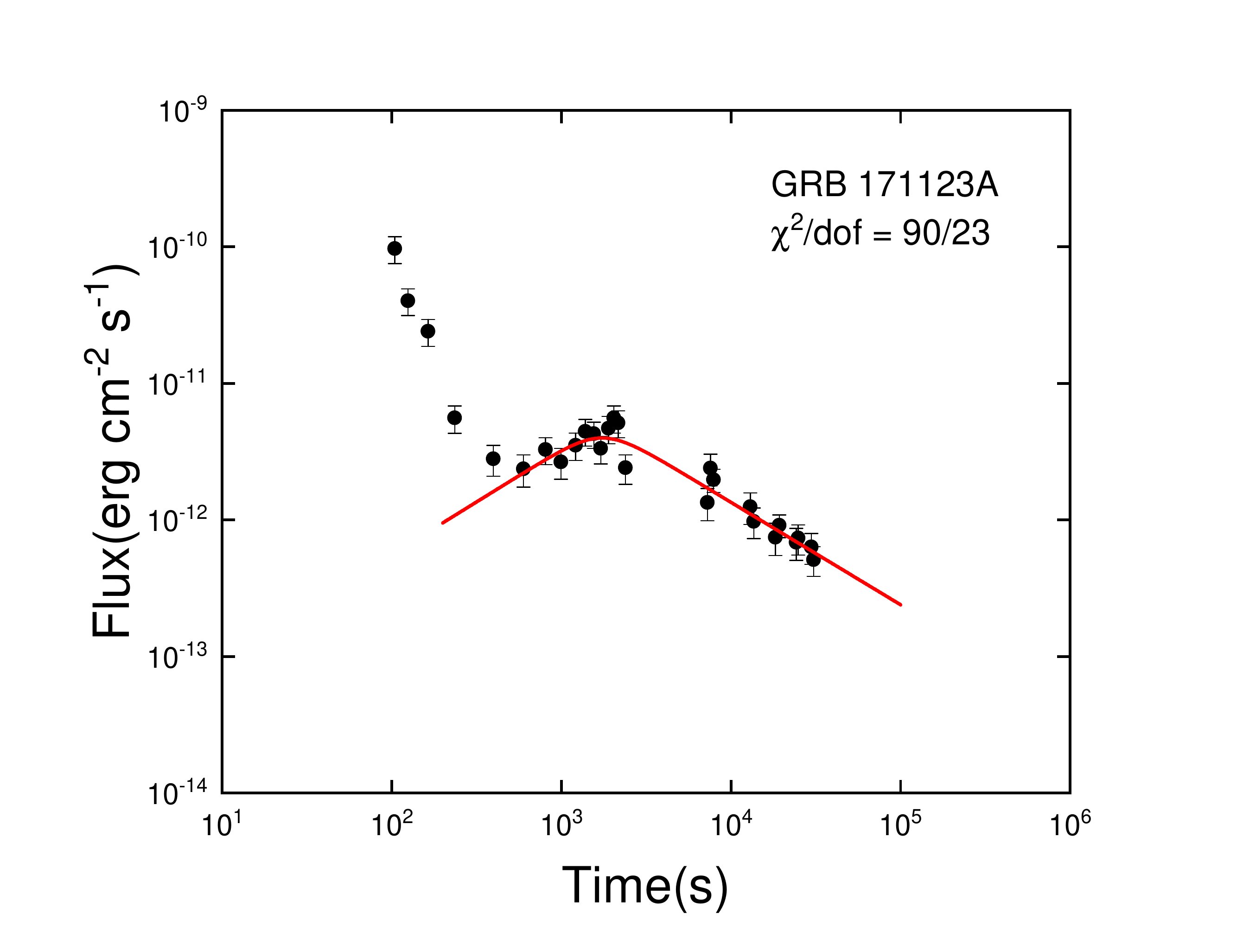}
\includegraphics[width=6cm,height=4.5cm]{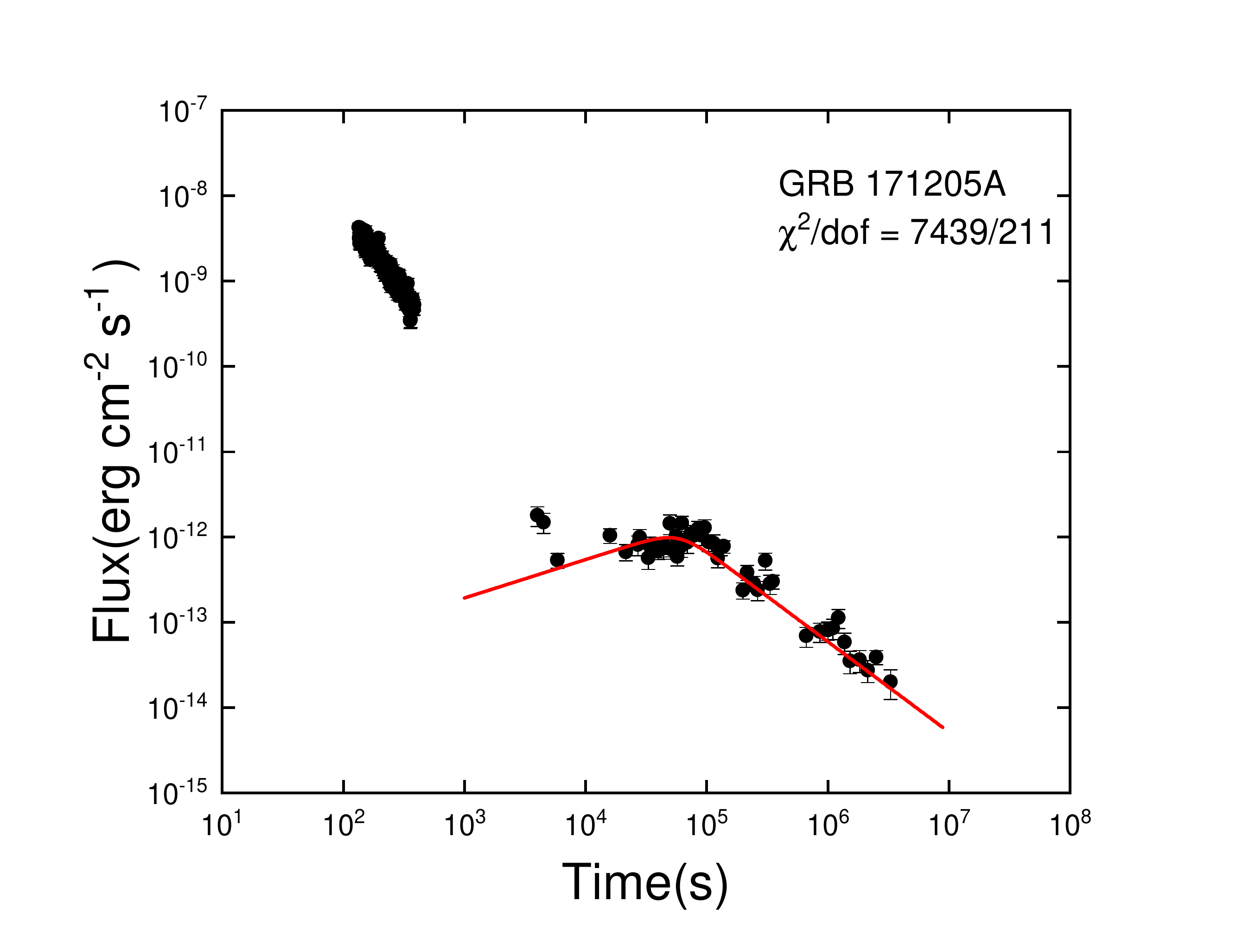}\includegraphics[width=6cm,height=4.5cm]{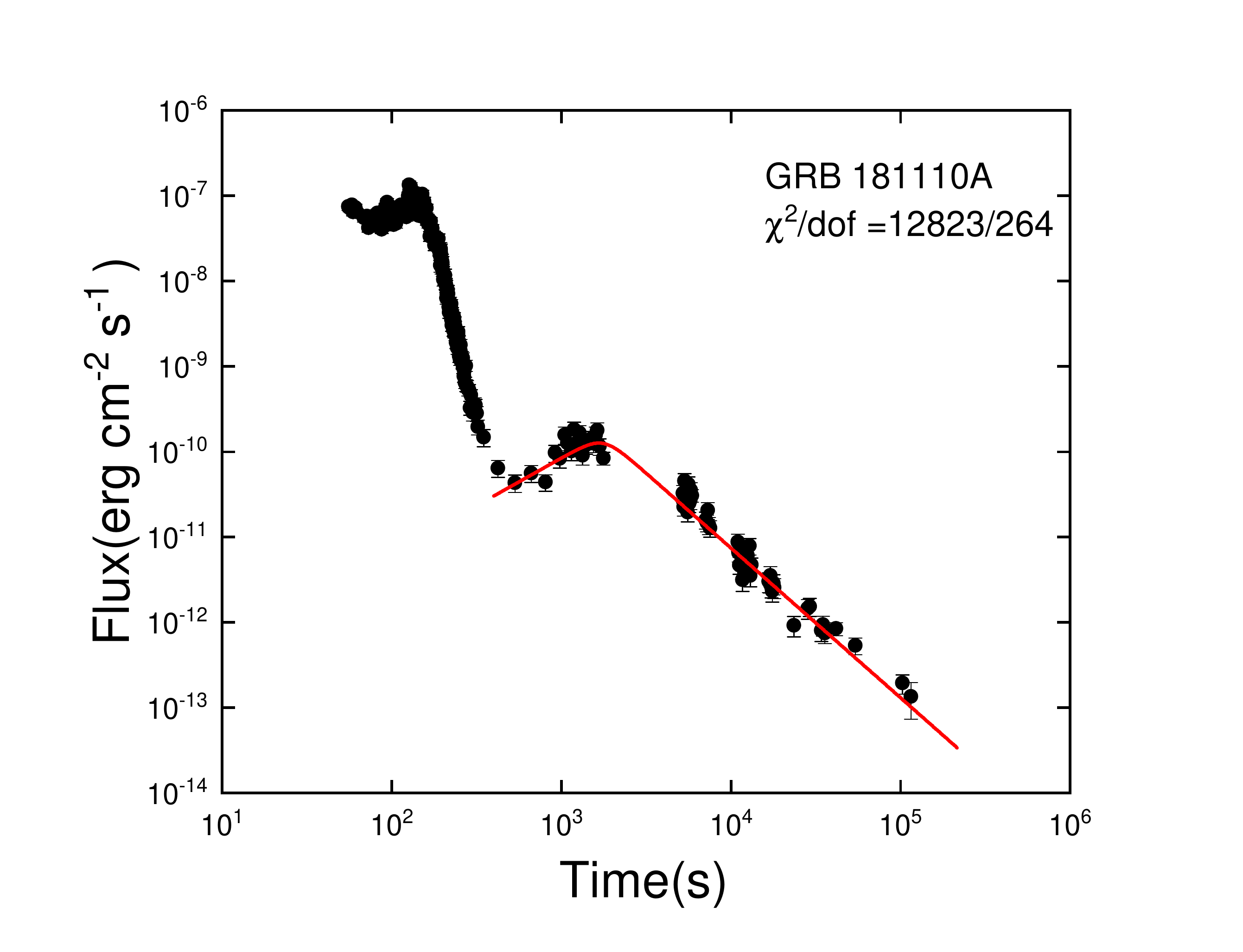}\includegraphics[width=6cm,height=4.5cm]{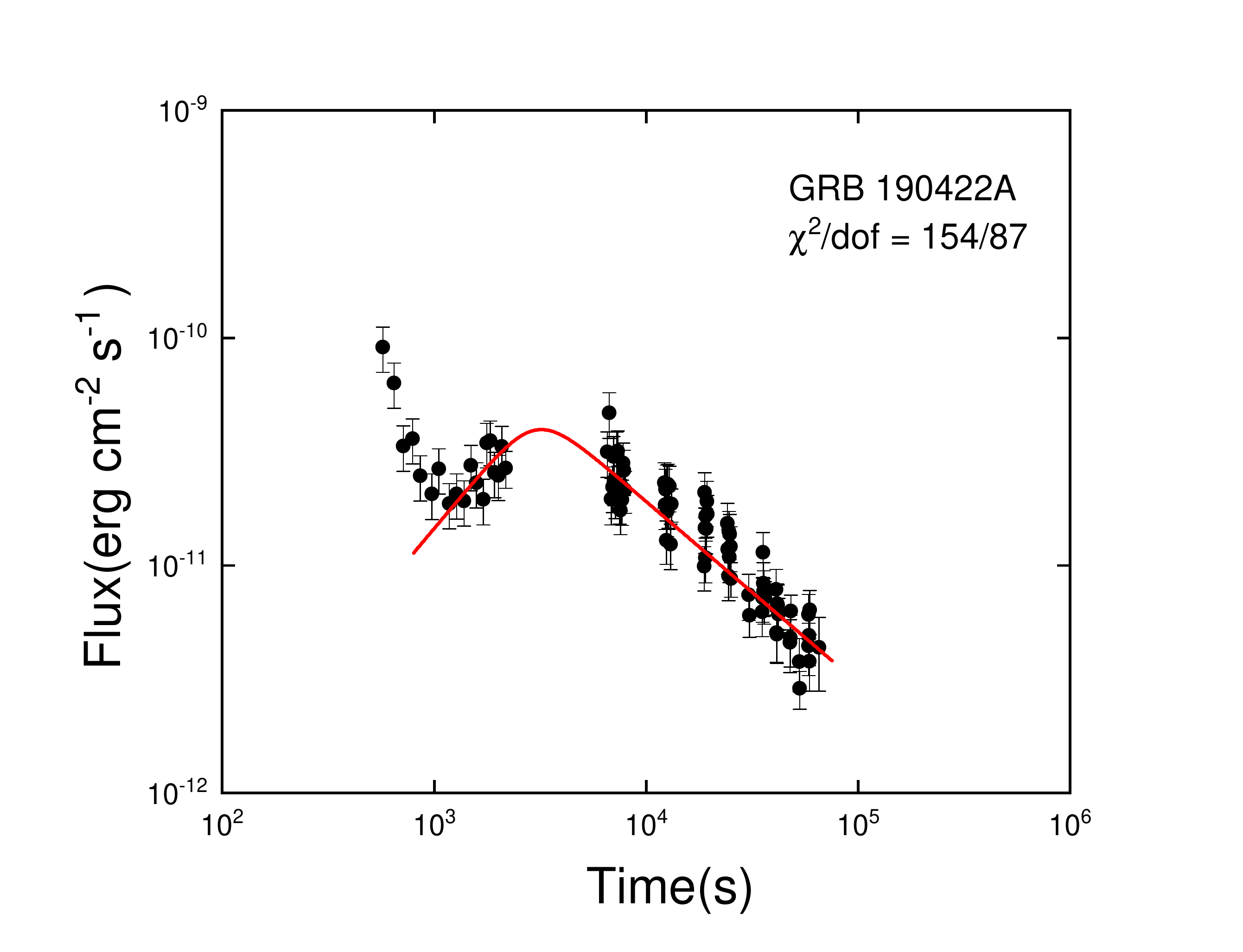}
\includegraphics[width=6cm,height=4.5cm]{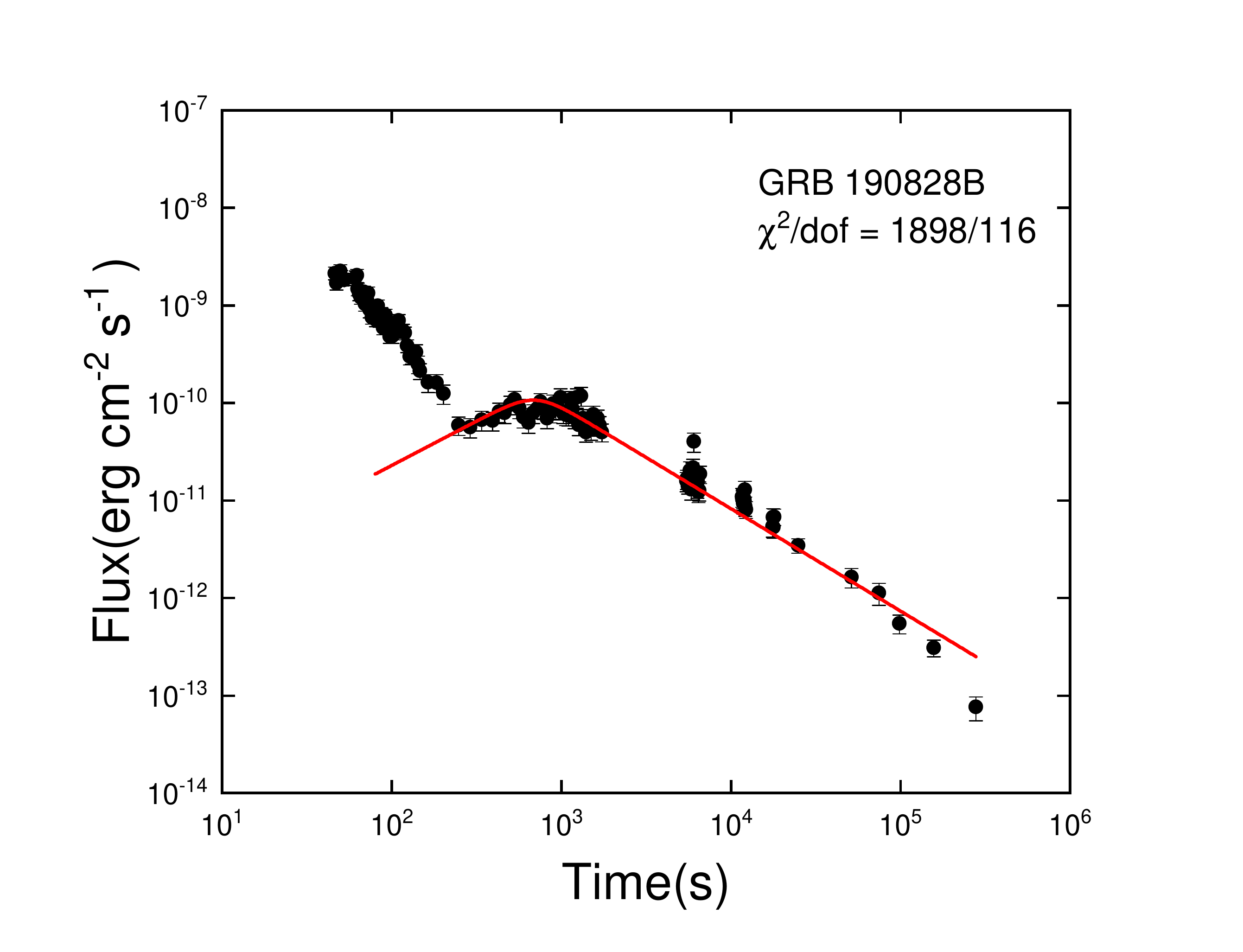}\includegraphics[width=6cm,height=4.5cm]{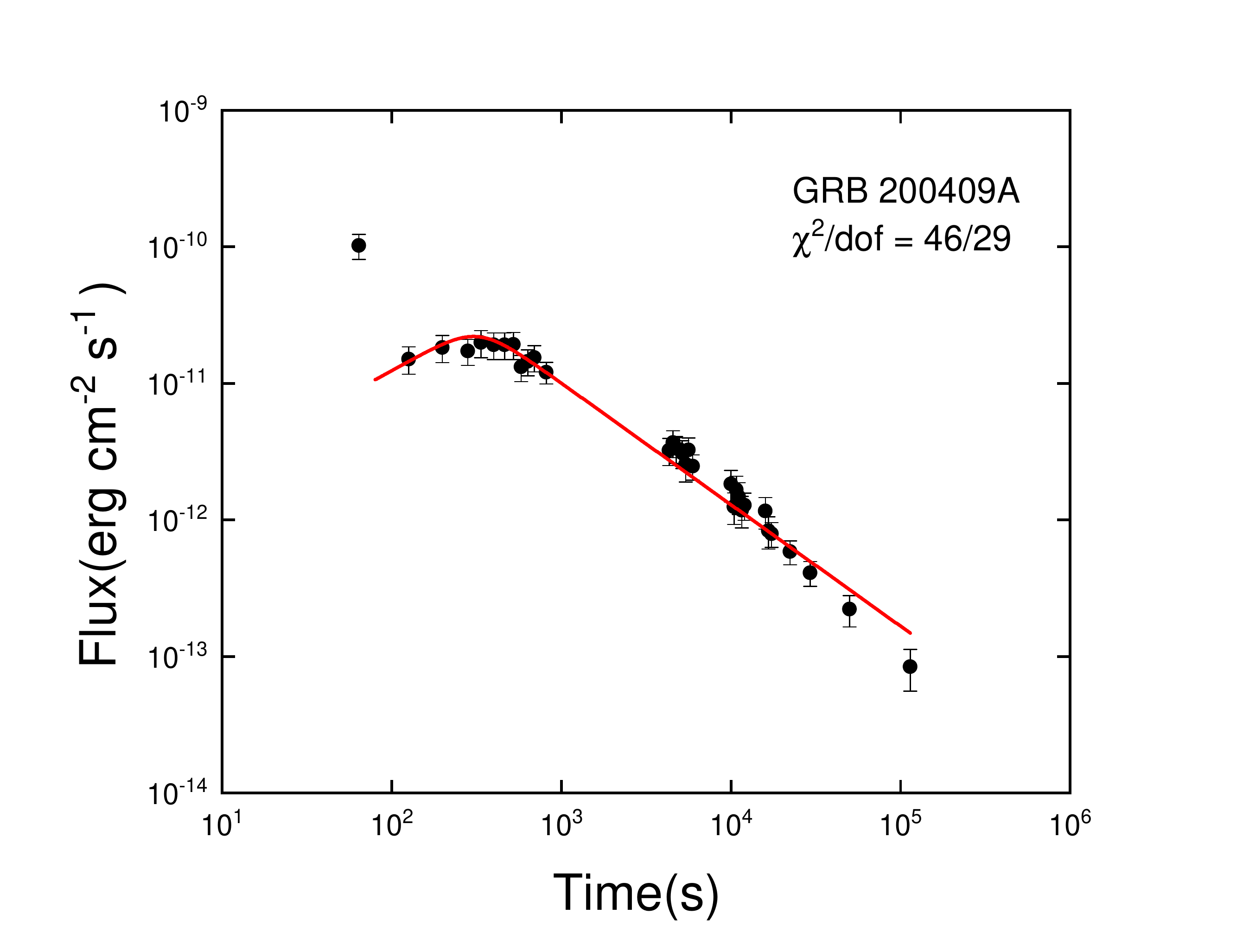}\includegraphics[width=6cm,height=4.5cm]{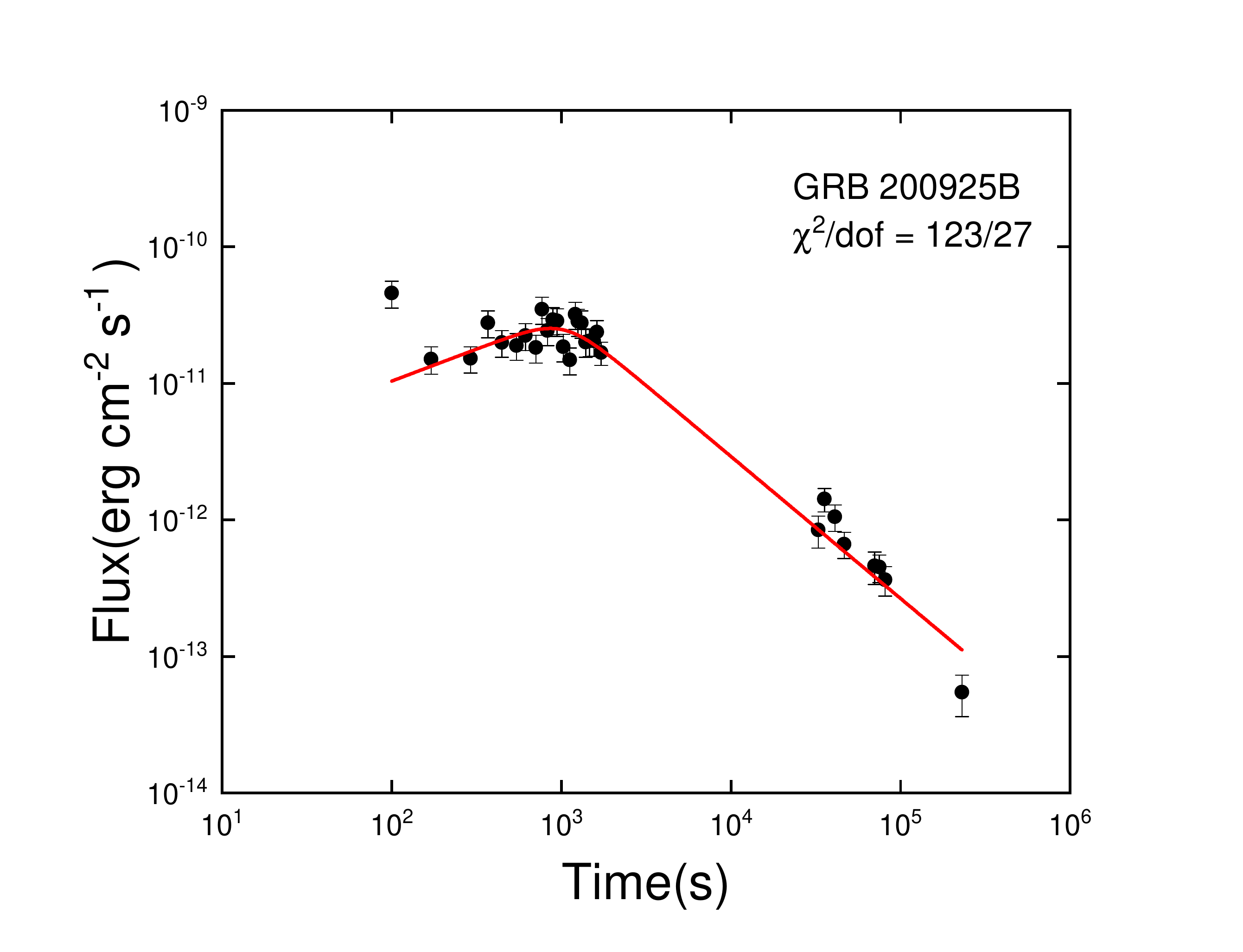}
\center{ Figure 1 (Continued)}
\end{figure*}

\section{CHARACTERISTICS OF THE ONSET BUMP AND THEIR CORRELATIONS}

The external FS model can well explain the optical afterglow with the onset characteristics, which is also shown in our currently selected X-ray sample. The rise index $\alpha_1$ of most GRBs is found in the range of $0.3 - 2.6$ and the central decay index $\alpha_2$ is around -$1.0$. We adopt the full width at half-maximum (FWHM) of the fitted curve as the feature width ($\omega$) of the bump. The rise and decay time scales ($T_{\rm r}$ and $T_{\rm d}$) are defined at FWHM. We find that the peak time $T_{\rm p}$ and the width $\omega$ have the same range, i.e., $10^2$$-$$10^4$s. In addition, the decay time ($T_{\rm d}$) is larger than the rise time ($T_{\rm r}$).

By using the Spearman pair correlation analysis, we obtain the corresponding coefficients.  In Figure 2, we present pair correlations for fitted parameters, as well as the best fitting result together with its 95\% credibility in red. Interestingly, we find tight correlations among rise time, decay time, the peak time and the feature width. These correlations are given as follows:

\begin{eqnarray}
\log T_{\rm d} = (0.44 \pm 0.11) + (1.00 \pm 0.03) \log T_{\rm r},
\\[5mm]
\log T_{\rm r} = (-0.10 \pm 0.07) + (0.95 \pm 0.02) \log T_{\rm p},
\\[5mm]
\log T_{\rm d} = (0.25 \pm 0.11) + (0.98 \pm 0.03) \log T_{\rm p},
\\[5mm]
\log \omega = (0.38 \pm 0.10) + (0.98 \pm 0.03) \log T_{\rm p}.
\end{eqnarray}

These pair correlations show well positive linear correlations, indicating that these X-ray bumps have a universal physical origin. We find that the results of these fitting coefficients are similar to those in \cite{2010ApJ...725.2209L}. This further indicates that the selected X-ray sample has the same onset characteristics as the optical light curves. Therefore, these X-ray bumps can also be explained by the external FS model. We also note that a wider X-ray bump ($\omega$) tends to peak at a later time ($T_{\rm p}$), which is consistent with the optical onset bumps.

The finding above is in line with the expectations of the external shock model, as the later deceleration time, corresponding to a smaller Lorentz factor, is equivalent to the peak time of onset bumps for the thin shell case. We then apply the theoretical FS model to these X-ray samples, and identify the type of circumburst medium in the following sections.

\begin{figure*}[htbp!]
\center
\includegraphics[angle=0,width=0.40\textwidth]{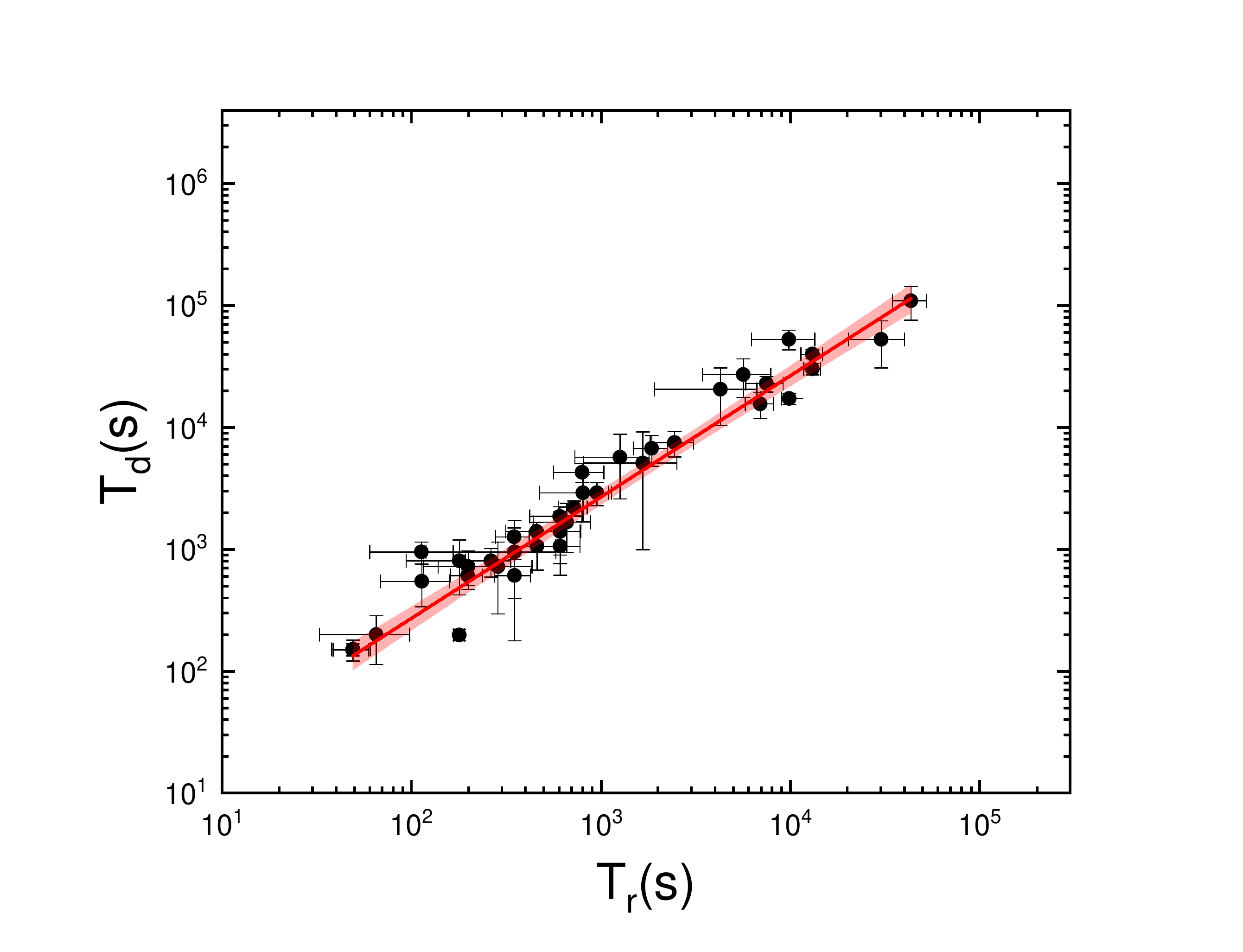}\includegraphics[angle=0,width=0.40\textwidth]{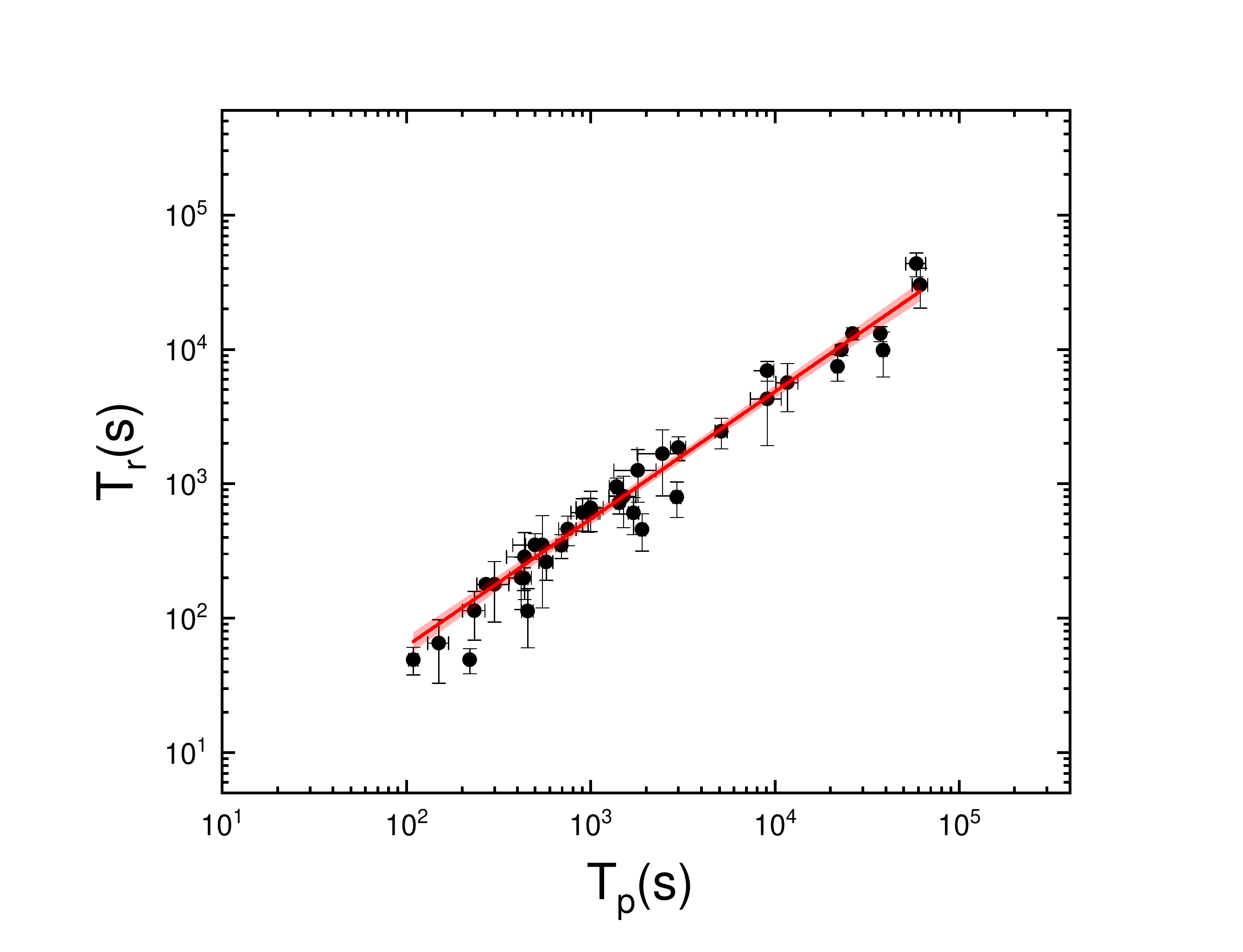}
\includegraphics[angle=0,width=0.40\textwidth]{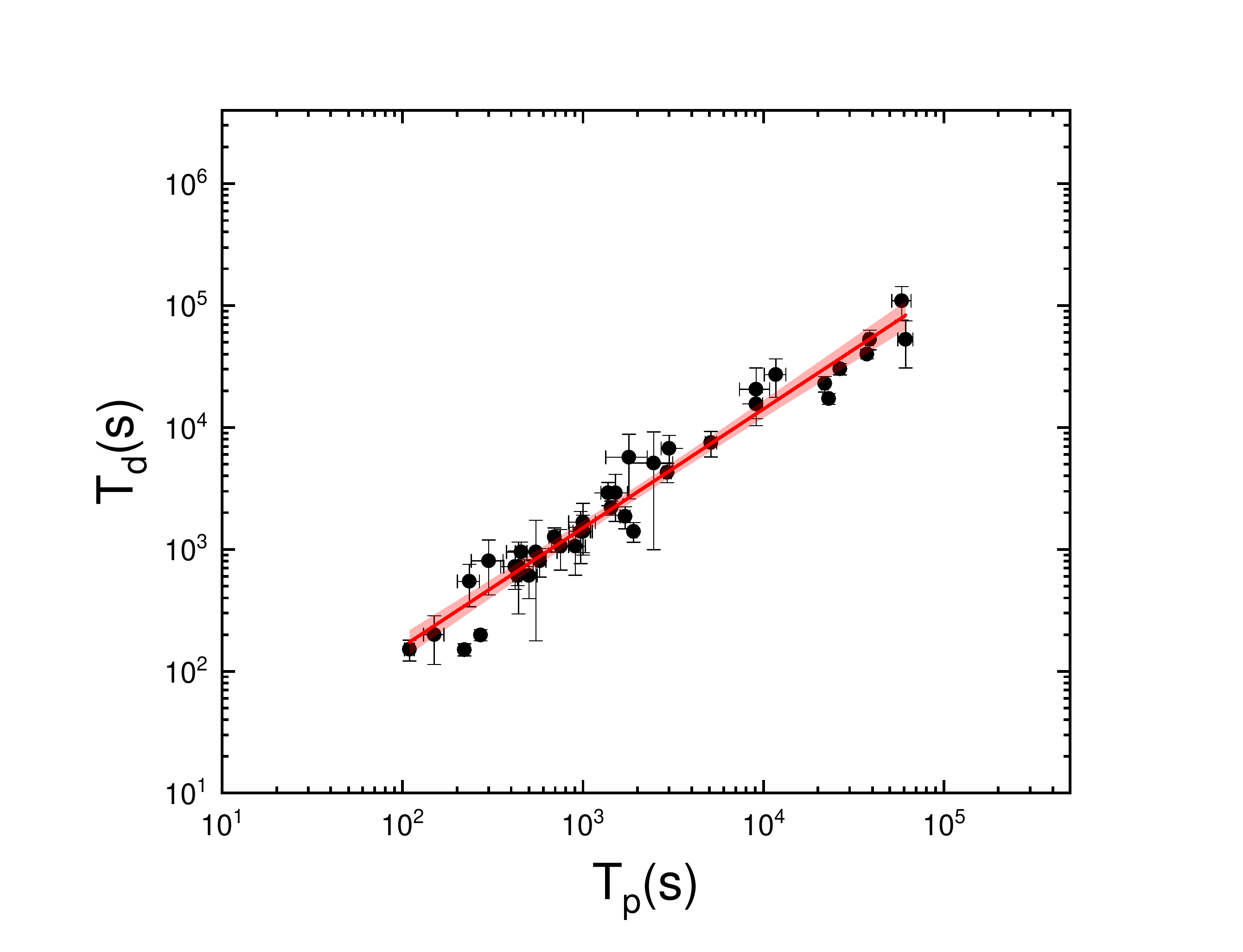}\includegraphics[angle=0,width=0.40\textwidth]{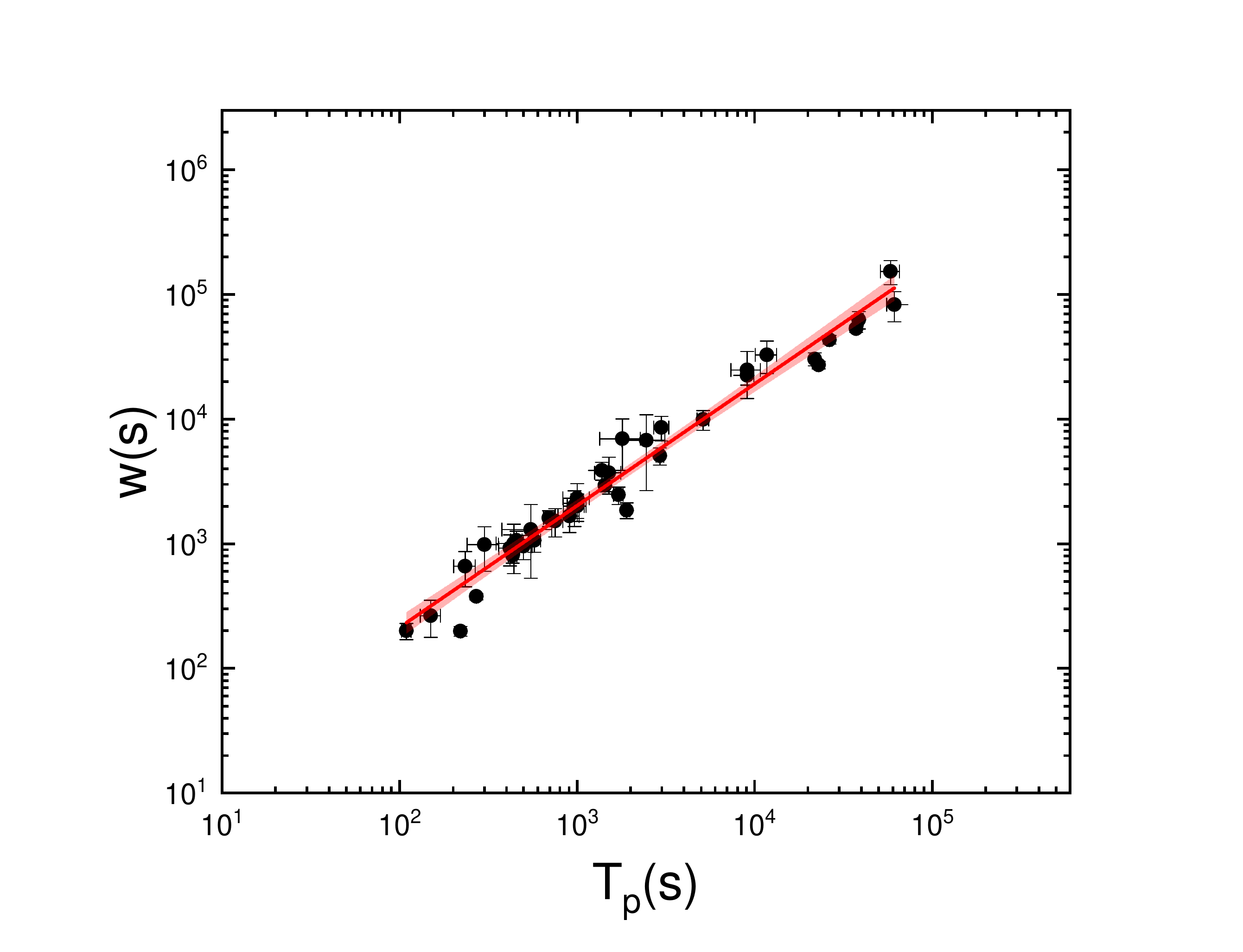}
\caption{ Pair correlations among the fitting parameters of the selected sample. Lines are the best fits. }
\end{figure*}

\section{ANALYSES AND RESULTS}
\label{Sec: }
As discussed above, the observational properties of the selected X-ray bumps could be produced by the deceleration of a fireball shell. Therefore, we apply the external FS model to those X-ray bumps. In the standard FS model, the theoretical rise index is 3 in the regime $\nu_m^f<\nu<\nu_c^f$ (or 2 for $\nu>max\{\nu_c^f,\nu_m^f\}$) for the ISM case, and the index is 0.5 for the interstellar wind environment. Since the rise index of most X-ray bumps is in the range of $0.3 - 2.6$, it's hardly to determine the real type of the circumburst medium of those GRBs. Therefore, we assume that the general density distribution of long GRB ambient medium is $n = AR^{\rm -k}$, where $n$ is the density of proton number, $R$ is the radius, and $k$ is the density distribution index of the medium. Through previous works, we obtain the value $k =0$ for the interstellar medium case and $k =2$ for the interstellar wind environment, respectively.

A large number of optical afterglows have been collected since 1997, some of {which} show clear smooth bumps and sharp RS signatures for different optical light curves \citep{2010ApJ...725.2209L, 2014ApJ...785...84J}. We successfully applied the external shock model of thin shell case to those optical afterglows with the onset bump features and signatures of dominant RS emission respectively, as well as identified the types of GRB ambient medium \citep{2013ApJ...776..120Y,2020ApJ...895...94Y,2020IJMPD..2950043Z}.
Following the previous works, we now continue to constrain the types of GRB circumburst medium with the selected X-ray onset bumps in this work. The method we adopt here is the same as that proposed before. The FS theoretical flux density of the thin shell before and after the peak time are \citep{2013ApJ...776..120Y,2020IJMPD..2950043Z}:

\begin{equation}
F_{\nu}^{FS}(t<T_{\Delta}) \propto \left\{\begin{array}{l}
t^{\frac{{{8-2k-kp}}}{{{4}}}}\nu^{-{\frac{{{p}}}{{{2}}}}},\,\,\nu>max\{\nu_c^f,\nu_m^f\}\\
t^{\frac{{{12-5k-kp}}}{{{4}}}}\nu^{-{\frac{{{p-1}}}{{{2}}}}},\,\,\nu_m^f<\nu<\nu_c^f\\
t^{\frac{{{8-3k}}}{{{4}}}}\nu^{-{\frac{{{1}}}{{{2}}}}},\,\,\nu_c^f<\nu<\nu_m^f
\end{array} \right.
\end{equation}

and

\begin{equation}
F_{\nu}^{FS}(t>T_{\Delta}) \propto \left\{\begin{array}{l}
t^{-{\frac{{{3p-2}}}{{{4}}}}}\nu^{-{\frac{{{p}}}{{{2}}}}},\,\,\nu>max\{\nu_c^f,\nu_m^f\}\\
t^{-{\frac{{{12p-3kp-12+5k}}}{{{4(4-k)}}}}}\nu^{-{\frac{{{p-1}}}{{{2}}}}},\,\,\nu_m^f<\nu<\nu_c^f\\
t^{-{\frac{{{1}}}{{{4}}}}}\nu^{-{\frac{{{1}}}{{{2}}}}},\,\,\nu_c^f<\nu<\nu_m^f
\end{array} \right.
\end{equation}
where the $\nu_m^f$ is the typical frequency, the $\nu_c^f$ is cooling frequency and $T_{\rm \Delta}$ is the RS crossing time. More details can be found in \cite{2013ApJ...776..120Y}. The real conditions of GRB ambient medium may be more complex, but it should be a viable approach when considering both temporal slopes and X-ray spectral indexes for determining the types of GRB circumburst medium.
Since the medium density distribution index $k$ and electron spectral index $p$ are two factors in determining the slopes before and after the peak time of light curves. Here, we do not consider the regime of $\nu<\{\nu_c^f,\nu_m^f\}$, as it is very rare to show in X-ray band. In the following, we will take GRB 070103 and GRB 090429B as two typical examples to elaborate how the index $k$  and $p$ are determined.

\subsection{GRB 070103}
GRB 070103 was detected by $Swift/BAT$ with a duration $T_{\rm 90} = 18.6$ s and thus belongs to a long GRB. Therefore, its circumburst medium is most likely to be a stellar wind environment, as long GRBs are widely believed to originate from the core collapse of massive stars. In the afterglow light curve, the flux increases with the time, i.e., $F_{\nu} \propto t^{0.55 \pm 0.16}$ before the peak time $T_{\rm p} = 905 \pm 124 $ s, and then decays after the peak, i.e., $F_{\nu} \propto t^{-1.32 \pm 0.05}$. We then consider both $\nu>max\{\nu_c^f,\nu_m^f\}$ and $\nu_m^f<\nu<\nu_c^f$ regimes to determine the circumburst medium of GRB 070103.

(1) $\nu>max\{\nu_c^f,\nu_m^f\}$. According to the Equations (6) and (7), we know that the temporal slopes of the onset bumps are determined by the medium density distribution index $k$ and the electron spectral index $p$. We then have the decay index $\alpha_2$ = $(3p-2)/4$ = $1.32 \pm 0.05$, and thus the value of $p$ = 2.43 $\pm$ 0.06. Similarly, the rise index $\alpha_1 = (8-2k-kp)/4$ = $0.55 \pm 0.16$, so $k = (8-4\alpha_1)/(2+p) = 1.31 \pm 0.14$. As we also have the expression $p = 2\beta_{\rm x}$ = 2.20 $\pm$ 0.54, then the X-ray spectral index $\beta_{\rm x} = 1.10 \pm 0.27$ in regime of $\nu>max\{\nu_c^f,\nu_m^f\}$, which is consistent with the value calculated by the temporal slopes. In this case, the values $k$ and $p$ are both reasonable (generally supposed 0 < $k$ < 2 and 2< $p$ < 3). We then apply the regime $\nu>max\{\nu_c^f,\nu_m^f\}$ to infer the ambient medium of GRB 070103, and obtain $k = 1.31 \pm0.14$ and $p =2.43\pm0.06$, respectively.

(2) $\nu_m^f<\nu<\nu_c^f$. According to the theoretical analysis above, we obtain $\alpha_1  = (12-5k-kp)/4$ and $\alpha_2 = (12p-3kp-12+5k)/(16-4k)$, as well as  $k = 1.32 \pm 0.09$ and $p = 2.43 \pm 0.07$ for the temporal slopes. In this case, $p = 2\beta_{\rm x}+1 = 3.20 \pm 0.54$, which is not consistent with that derived from the temporal indexes. As a result, this case $\nu_m^f<\nu<\nu_c^f$ may not be suitable to identify the type of the ambient medium for GRB 070103.

We conclude that the $\nu>max\{\nu_c^f,\nu_m^f\}$ case might be used to explain the ambient medium of GRB 070103. We then obtain the medium density profile of the GRB 070103, i.e., $n \propto R^{\rm -1.31}$. This indicates that the ambient medium of GRB 070103 is neither a homogeneous ISM for $k = 0$, nor the typical interstellar wind environment for $k = 2$. It indicates that the ambient medium of GRB 070103 may be in the intermediate regime, between ISM and the stellar wind.

\subsection{GRB 090429B}

The GRB 090429B has a duration $T_{\rm 90} = 5.5 s $. In our study, we fit the afterglow by using an empirical smooth broken power-law function.  Then we have the rise index $\alpha_1 = 1.57 \pm 0.34$, and the decay index $\alpha_2 = -1.40 \pm 0.06$. We also consider the two following regimes $\nu>max\{\nu_c^f,\nu_m^f\}$ and $\nu_m^f<\nu<\nu_c^f$, respectively.

(1) $\nu>max\{\nu_c^f,\nu_m^f\}$. In this case, we obtain $k = 0.38 \pm 0.29$ and $p = 2.53 \pm 0.07$ for rise-decay indices. Combining the theoretical spectral index, we obtain $p = 2\beta_{\rm x} = 1.74 \pm 0.44$ with $\beta_{\rm x}= 0.87 \pm 0.22$ for GRB 090429B. This result is inconsistent with the constraint of the temporal slopes. Therefore, the regime of $\nu>max\{\nu_c^f,\nu_m^f\}$ may be not suitable for the X-ray light curve of GRB 090429B.

(2) $\nu_m^f<\nu<\nu_c^f$. In this case, $p = 2\beta_{\rm x}+1 = 2.74 \pm0.44$. Combining with the rise-decay indices, we have the rise index $\alpha_1  = (12-5k-kp)/4$ and the decay index $\alpha_2 = (12p-3kp-12+5k)/(16-4k)$, respectively.  We then obtain $k = 0.74 \pm 0.17$ and $p = 2.71 \pm 0.09$, respectively. Both $k = 0.74 \pm 0.17$ and $p = 2.71 \pm 0.09$ are in a reasonable range. Therefore, the $\nu_m^f<\nu<\nu_c^f$ case of the FS model can be used to explain the X-ray light curve of GRB 090429B.  And we here adopt the values of $k = 0.74 \pm 0.17$ and $p = 2.71 \pm 0.09$ for GRB 090429B.

After applying the FS model with the uncertainty index $k$ to the rest of X-ray samples, we find that the value $k$ is centered at around 1.0, with a range of $0.2$ and $1.8$. Figure 3 shows the distribution of $k$ and $p$ features. The values of $p$ is between $1.7$ and $3.8$. Our results indicate that the X-ray onset bumps could also be applied to constrain the circumburst medium of long GRBs, further implying that GRBs may have diverse ambient media.

\begin{figure*}[htbp!]
\center
\includegraphics[angle=0,width=0.45\textwidth]{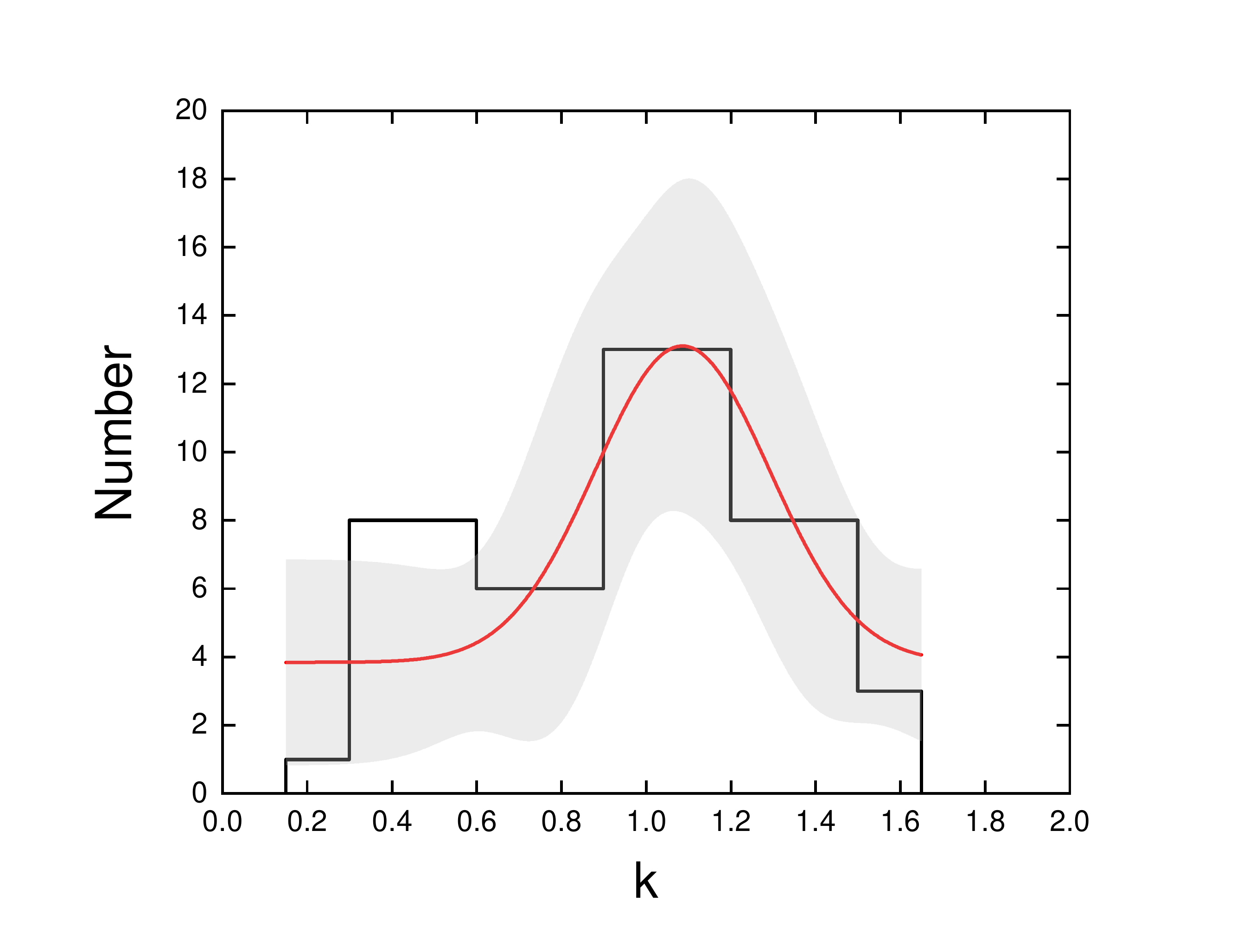}\includegraphics[angle=0,width=0.45\textwidth]{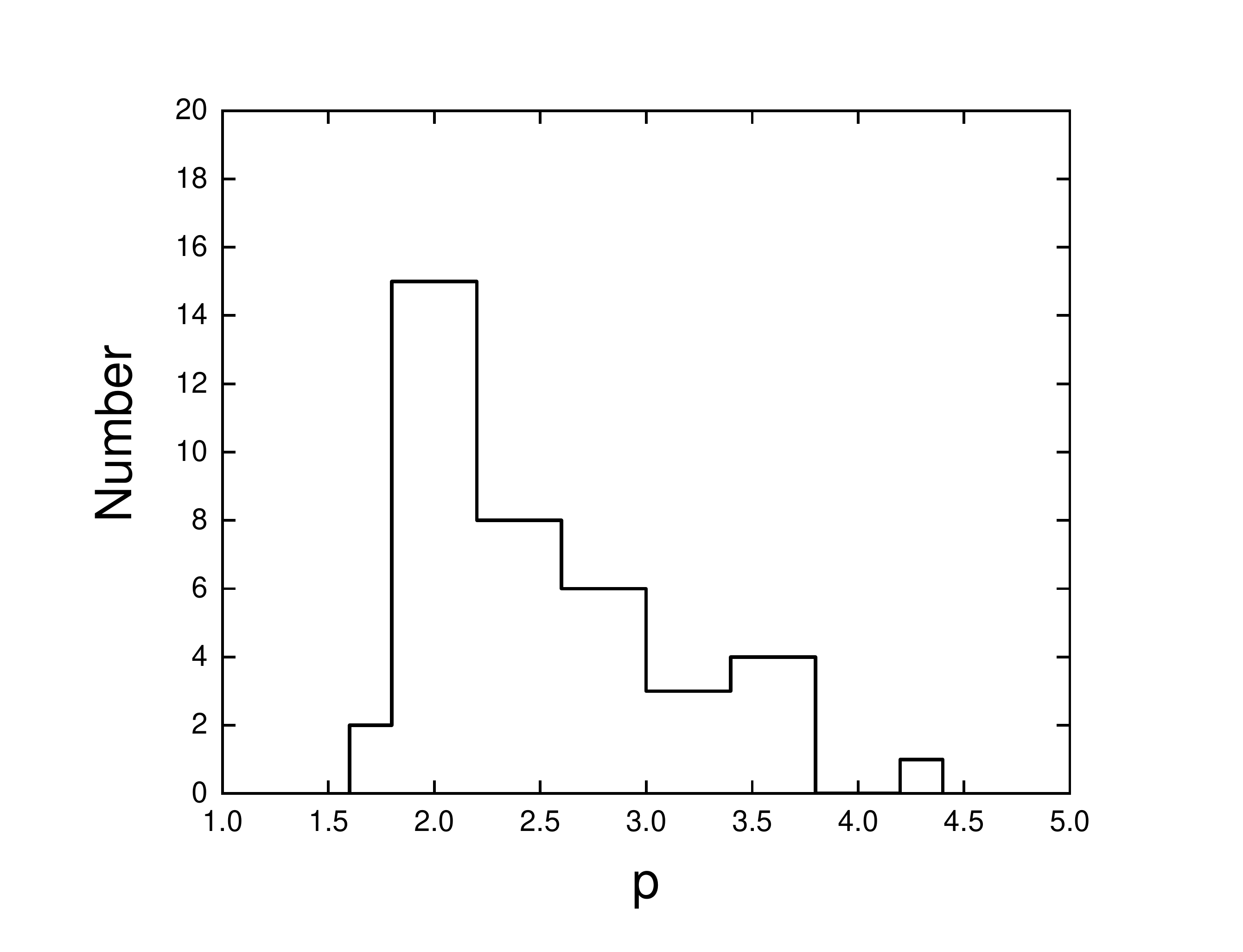}
\caption{ Distributions of the values $k$ and $p$ of our sample. The red line is the Gauss fitting line with 1$\sigma$ error bars.}
\end{figure*}

\section{Discussions}
\label{Sec: conclusions}

The afterglows of GRBs are generally considered as efficient tools to probe the properties of the circumburst medium (\citealt{2013ApJ...776..120Y}; \citealt{2013ApJ...774...13L}). Different authors have reached a consistent result, but with a slight deviation mainly due to the different sample selections. These works involve the constraints from the GRB 970616 \citep{1998MNRAS.298...87D}, from a subsample of ten $BeppoSAX$ GRBs with the simultaneous spectra of X-ray, optical, and NIR afterglow data \citep{2008ApJ...672..433S}, from $Swift$ GRBs with late-time optical and $Swift$/X-ray afterglow \citep{2011A&A...526A..23S}, as well as from a subsample of GRB optical afterglows with onset bump features \citep{2013ApJ...776..120Y,2013ApJ...774...13L,2020IJMPD..2950043Z}. The results from these studies indicate that the circumburst medium of GRBs is not simply a wind environment, or an ISM.

From a theoretical point of view, The circumburst medium of GRBs, in general, is complex. It is accepted that WR stars are the progenitors of long GRBs, however, the mass-loss rates of WR stars are still one of the major uncertainties in modern astrophysics \citep{2020MNRAS.499..873S}. These progenitor stars are expected to have significant stellar winds throughout the last stages of their evolution, which will thus play a key role in the conditions of the ambient environment. Information for the environment around the star after exploding, can be obtained with the observed properties of the afterglow. Two types are generally assumed, i.e., a wind environment ($k$ = 2) and a homogeneous environment (ISM, $k$ = 0). It is expected that the former case is favored based on our current understanding of stellar evolution. However, this is challenged by the observed properties of GBR afterglows.

\section{Conclusions}

In our extensive search for the onset bump characteristics in the early X-ray light curves, we have collected 39 GRBs, 20 of which have the redshift measurements. By applying the publicly-accepted empirical broken power-law function to fit the X-ray onset bumps, we derive the corresponding indices $\alpha_1$ and $\alpha_2$ for the temporal before and after the peak time $T_{\rm p}$, as well as the peak flux time $F_{\rm p}$. We find that the rise index $\alpha_1$ for most GRBs falls in the range of $0.3 - 2.6$, and that the central value of the decay index $\alpha_2$ is around -$1.0$. Interestingly, we find that the characteristic time scales of the X-ray onset bumps have tight correlations, which are the same as those in the optical onset bumps.

In the standard collapsar model, the afterglow is believed to originate in the external shock produced as the blast wave collides with and sweeps up material in the ambient interstellar medium. In this work, by applying the external FS model to the onset bumps of X-ray light curves, we identify the circumburst medium of our target GRB samples. A systematic investigation of the temporal and spectral shows that the center value of the new inferred $k$ is around 1.0, with a range from $0.2$ to $1.8$. This indicates that GRBs circumburst environment may be miscellaneous, neither the ISM nor the typical interstellar wind.

\section{Acknowledgments}
We thank the anonymous referee for the helpful comments and suggestions.
We also thank Yuan-Chuan Zou, Xue-Feng Wu, En-Wei Liang and Qing-Wen Tang for helpful discussions. This work was supported by the National Natural Science Foundation of China (Grant Nos. U2038106, U1931106), Ying Qin acknowledges the support from the Doctoral research start-up funding of Anhui Normal University and the funding from Key Laboratory for Relativistic Astrophysics in Guangxi University, Shandong Provincial Natural Science Foundation of China(grant No. ZR2019YQ03).

{}

\begin{thebibliography}{}

\bibitem[Atwood et al.(2009)]{2009ApJ...697.1071A} Atwood, W.~B., Abdo, A.~A., Ackermann, M., et al.\ 2009, \apj, 697, 1071. doi:10.1088/0004-637X/697/2/1071
\bibitem[Beuermann et al.(1999)]{1999A&A...352L..26B} Beuermann, K., Hessman, F.~V., Reinsch, K., et al.\ 1999, \aap, 352, L26
\bibitem[Burrows et al.(2005)]{2005SSRv..120..165B} Burrows, D.~N., Hill, J.~E., Nousek, J.~A., et al.\ 2005, \ssr, 120, 165. doi:10.1007/s11214-005-5097-2
\bibitem[Cantiello et al.(2007)]{2007A&A...465L..29C} Cantiello, M., Yoon, S.-C., Langer, N., et al.\ 2007, \aap, 465, L29. doi:10.1051/0004-6361:20077115
\bibitem[Dai \& Lu(1998)]{1998MNRAS.298...87D} Dai, Z.~G. \& Lu, T.\ 1998, \mnras, 298, 87. doi:10.1046/j.1365-8711.1998.01681.x
\bibitem[Du et al.(2021)]{2021ApJ...908..242D} Du, M., Yi, S.-X., Liu, T., et al.\ 2021, \apj, 908, 242. doi:10.3847/1538-4357/abd6bd
\bibitem[Detmers et al.(2008)]{2008A&A...484..831D} Detmers, R.~G., Langer, N., Podsiadlowski, P., et al.\ 2008, \aap, 484, 831. doi:10.1051/0004-6361:200809371
\bibitem[Gehrels et al.(2004)]{2004ApJ...611.1005G} Gehrels, N., Chincarini, G., Giommi, P., et al.\ 2004, \apj, 611, 1005. doi:10.1086/422091
\bibitem[Ghirlanda et al.(2012)]{2012MNRAS.420..483G} Ghirlanda, G., Nava, L., Ghisellini, G., et al.\ 2012, \mnras, 420, 483. doi:10.1111/j.1365-2966.2011.20053.x
\bibitem[Japelj et al.(2014)]{2014ApJ...785...84J} Japelj, J., Kopa{\v{c}}, D., Kobayashi, S., et al.\ 2014, \apj, 785, 84. doi:10.1088/0004-637X/785/2/84
\bibitem[Li et al.(2020b)]{2020ApJ...900..176L} Li, L., Wang, X.-G., Zheng, W., et al.\ 2020, \apj, 900, 176. doi:10.3847/1538-4357/aba757
\bibitem[Li et al.(2020a)]{2020SSPMA..50l9508L} Li, Y., Wen, X., Sun, X., et al.\ 2020, Scientia Sinica Physica, Mechanica \& Astronomica, 50, 129508. doi:10.1360/SSPMA-2019-0417
\bibitem[Liang et al.(2013)]{2013ApJ...774...13L} Liang, E.-W., Li, L., Gao, H., et al.\ 2013, \apj, 774, 13. doi:10.1088/0004-637X/774/1/13
\bibitem[Liang et al.(2010)]{2010ApJ...725.2209L} Liang, E.-W., Yi, S.-X., Zhang, J., et al.\ 2010, \apj, 725, 2209. doi:10.1088/0004-637X/725/2/2209
\bibitem[Liu et al.(2017)]{2017NewAR..79....1L} Liu, T., Gu, W.-M., \& Zhang, B.\ 2017, \nar, 79, 1. doi:10.1016/j.newar.2017.07.001
\bibitem[Liu et al.(2018)]{2018ApJ...852...20L} Liu, T., Song, C.-Y., Zhang, B., et al.\ 2018, \apj, 852, 20. doi:10.3847/1538-4357/aa9e4f
\bibitem[MacFadyen \& Woosley(1999)]{1999ApJ...524..262M} MacFadyen, A.~I. \& Woosley, S.~E.\ 1999, \apj, 524, 262. doi:10.1086/307790
\bibitem[Meegan et al.(2009)]{2009ApJ...702..791M} Meegan, C., Lichti, G., Bhat, P.~N., et al.\ 2009, \apj, 702, 791. doi:10.1088/0004-637X/702/1/791
\bibitem[Meszaros \& Rees(1993)]{1993ApJ...405..278M} Meszaros, P. \& Rees, M.~J.\ 1993, \apj, 405, 278. doi:10.1086/172360
\bibitem[M{\'e}sz{\'a}ros \& Rees(1997)]{1997ApJ...476..232M} M{\'e}sz{\'a}ros, P. \& Rees, M.~J.\ 1997, \apj, 476, 232. doi:10.1086/303625
\bibitem[M{\'e}sz{\'a}ros(2002)]{2002ARA&A..40..137M} M{\'e}sz{\'a}ros, P.\ 2002, \araa, 40, 137. doi:10.1146/annurev.astro.40.060401.093821
\bibitem[Molinari et al.(2007)]{2007A&A...469L..13M} Molinari, E., Vergani, S.~D., Malesani, D., et al.\ 2007, \aap, 469, L13. doi:10.1051/0004-6361:20077388
\bibitem[Nousek et al.(2006)]{2006ApJ...642..389N} Nousek, J.~A., Kouveliotou, C., Grupe, D., et al.\ 2006, \apj, 642, 389. doi:10.1086/500724
\bibitem[Oates et al.(2009)]{2009MNRAS.395..490O} Oates, S.~R., Page, M.~J., Schady, P., et al.\ 2009, \mnras, 395, 490. doi:10.1111/j.1365-2966.2009.14544.x
\bibitem[Panaitescu \& Kumar(2002)]{2002ApJ...571..779P} Panaitescu, A. \& Kumar, P.\ 2002, \apj, 571, 779. doi:10.1086/340094
\bibitem[Piran(2004)]{2004RvMP...76.1143P} Piran, T.\ 2004, Reviews of Modern Physics, 76, 1143. doi:10.1103/RevModPhys.76.1143
\bibitem[Qin et al.(2013)]{2013ApJ...763...15Q} Qin, Y., Liang, E.-W., Liang, Y.-F., et al.\ 2013, \apj, 763, 15. doi:10.1088/0004-637X/763/1/15
\bibitem[Qin et al.(2018)]{2018A&A...616A..28Q} Qin, Y., Fragos, T., Meynet, G., et al.\ 2018, \aap, 616, A28. doi:10.1051/0004-6361/201832839
\bibitem[Rees \& Meszaros(1994)]{1994ApJ...430L..93R} Rees, M.~J. \& Meszaros, P.\ 1994, \apjl, 430, L93. doi:10.1086/187446
\bibitem[Roming et al.(2005)]{2005SSRv..120...95R} Roming, P.~W.~A., Kennedy, T.~E., Mason, K.~O., et al.\ 2005, \ssr, 120, 95. doi:10.1007/s11214-005-5095-4
\bibitem[Rykoff et al.(2009)]{2009ApJ...702..489R} Rykoff, E.~S., Aharonian, F., Akerlof, C.~W., et al.\ 2009, \apj, 702, 489. doi:10.1088/0004-637X/702/1/489
\bibitem[Sari \& Piran(1999)]{1999ApJ...520..641S} Sari, R. \& Piran, T.\ 1999, \apj, 520, 641. doi:10.1086/307508
\bibitem[Sari et al.(1998)]{1998ApJ...497L..17S} Sari, R., Piran, T., \& Narayan, R.\ 1998, \apjl, 497, L17. doi:10.1086/311269
\bibitem[Si et al.(2018)]{2018ApJ...863...50S} Si, S.-K., Qi, Y.-Q., Xue, F.-X., et al.\ 2018, \apj, 863, 50. doi:10.3847/1538-4357/aad08a
\bibitem[Schulze et al.(2011)]{2011A&A...526A..23S} Schulze, S., Klose, S., Bj{\"o}rnsson, G., et al.\ 2011, \aap, 526, A23. doi:10.1051/0004-6361/201015581
\bibitem[Starling et al.(2008)]{2008ApJ...672..433S} Starling, R.~L.~C., van der Horst, A.~J., Rol, E., et al.\ 2008, \apj, 672, 433. doi:10.1086/521975
\bibitem[Sander \& Vink(2020)]{2020MNRAS.499..873S} Sander, A.~A.~C. \& Vink, J.~S.\ 2020, \mnras, 499, 873. doi:10.1093/mnras/staa2712
\bibitem[Wang et al.(2018)]{2018ApJ...859..160W} Wang, X.-G., Zhang, B., Liang, E.-W., et al.\ 2018, \apj, 859, 160. doi:10.3847/1538-4357/aabc13
\bibitem[Wang et al.(2020)]{2020ApJ...893...77W} Wang, F., Zou, Y.-C., Liu, F., et al.\ 2020, \apj, 893, 77. doi:10.3847/1538-4357/ab0a86
\bibitem[Woosley(1993)]{1993ApJ...405..273W} Woosley, S.~E.\ 1993, \apj, 405, 273. doi:10.1086/172359
\bibitem[Yi et al.(2020)]{2020ApJ...895...94Y} Yi, S.-X., Wu, X.-F., Zou, Y.-C., et al.\ 2020, \apj, 895, 94. doi:10.3847/1538-4357/ab8a53
\bibitem[Yi et al.(2016)]{2016ApJS..224...20Y} Yi, S.-X., Xi, S.-Q., Yu, H., et al.\ 2016, \apjs, 224, 20. doi:10.3847/0067-0049/224/2/20
\bibitem[Yi et al.(2013)]{2013ApJ...776..120Y} Yi, S.-X., Wu, X.-F., \& Dai, Z.-G.\ 2013, \apj, 776, 120. doi:10.1088/0004-637X/776/2/120
\bibitem[Yi et al.(2021a)]{2021MNRAS.507.1047Y} Yi, S.-X., Xie, W., Ma, S.-B., et al.\ 2021, \mnras, 507, 1047. doi:10.1093/mnras/stab2186
\bibitem[Yi et al.(2017)]{2017JHEAp..13....1Y} Yi, S.-X., Lei, W.-H., Zhang, B., et al.\ 2017, Journal of High Energy Astrophysics, 13, 1. doi:10.1016/j.jheap.2017.01.001
\bibitem[Yi et al.(2021b)]{2021arXiv211101041Y} Yi, S.-X., Du, M., \& Liu, T.\ 2021, arXiv:2111.01041
\bibitem[Zhang \& M{\'e}sz{\'a}ros(2004)]{2004IJMPA..19.2385Z} Zhang, B. \& M{\'e}sz{\'a}ros, P.\ 2004, International Journal of Modern Physics A, 19, 2385. doi:10.1142/S0217751X0401746X
\bibitem[Zhang et al.(2006)]{2006ApJ...642..354Z} Zhang, B., Fan, Y.~Z., Dyks, J., et al.\ 2006, \apj, 642, 354. doi:10.1086/500723
\bibitem[Zhang et al.(2020)]{2020SCPMA..6349502Z} Zhang, S.-N., Li, T., Lu, F., et al.\ 2020, Science China Physics, Mechanics, and Astronomy, 63, 249502. doi:10.1007/s11433-019-1432-6
\bibitem[Zhao et al.(2020)]{2020ApJ...900..112Z} Zhao, W., Zhang, J.-C., Zhang, Q.-X., et al.\ 2020, \apj, 900, 112. doi:10.3847/1538-4357/aba43a
\bibitem[Zhou et al.(2020)]{2020IJMPD..2950043Z} Zhou, Q.-Q., Yi, S.-X., Huang, X.-L., et al.\ 2020, International Journal of Modern Physics D, 29, 2050043. doi:10.1142/S0218271820500431


\end{thebibliography}
\end{document}